\documentclass{aastex63}

\received{April 3, 2020}
\revised{August 27, 2020}
\accepted{August 31, 2020}
\submitjournal{ApJ}

\shorttitle{Herschel Observations of Cepheus Flare Clouds}
\shortauthors{Di~Francesco et al.}
\graphicspath{{./}{figures/}}

\begin{document}

\title{Herschel Gould Belt Survey Observations of Dense Cores in the Cepheus Flare Clouds}

\correspondingauthor{James Di Francesco}
\email{james.difrancesco@nrc-cnrc.gc.ca}

\author[0000-0002-9289-2450]{James Di Francesco}
\affil{National Research Council of Canada, Herzberg Astronomy \& Astrophysics Research Centre, 5071 West Saanich Road, Victoria, BC V9E 2E7, Canada}

\author{Jared Keown}
\affiliation{University of Victoria, Department of Physics \& Astronomy, PO Box 1700 STN CSC, Victoria, BC V8W 2Y2, Canada}

\author{Cassandra Fallscheer}
\affiliation{Central Washington University, Department of Physics, 400 E. University Way, Ellensburg, WA 98926-7422, U.S.A.}

\author{Philippe Andr\'e}
\affiliation{Laboratoire AIM Paris-Saclay, CEA/IRFU CNRS/INSU Universit\'e Paris Diderot, 91191 Gif-sur-Yvette, France}

\author{Bilal Ladjelate}
\affiliation{Instituto de Radioastronomía Milim\'etrica, Avenida Divina Pastora, 7, Núcleo Central, E 18012 Granada, Spain}

\author{Vera K\"onyves}
\affiliation{Jeremiah Horrocks Institute, University of Central Lancashire, Preston, PR1 2HE, UK}

\author{Alexander Men'shchikov}
\affiliation{Laboratoire AIM Paris-Saclay, CEA/IRFU CNRS/INSU Universit\'e Paris Diderot, 91191 Gif-sur-Yvette, France}

\author{Shaun Stephens-Whale}
\affiliation{University of Victoria, Department of Physics \& Astronomy, PO Box 1700 STN CSC, Victoria, BC V8W 2Y2, Canada}

\author{Quang Nguyen-Luong}
\affiliation{Graduate School of Science, Nagoya City University, Mizuho-ku, Nagoya, 467-8501, Japan}

\author{Peter Martin}
\affiliation{University of Toronto, Department of Astronomy \& Astrophysics, 50 St. George Street, Toronto, ON M5S 3H4, Canada}

\author{Sarah Sadavoy}
\affiliation{Queen's University, Department of Physics, Engineering and Astronomy, 64 Bader Lane, Kingston, ON, K7L 3N6, Canada}

\author{Stefano Pezzuto}
\affiliation{Istituto di Astrofisica e Planetologia Spaziali, via del Fosso del Cavaliere 100, 00133 Roma, Italy}

\author{Eleonora Fiorellino}
\affiliation{Istituto di Astrofisica e Planetologia Spaziali, via del Fosso del Cavaliere 100, 00133 Roma, Italy}

\author{Milena Benedettini}
\affiliation{Istituto di Astrofisica e Planetologia Spaziali, via del Fosso del Cavaliere 100, 00133 Roma, Italy}

\author{Nicola Schneider}
\affiliation{I. Physikalisches Institut, Universität zu Köln, Zülpicher Str. 77, 50937, Köln, Germany}
\affiliation{OASU/LAB, Universit\'e de Bordeaux, 33615, Pessac, France}

\author{Sylvain Bontemps}
\affiliation{OASU/LAB, Universit\'e de Bordeaux, 33615, Pessac, France}

\author{Doris Arzoumanian}
\affiliation{Instituto de Astrofisica e Ciencias do Espaco, Universidade do Porto, CAUP, Rua das Estrelas, PT4150-762, Porto, Portugal}

\author{Pedro Palmeirim}
\affiliation{Instituto de Astrofisica e Ciencias do Espaco, Universidade do Porto, CAUP, Rua das Estrelas, PT4150-762, Porto, Portugal}

\author{Jason M. Kirk}
\affiliation{Jeremiah Horrocks Institute, University of Central Lancashire, Preston, PR1 2HE, UK}

\author{Derek Ward-Thompson}
\affiliation{Jeremiah Horrocks Institute, University of Central Lancashire, Preston, PR1 2HE, UK}






\begin{abstract}
We present {\it Herschel} SPIRE and PACS maps of the Cepheus Flare clouds L1157, L1172, L1228, L1241, and L1251, observed by the {\it Herschel} Gould Belt Survey (HGBS) of nearby star-forming molecular clouds.  Through modified blackbody fits to the SPIRE and PACS data, we determine typical cloud column densities of 0.5-1.0 $\times$ 10$^{21}$ cm$^{-2}$ and typical cloud temperatures of 14-15 K.  Using the {\it getsources} identification algorithm, we extract 832 dense cores from the SPIRE and PACS data at 160-500 $\mu$m.  From placement in a mass vs.\ size diagram, we consider 303 to be candidate prestellar cores, and 178 of these to be ``robust" prestellar cores.  From an independent extraction of sources at 70 $\mu$m, we consider 25 of the 832 dense cores to be protostellar.  The distribution of background column densities coincident with candidate prestellar cores peaks at 2-4 $\times$ 10$^{21}$ cm$^{-2}$.  About half of the candidate prestellar cores in Cepheus may have formed due to the widespread fragmentation expected to occur within filaments of ``transcritical" line mass.  The lognormal robust prestellar core mass function (CMF) drawn from all five Cepheus clouds peaks at 0.56 M$_{\odot}$ and has a width of $\sim$0.5 dex, similar to that of Aquila's CMF.  Indeed, the width of Cepheus' aggregate CMF is similar to the stellar system Initial Mass Function (IMF).  The similarity of CMF widths in different clouds and the system IMF suggests a common, possibly turbulent origin for seeding the fluctuations that evolve into prestellar cores and stars.
\end{abstract}


~

~


\section{Introduction}\label{sec:intro}


Understanding the process of star formation is a major cornerstone of
modern astrophysics.  Stars form primarily in giant molecular clouds, when 
gas over-densities called ``dense cores" become unstable to gravitational
collapse.  Indeed, the process of star formation is likely inextricably linked
to the process of dense core formation in molecular clouds.  Dense cores are
generally 0.1 pc in size or less, 10 K in temperature, and 10$^{4-5}$ cm$^{-3}$
in density.  Dense cores that have not formed a young stellar object (YSO) and
are arguably not bound by their own gravity are called ``starless cores.'' Those
dense cores that are arguably bound and hence more likely to collapse are called
``prestellar cores" \citep[cf.][]{DiFrancesco07,Ward-Thompson07}.  By
examining the populations of starless and prestellar cores in molecular
clouds, we can gain insight into how star formation is proceeding in those
clouds, i.e., the potential yield of new stars into the Galaxy.

The {\it Herschel} Gould Belt Survey \citep[HGBS;][]{Andre10} is a key program
to use the PACS and SPIRE continuum instruments of the ESA {\it Herschel Space 
Observatory} to map nearby molecular clouds within 500 pc distance and identify
their populations of cores via emission from dust mixed with the dense gas.
Given the low temperatures of cores, they emit most brightly at the far-infrared
and submillimeter wavelengths {\it Herschel} was designed to observe.  Indeed,
the location of {\it Herschel} at the Sun-Earth L2 point allowed it to map the
emission from such wavelengths without a bright and highly opaque atmosphere
being simultaneously observed.  The resulting {\it Herschel} maps have been
unprecedented in their sensitivity to faint, diffuse emission from cold dust
in molecular clouds.

Key results from the HGBS so far include the identification of the ubiquity of
filamentary substructure in molecular clouds, the close association of prestellar 
cores in some clouds with filaments of average column density greater than a 
fiducial threshold or transition of $\sim$7 $\times$ 10$^{21}$ cm$^{-2}$, and
the universality of the lognormal morphology of the prestellar core mass function 
\citep{Andre14}.  Much of these results have come from observation of the relatively
active Aquila Rift star-forming cloud \citep[]{Andre10,Konyves15}.  The
Gould Belt consists of a variety of star-forming clouds, however, and it is
important to explore star formation in different environments to gauge how
universal the early findings of the HGBS are.

In this paper, we present the HGBS observations of the Cepheus Flare clouds,
L1157, L1172/74, L1228, L1241, and L1251.  These clouds are located in a loose
association of compact dark clouds scattered across $\sim$10\degr\ of sky at
high declination ($\sim$68-78\degr).  These Cepheus clouds were selected to be
part of the HGBS given their known star-formation activity \citep[see][for a review]{Kun08}
and relatively close distances to the Sun.  L1172/74 (hereafter,
L1172) in particular is home to the bright NGC 7023 nebula (a.k.a. the Iris
Nebula) that is illuminated by the Herbig Ae star HD200775.  L1241 and L1251
lie within the Cepheus Flare Shell, an expanding supernova bubble about 10\degr\
in radius that may have enhanced star formation in those clouds.
Indeed, L1228 may be coincident with the edge of the Cepheus Flare Shell itself.
L1157 and L1172,  however, appear to be exterior to the Shell.  At the time of
their selection, the Cepheus clouds were estimated to be 200-300 pc distant.
\cite{Dzib18}, however, recently used GAIA DR2 data to determine a new distance
of 358 pc $\pm$ 32 pc to the Cepheus Flare, which we adopt for all five clouds
examined in this paper.

\cite{Kirk09} presented the results of the {\it Spitzer} Gould Belt Survey
observations of several clouds in the Cepheus Flare, including L1172, L1228,
L1241, and L1251 but not L1157.  The {\it Spitzer} data from the near- to 
mid-infrared IRAC and mid-infrared MIPS instruments largely sampled the
more-evolved YSO population of the Cepheus clouds, i.e., Class I, Flat, II
(T-Tauri), and III objects.  Notably, 93 YSOs were found in the L1172, L1228,
and L1251 clouds in total.  Beyond a single Class III object, L1241 was found
to be without YSOs down to a limit of 0.06 L$_{\odot}$.  
More recently, \cite{Pattle17} presented the results of the JCMT Gould Belt
Survey of the Cepheus clouds, including L1172, L1228, and L1251 but not L1157
or L1241.  This survey included SCUBA-2 observations of high column density
regions at 850 $\mu$m, and largely sampled the most compact, cold structures
in these clouds, such as prestellar cores and Class 0 objects.  From ratios
of numbers of starless cores to Class II objects, \cite{Pattle17} suggested 
that low-ratio L1228  was a less active star-forming cloud while high-ratio
L1172 and L1251 were more active.

With {\it Herschel} data, we have access to far-infrared/submillimeter 
emission from the Cepheus clouds of high sensitivity and resolution over a
wide range of spatial scales, sampling both faint and diffuse and bright and
compact emission sources.  With such capability, we can examine filamentary
structure in these clouds previously undetected from ground-based emission
or extinction map studies.  Moreover, we can provide a census of the starless
cores, prestellar cores, and protostellar cores in these five clouds, and
corresponding catalogues of their observed and physical properties.  Moreover,
given the lower column densities previously estimated for these clouds, these 
observations provide a counterpoint to observations of more active star-forming 
clouds in the Gould Belt.  A recent analysis of other lower column density 
star-forming clouds in Lupus was recently presented by \cite{Benedettini18}.

This paper is organized as follows: in $\S$2, we describe the observations
and data reduction performed on the {\it Herschel} data; in $\S$3, we present
the results of the data, including column density
and temperature maps, and source extractions; in $\S$4, we discuss core formation
in the low column density regime of the Cepheus clouds, and discuss the core
mass functions of the Cepheus clouds as a whole and separately; in $\S$5, we
provide a summary and conclusions.  The paper also contains three Appendices:
in A, we provide the images of all five clouds at 70 $\mu$m, 160 $\mu$m, 250 $\mu$m, 
350 $\mu$m and 500 $\mu$m; in B, we provide the criteria applied to sources
detected by the {\it getsources} algorithm for reliability; and in C, we list
the information provided in online material, including catalogues of the 
observed and derived physical properties of all dense cores identified in the
Cepheus clouds studied and thumbnail images of each core at 70-500 $\mu$m and
in H$_{2}$ column density.

\begin{figure}
\plotone{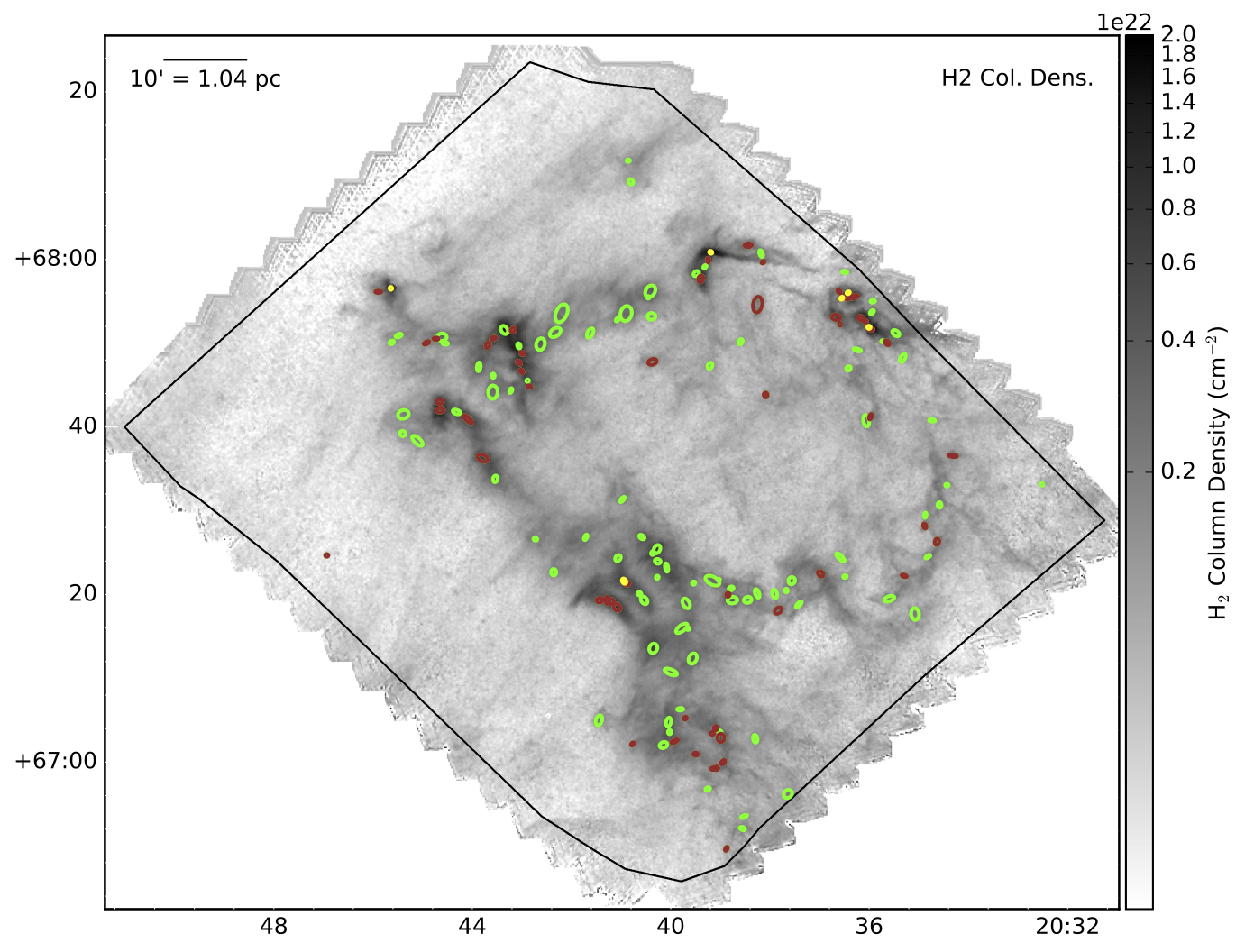}
\caption{{\it Herschel}\/-derived high-resolution H$_{2}$ column densities for L1157.  Column densities are shown on a logarithmic scale between 2 $\times$ 10$^{21}$ cm$^{-2}$ and 2 $\times$ 10$^{23}$ cm$^{-2}$.  Ellipses correspond to cores identified in the cloud via {\it getsources}.  Green, red, and yellow ellipses are cores we identify as ``starless cores," ``candidate prestellar cores," and ``protostellar cores," respectively (see $\S$3.2).  The size of an ellipse corresponds to the measured extent of its respective core.  The solid black border excludes noisy map edges and delineates the region in the map over which statistics are calculated.
\label{fig:fig1}}
\end{figure}

\begin{figure}
\plotone{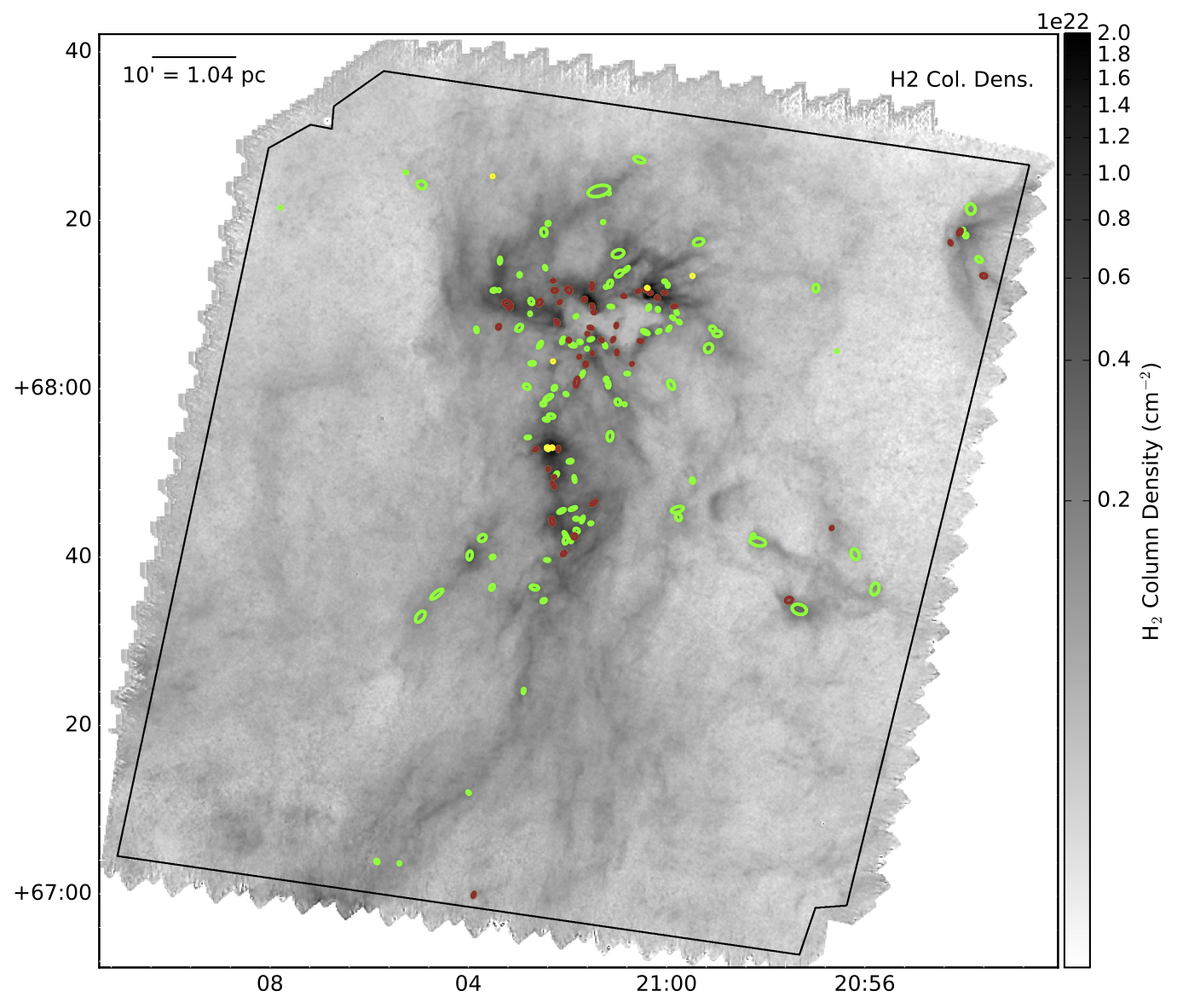}
\caption{{\it Herschel}\/-derived high-resolution H$_{2}$ column densities for L1172.  The greyscale range and symbols are defined as for Figure \ref{fig:fig1}.
\label{fig:fig2}}
\end{figure}

\begin{figure}
\plotone{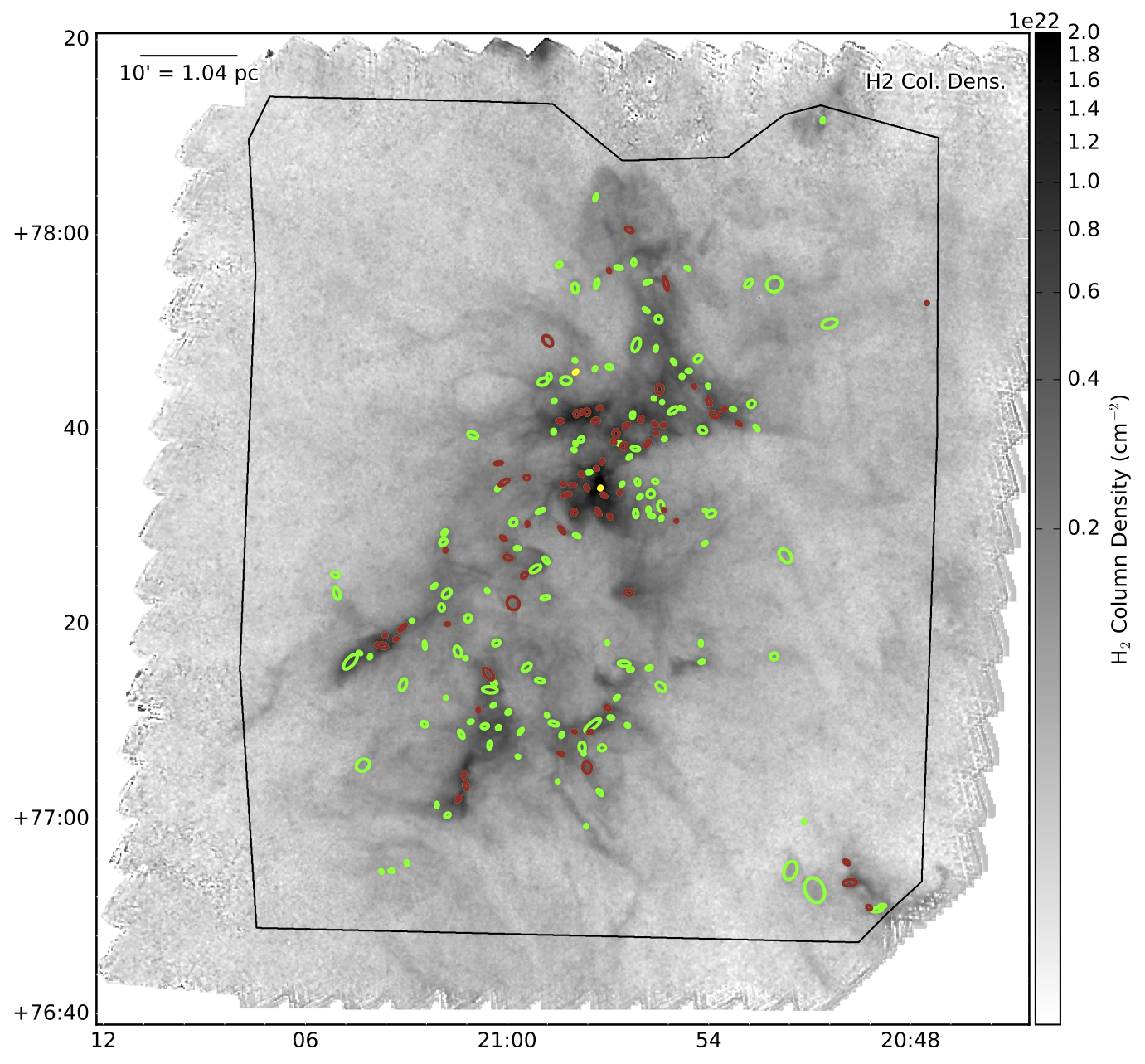}
\caption{{\it Herschel}\/-derived high-resolution H$_{2}$ column densities for L1228.  The greyscale range and symbols are defined as for Figure \ref{fig:fig1}.
\label{fig:fig3}}
\end{figure}

\begin{figure}
\plotone{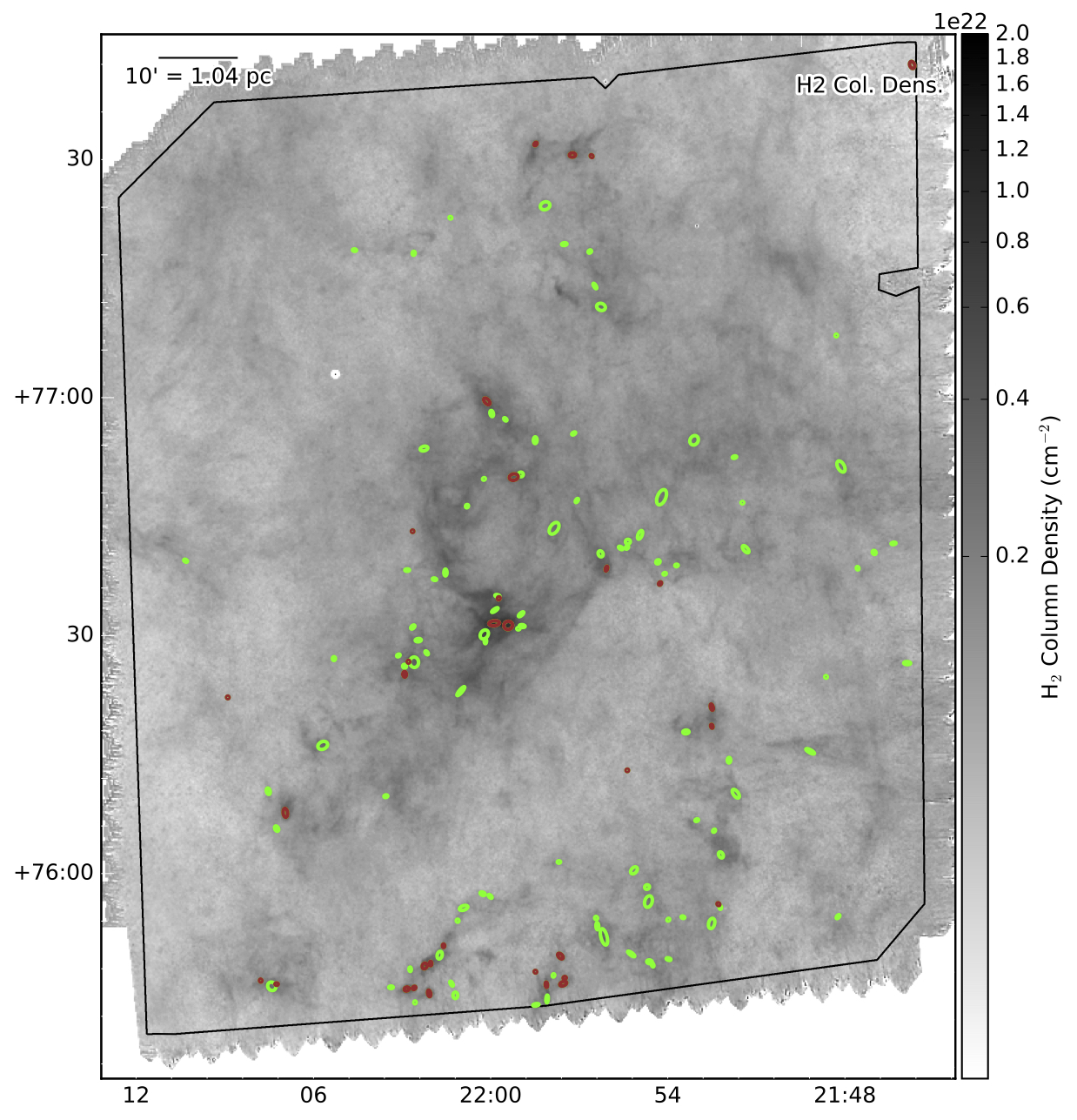}
\caption{{\it Herschel}\/-derived high-resolution H$_{2}$ column densities for L1241.  The greyscale range and symbols are defined as for Figure \ref{fig:fig1}.
\label{fig:fig4}}
\end{figure}

\begin{figure}
\plotone{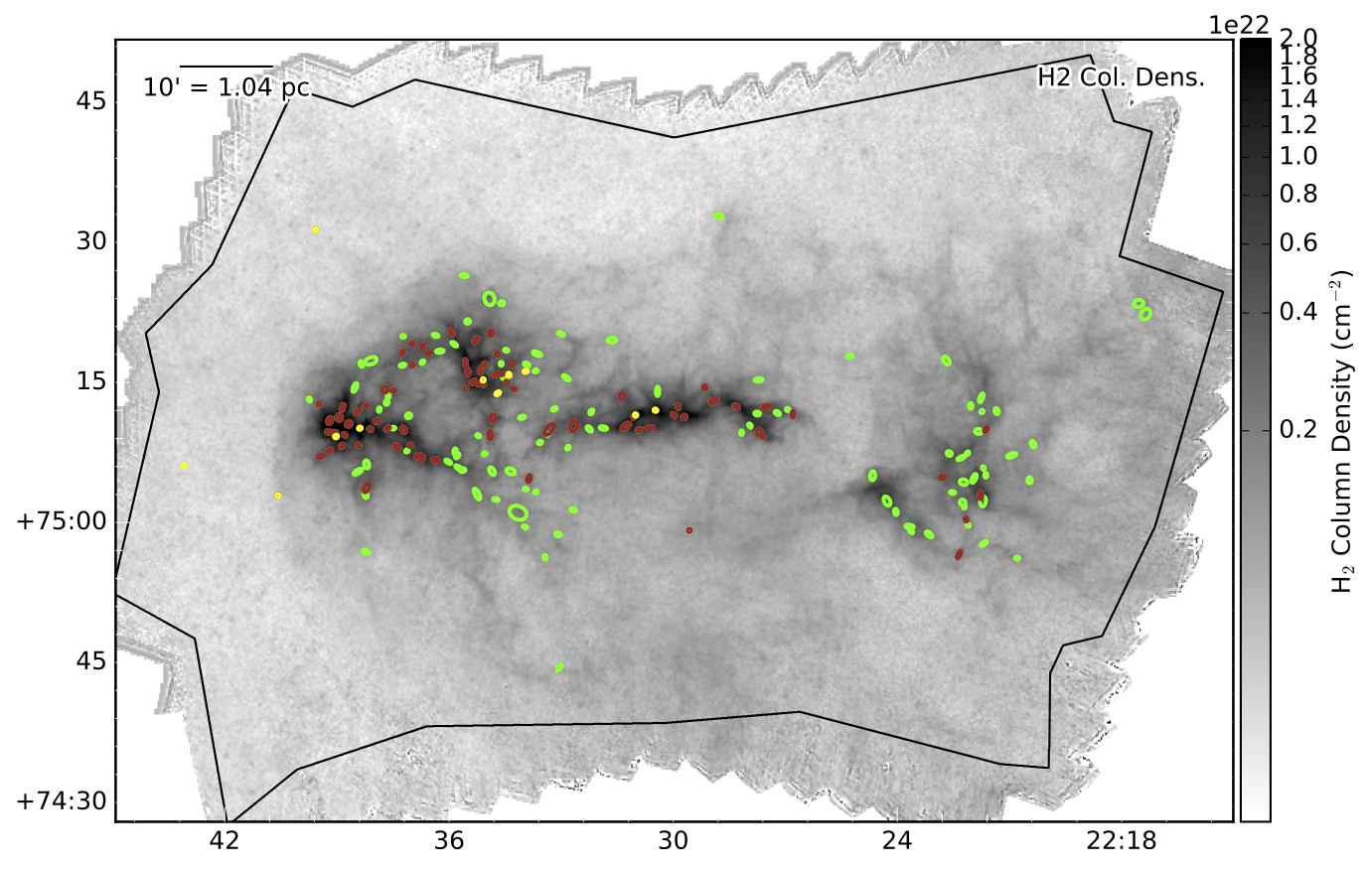}
\caption{{\it Herschel}\/-derived high-resolution H$_{2}$ column densities for L1251.  The greyscale range and symbols are defined as for Figure \ref{fig:fig1}.
\label{fig:fig5}}
\end{figure}

\section{Observations and Data Reduction} \label{sec:obs}

Each Cepheus field was observed simultaneously with the {\it Herschel}
Space Observatory's Spectral and Photometric Imaging Receiver (SPIRE) 
and Photodetector Array Camera and Spectrometer (PACS) instruments in
parallel mode, as part of the {\it Herschel}\/ Gould Belt Survey (HGBS)
Key Program \citep{Andre10}\footnote{For information on the HGBS,
see http://gouldbelt-herschel.cea.fr.}.  Further information on {\it
Herschel}, SPIRE, and PACS can be found in papers by \cite{Pilbratt10}, \cite{Griffin10}, and \cite{Poglitsch10}, respectively.  Table 1 summarizes
the {\it Herschel}\/ observations of the five Cepheus fields, giving the
name of each field, the J2000 coordinates of the reference position of
each map, the observational date (in UT), and the Observation IDs (OBSIDs),
respectively. Each field was observed twice, with scans made in roughly
perpendicular position angles, at a scanning speed of 60$\arcsec$~s$^{-1}$.
Fortuitously, L1251 was actually observed four times, i.e., two pairs of
orthogonal scans.

The {\it Herschel} data were reduced using HIPE \citep{Ott11}, following
the standard HGBS ``first generation catalogue" prescriptions.  We refer
readers to the first HGBS data catalogue paper \citep{Konyves15} for details
on the data reduction steps.  (See also other recent HGBS data catalogue
papers by \cite{Marsh16}, \cite{Bresnahan18}, \cite{Benedettini18},
\cite{Konyves20}, and Ladjelate et al.\/ (2020, in press).)  We provide a brief sketch of the steps below.

For the PACS data at 70 $\mu$m and 160 $\mu$m, HIPE version 9.0.3063 was used.
After standard steps of masking bad pixels, applying flat-field corrections 
and non-linear responsivity corrections to the data, deglitching cosmic ray 
hits, and applying a high-pass filter of scan-leg length, the PACS images were 
produced with the IDL-based map-maker, Scanamorphos version 20 
\citep{Roussel13}. As in \cite{Konyves15}, the absolute point source flux 
accuracies in the PACS images are 3$\%$ at 70 $\mu$m and $<$5$\%$ at 160 
$\mu$m, with the extended source calibration flux accuracies being 
uncertain.  We also adopt here the more conservative absolute calibration 
uncertainties of 10$\%$ and 20$\%$ for integrated source flux densities at 70 
$\mu$m and 160 $\mu$m, respectively. 

\startlongtable
\begin{deluxetable*}{cccc}
\tablecaption{Details of {\it Herschel}\/ Observations\label{tab:table}}
\tablehead{
\colhead{Field} & \colhead{Reference Coordinates} & \colhead{Observation 
Date} & \colhead{OBSID}\\
\colhead{} & \colhead{(J2000)} & \colhead{} & \colhead{}
}
\startdata
L1157 &  310.260070672, +67.7461347933 & 2010-01-28 & 1342189843, 1342189844 \\
L1172 &  315.713555697. +67.9081656174 & 2009-12-28 & 1342188652, 1342188653 \\
L1228 &  314.040903913, +77.4053720025 & 2010-06-21 & 1342198861, 1342198862 \\
L1241 &  330.152030576, +76.8345462465 & 2009-12-28 & 1342188679, 1342186680 \\
L1251 &  338.041909837, +75.2584297386 & 2009-12-28 & 1342188654, 1342186655 \\
L1251 &  338.041909837, +75.2584297386 & 2010-01-25 & 1342189663, 1342189664 \\
\enddata
\end{deluxetable*}

For the SPIRE data at 250 $\mu$m, 350 $\mu$m, and 500 $\mu$m, HIPE version 
10.0.2751 was used.  The data were calibrated and appropriate corrections made
to account for several issues, including electrical crosstalk, temperature 
drifts, cosmic ray hits, ``cooler burps," and relative gain factors between
bolometers.  A na\"ive mapmaker in HIPE was used to produce preliminary images
that were destriped.  Subsequent iterations of offset fitting and subtraction
and destriping were performed until convergence was reached.  As in
\cite{Konyves15}, the absolute flux accuracy is considered to be $<$5\%
for point sources \citep[cf.][]{Bendo13} and $<$10\% for extended sources
\citep[cf.][]{Griffin13} in the three SPIRE bands.

Highly smoothed versions of the {\it Herschel} images at each wavelength were
compared with images made by extrapolating low-resolution data of the same sky 
locations obtained with the ESA {\it Planck}\/ observatory to the {\it Herschel}
wavelengths after adopting a dust model \citep{Bernard10}.  This comparison
allowed appropriate values of the background emission at each wavelength not
included in the {\it Herschel} data to be determined.   

The intensities in all {\it Herschel}\/ PACS/SPIRE maps were converted to MJy 
sr$^{-1}$ and reprojected onto a common grid of 3$\arcsec$ $\times$ 3$\arcsec$ 
pixels.  Following extensive tests conducted by \cite{Konyves15} on the HGBS data
of Aquila, we expect the absolute astrometric accuracy to be $<$3$\arcsec$.  The
half-power beam width (HPBW) resolutions of the maps are 8$\farcs$4, 13$\farcs$5, 
18$\farcs$2, 24$\farcs$9, and 36$\farcs$2 at 70 $\mu$m, 160 $\mu$m, 250 $\mu$m,
350 $\mu$m, and 500 $\mu$m, respectively.  The pixel sizes of the final images
at each wavelength have been set to 3$\arcsec$, 3$\arcsec$, 6$\arcsec$, 10$\arcsec$,
and 14$\arcsec$, respectively.  The HGBS PACS and SPIRE maps of the Cepheus Flare
fields data are publicly available on the HGBS Archive (http://gouldbelt-herschel.cea.fr/archives)
in standard FITS format.  Note that {\it Planck}-derived offsets have not
been added to these files.  Instead, these offsets can be found in the
respective header of each map.

\section{Results} \label{sec:results}

\subsection{Column Densities and Temperatures}

\begin{figure*}
\gridline{\fig{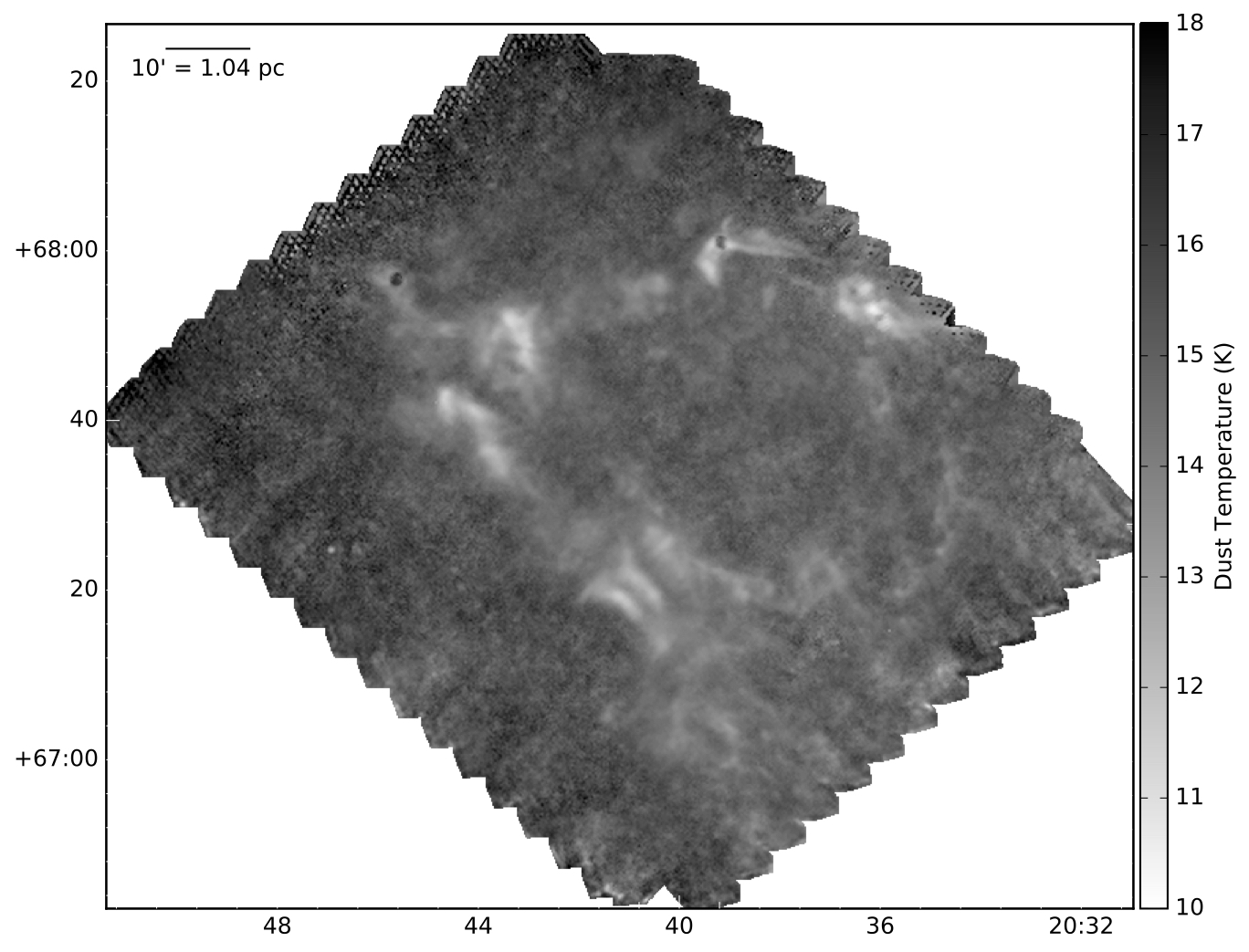}{0.4\textwidth}{(a)}
          \fig{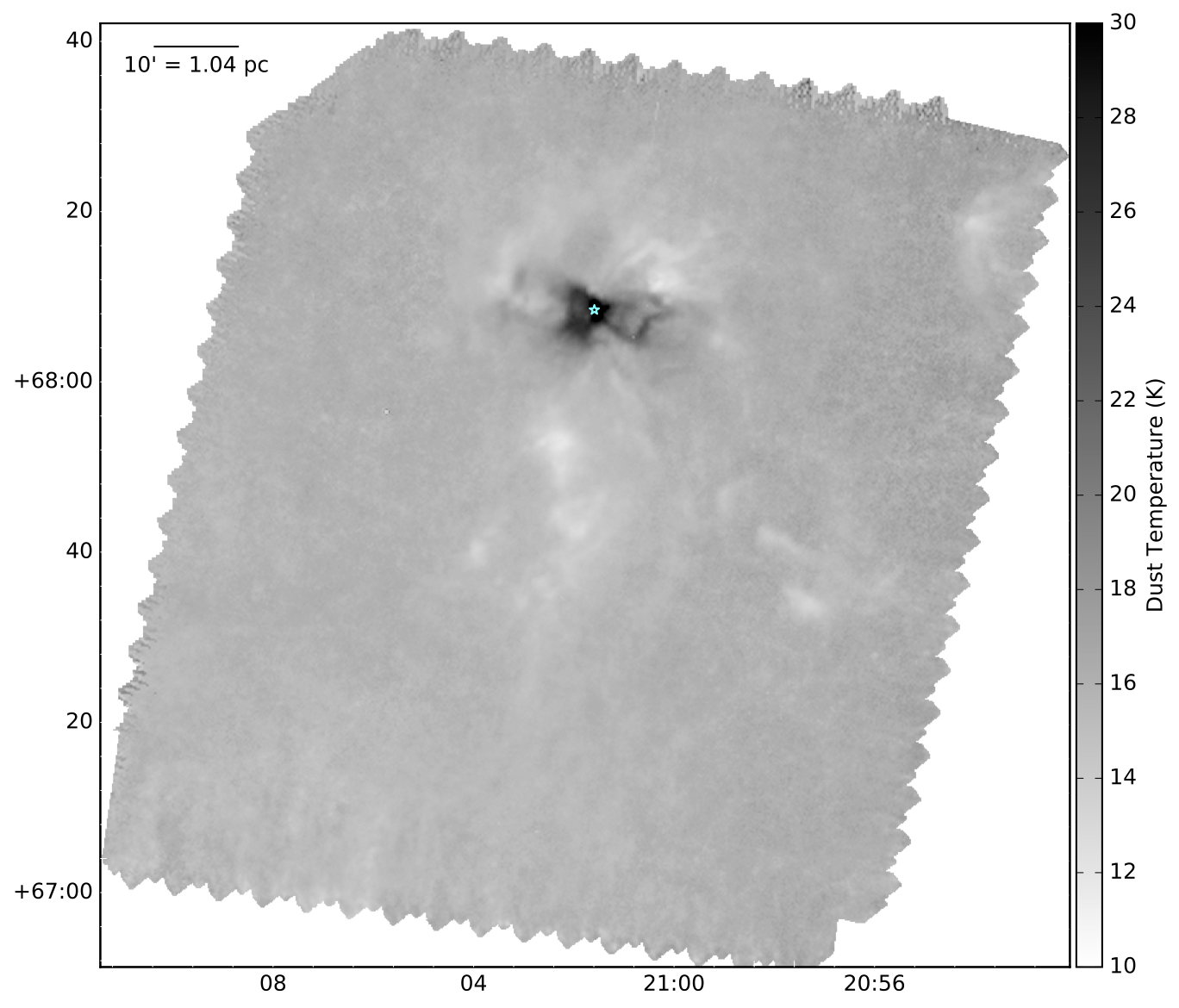}{0.4\textwidth}{(b)}
          }
\gridline{\fig{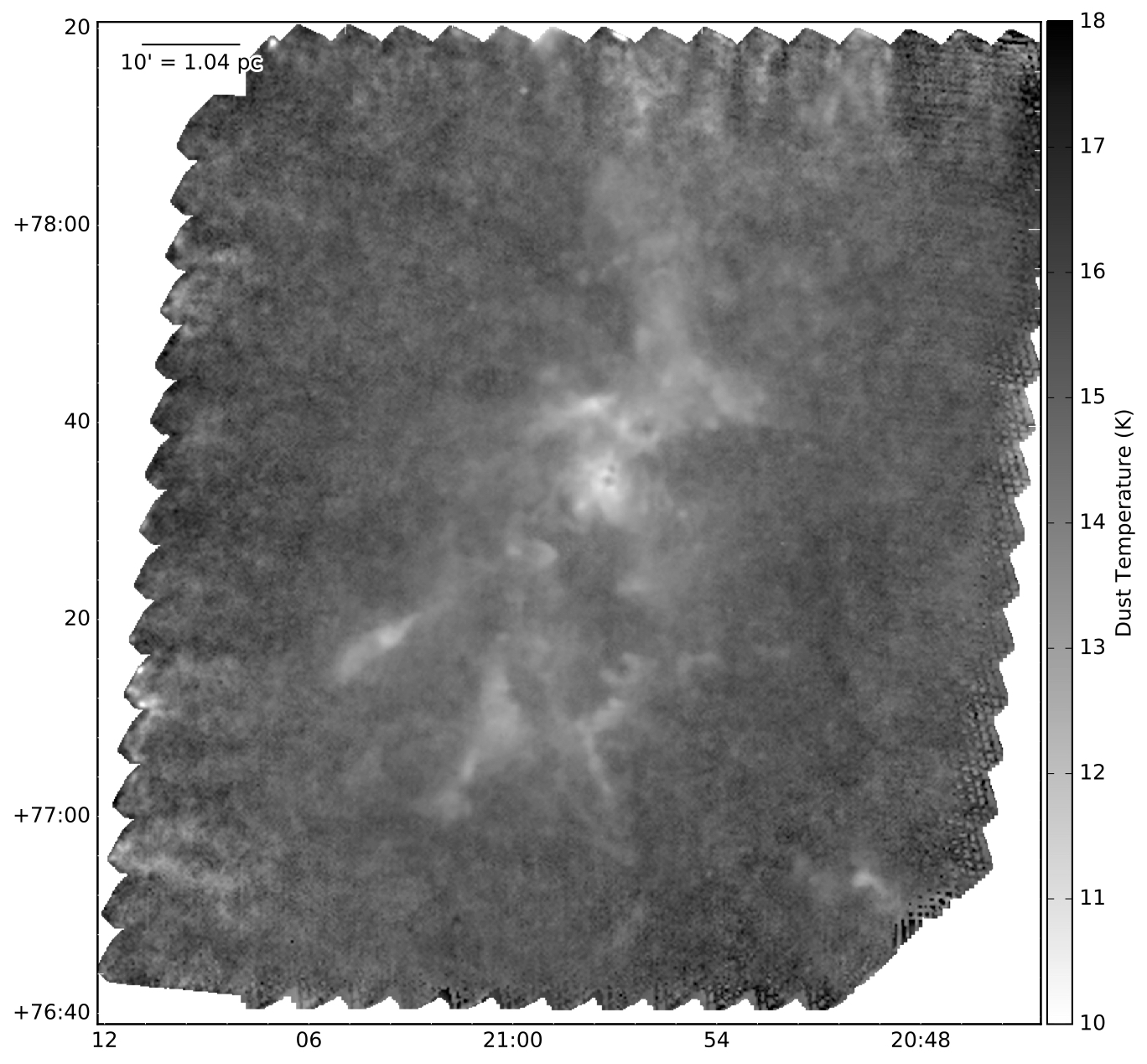}{0.4\textwidth}{(c)}
          \fig{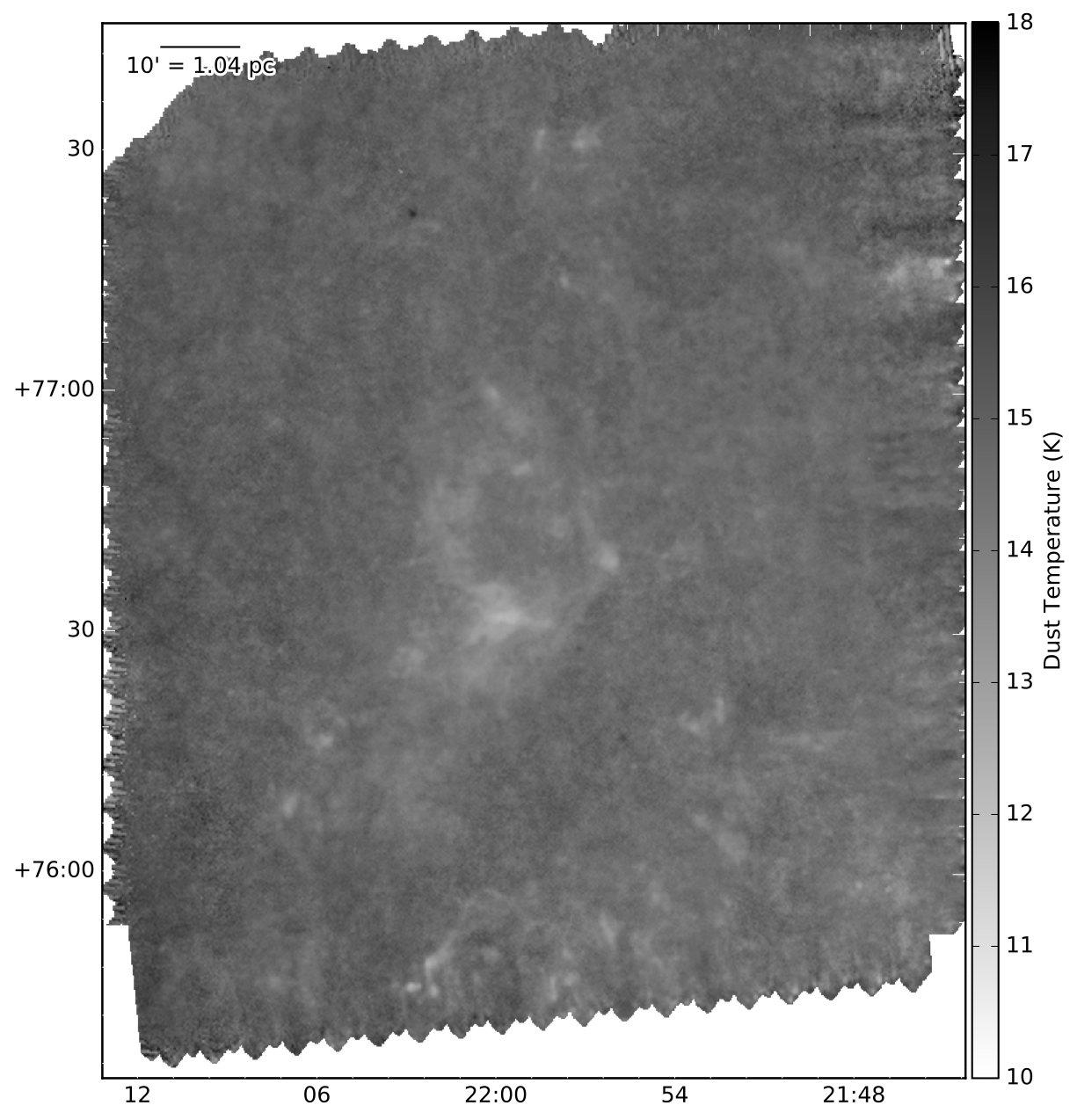}{0.4\textwidth}{(d)}
          }
\gridline{\fig{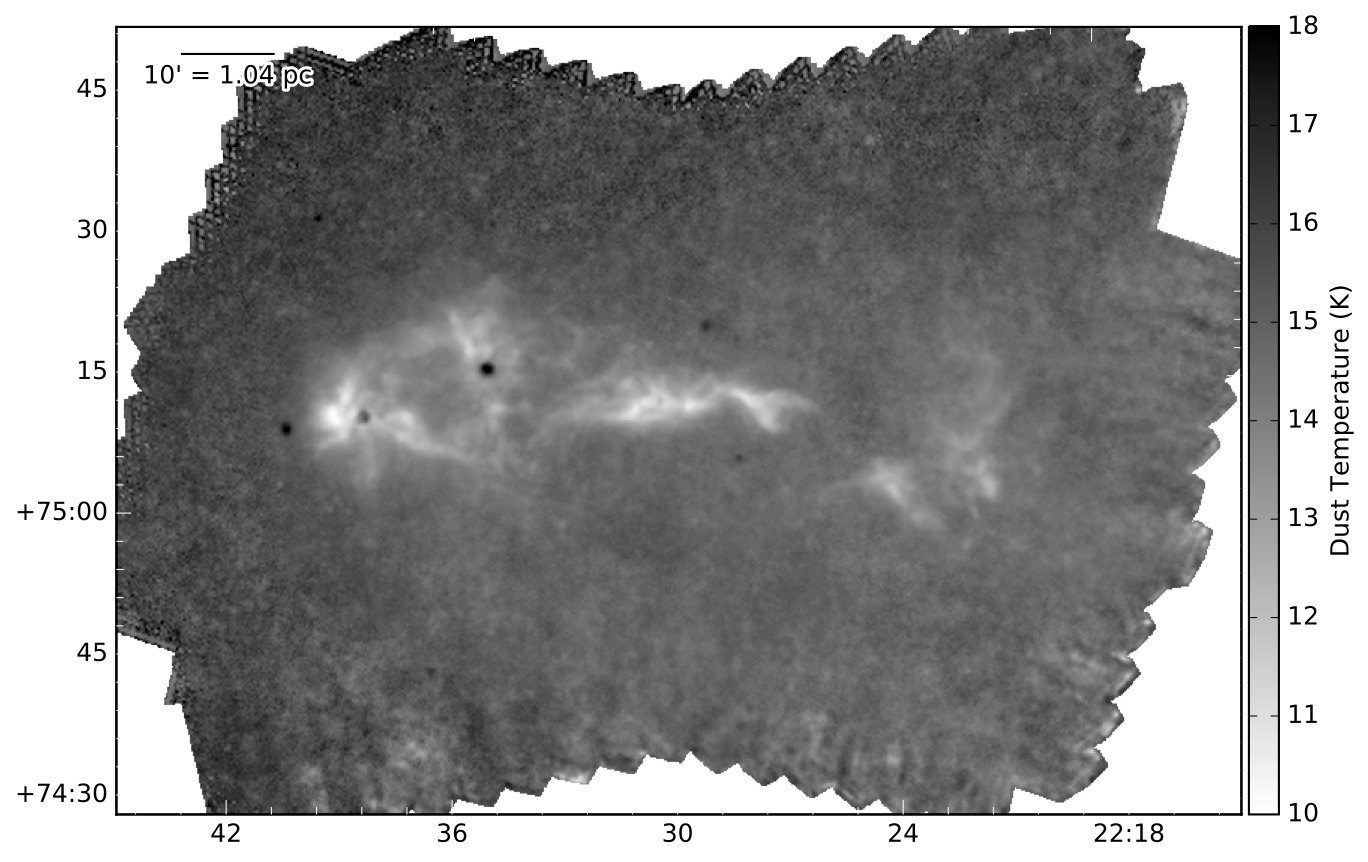}{0.4\textwidth}{(e)}
          }
\caption{{\it Herschel}\/-derived temperatures for (a) L1157, (b) 1172, (c) L1228, (d) L1241, and (e) L1251. The temperatures are shown on a linear scale between 10~K and 18~K, except for L1172 which scale linearly between 10~K and 30~K.  The cyan star shown in panel {\it b} denotes the position of the Herbig Ae star HD 200775.}
\label{fig:fig6}
\end{figure*}

Figures \ref{fig:fig1}-\ref{fig:fig5} show H$_{2}$ column density maps
of L1157, L1172, L1228, L1241, and L1251, respectively, obtained by
fitting the {\it Herschel}\/ spectral energy distributions (SEDs) 
of each pixel (adding {\it Planck} offsets at each wavelength) over 
a wavelength range of 160 $\mu$m to 500 $\mu$m with a modified blackbody
function.  The maps were produced at a resolution of 18$\farcs$2 following
the ``high-resolution" method described in Appendix A of \cite{Palmeirim13}
that is standard for HGBS catalogue papers.  The modified blackbody used
in the SED fitting includes a dust opacity $\kappa_{\nu}$ = 0.144 cm$^{2}$
g$^{-1}$ at 250~$\mu$m (incorporating a dust-to-gas ratio of 100) with a
power-law dependence with wavelength of index $\beta$ = 2.0 
\citep{Hildebrand83}.  In addition, the same mean molecular weight per
H$_{2}$ molecule of $\mu$ = 2.8 is assumed here \citep{Kauffmann08} to
convert gas surface density into H$_{2}$ column density.  (To determine
the isothermal sound speed, $\mu$ = 2.33 is assumed.)  Figure \ref{fig:fig6}
shows the dust temperature maps for each field obtained simultaneously from
the fitting of modified blackbodies to the multi-wavelength {\it Herschel}
data. In Appendix A, we present the actual {\it Herschel} images of each
field at 70~$\mu$m, 160~$\mu$m, 250~$\mu$m, 350~$\mu$m, and 500~$\mu$m
in their native resolutions and without the {\it Planck} offsets added.

The column densities seen in each map are typical of those seen in
other HGBS fields of nearby star-forming molecular clouds, e.g., non-zero
column densities are found in every pixel.  Similarly, each Cepheus cloud
field displays a multitude of substructure on many scales, including
compact knots amidst longer filaments. 

\startlongtable
\begin{deluxetable*}{ccccc}
\tablecaption{Median and Standard Deviation Values of Column Density and Temperature in Cepheus Flare clouds\label{tab:tab2}}
\tablehead{
& & \colhead{Standard} & & \colhead{Standard}\\
& \colhead{Median} & \colhead{Deviation of} & \colhead{Median} & \colhead{Deviation of}\\ 
\colhead{Field} & \colhead{Column Density} & \colhead{Column Density} & \colhead{Temperature} & \colhead{Temperature}\\
& \colhead{(cm$^2$)} & \colhead{(cm$^2$)} & \colhead{(K)} & \colhead{(K)}
}
\startdata
L1157 & 6.02E+20 & 9.60E+20 & 14.2 & 0.56 \\
L1172 & 7.14E+20 & 9.09E+20 & 14.7 & 0.85 \\
L1228 & 6.37E+20 & 1.06E+21 & 14.0 & 0.33 \\
L1241 & 9.95E+20 & 5.40E+20 & 14.3 & 0.54 \\
L1251 & 6.36E+20 & 1.55E+21 & 14.3 & 0.58 \\
\enddata
\end{deluxetable*}

\begin{figure*}
\gridline{\fig{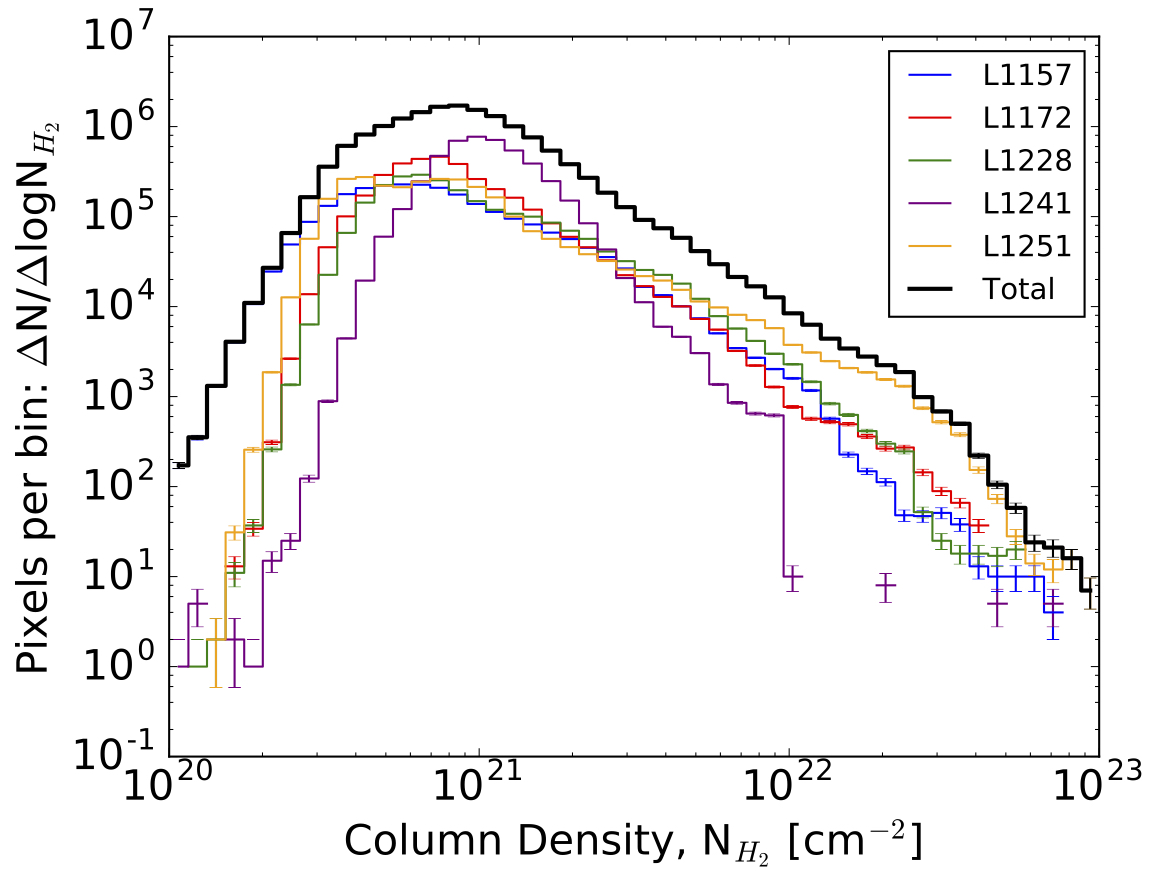}{0.5\textwidth}{(a)}
          \fig{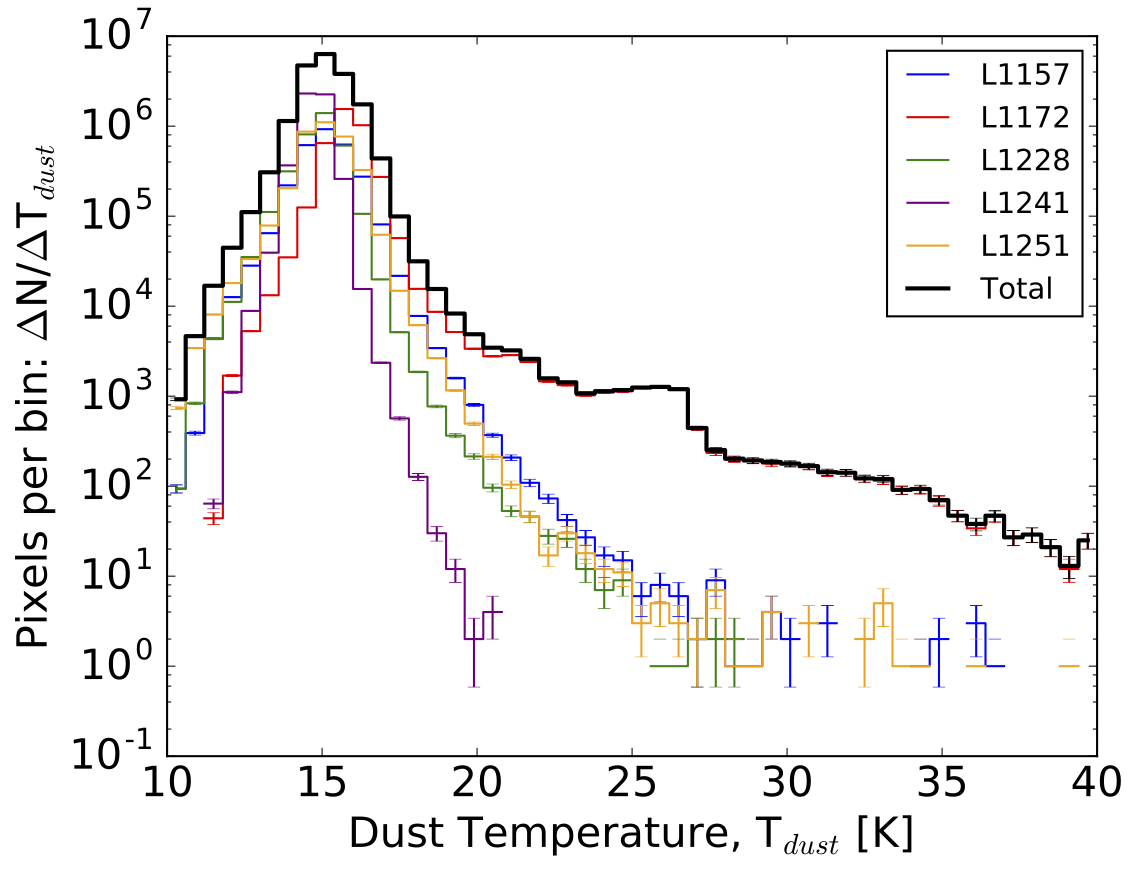}{0.5\textwidth}{(b)}
          }
\caption{(a) Log-log histogram of {\it Herschel}\/-derived column densities
and (b) log-linear histogram of {\it Herschel}\/-derived temperatures for
L1157 (blue), L1172 (red), L1228 (green), L1241 (purple), and L1251 (yellow)
and all five clouds together (black lines).}
\label{fig:fig7}
\end{figure*}

Figure \ref{fig:fig7} shows histograms of the column densities and temperatures
of each field, i.e., their respective probability density functions (PDFs).
(Note that the column density PDFs are constructed from all data shown in Figures
\ref{fig:fig1}-\ref{fig:fig5} and not only in the last closed contour, as shown
recently by \cite{Soler19}.)  Table \ref{tab:tab2} lists the median column density
and median temperature of each field shown in Figures \ref{fig:fig1}-\ref{fig:fig6},
with associated standard deviations.  In terms
of column density, L1157, L1172, and L1228 have very similar PDFs, peaking at 6-7
$\times$ 10$^{20}$ cm$^{-2}$ and decreasing to smaller and larger column densities
at similar rates.  Interestingly, L1241 has column densities that peak at a slightly
larger value, about twice that of the others, but its PDF falls off faster at smaller
and larger column densities than those of L1157, L1172, and L1228 do.   Indeed, the width of
L1241's column density PDF is about half that observed in other clouds, and no column
densities above 10$^{22}$ cm$^{-2}$ are found.  In addition, the column
densities of L1251 peak at a similar value to that of L1157, L1172, and L1228 but its
PDF falls off slightly more slowly at high column densities.  In terms of temperature,
all five clouds have distributions that peak at around 14 K.  The temperature PDFs of
L1157, L1228, and L1251 are similar with very few pixels with temperatures above 25 K.
L1241 has a temperature PDF that is significantly narrower than the others, with few
pixels with values above 20 K.  In $\S$3.2 below, we report that no protostellar cores
are found in L1241.  Such objects would otherwise provide internal heating to the cloud,
leading to higher temperatures.  In contrast, a long tail to 20-40 K is seen in the
temperature PDF of L1172.  These high temperatures are found in pixels adjacent to the
Herbig Ae star HD 200775 (see Figure \ref{fig:fig6}b), and hence likely arise from the
radiative heating of dust by that particularly luminous star.  

Table \ref{tab:tab3} lists the estimated masses of each Cepheus cloud given
the column densities derived here. L1157, L1172, L1228, and L1251 have similar
masses of 1400-1900 M$_{\odot}$, while L1241 is about double that at 
3200 M$_{\odot}$.  Table \ref{tab:tab3} also gives the area of each 
cloud in pc$^{2}$ given the $\sim$360 pc distance to the clouds.  In addition,
Table \ref{tab:tab3} gives the amount of mass above extinction levels of
$A_{V}$ = 1 and 7, where we use the conversion between H$_{2}$ column density
and extinction of \cite{Bohlin78}.  The masses of each cloud at $A_{V}$ $>$ 1
are on average similar (i.e., within 5-40\%) to those reported by \cite{Yonekura97}
from $^{13}$CO (1-0) observations after adjustment to a distance of 360 pc.  From
this comparison, we see that L1241 has the largest amount of material at $A_{V}$
$>$ 1 but curiously also the lowest amount of mass (and lowest fraction of total
mass) at $A_{V}$ $>$ 7.  L1251, however, has the most amount of mass (and highest
fraction) of mass at $A_{V}$ $>$ 7.  In total, the Cepheus clouds have $\sim$500
M$_{\odot}$ at $A_{V}$ $>$ 7, i.e., $\sim$8\% of the mass of all five clouds at
$A_{V}$ $>$ 1.

\startlongtable
\begin{deluxetable*}{ccccc}
\tablecaption{Mass Distributions of Cepheus Flare Clouds\label{tab:tab3}}
\tablehead{
& \colhead{Total} & \colhead{Total} & \colhead{Mass at} & \colhead{Mass at}\\ 
\colhead{Cloud} & \colhead{Mass} & \colhead{Area} & \colhead{$A_{V}$ $>$ 1} & \colhead{$A_{V}$ $>$ 7}\\
& \colhead{(M$_{\odot}$)} & \colhead{(pc$^2$)} & \colhead{(M$_{\odot}$)} & \colhead{(M$_{\odot}$)}
}
\startdata
L1157 & 1400 & 67 & 790  & 66 \\
L1172 & 1900 & 91 & 1000  & 61 \\
L1228 & 1600  & 64 & 1000  & 100 \\
L1241 & 3200 & 120 & 2300 & 10 \\
L1251 & 1800 & 77 & 1100  & 270 \\
\hline
Total & 9800 & 420 & 6300 & 510 \\
\enddata
\end{deluxetable*}

\subsection{Source Extraction}

Compact objects in each Cepheus field were extracted using version 1.140127
of the source identification algorithm {\it getsources} \citep{Men'shchikov12}.
For consistency, this approach is the same as that used to produce other HGBS
catalogues of dense cores and protostars, e.g., \cite{Konyves15}.  The {\it
getsources} algorithm was specifically developed to identify compact objects
within non-uniform emission across many wavelengths and scales, e.g., in
{\it Herschel} observations of molecular clouds.

The {\it getsources} algorithm consists of two distinct ``detection" and  
``measurement" stages.  For the first ``detection" stage, {\it getsources}
successively smooths input maps to ever lower resolutions, subtracts maps
of adjacent resolutions, and identifies positions of significant emission
in the difference maps.  Comparing these difference maps over many scales
and wavelengths (if available) allows sources to be built up and identified
over ranges of scale and wavelength.  For the second ``measurement" stage,
{\it getsources} determines fluxes and sizes of sources using the original
input images at each wavelength.  Overlapping sources are intelligently
deblended.  Background levels are subtracted after being determined by linear 
interpolation under the source footprints, taking into consideration the
different native angular resolutions at each wavelength.  Finally, {\it 
getsources} applies aperture corrections at each wavelength obtained from
the PACS and SPIRE Instrument Control Centres (e.g., see \citep[e.g., see][respectively]{Balog14,Bendo13}.

We ran {\it getsources} to extract dense cores and YSOs/protostars separately
in the Cepheus fields, using different input maps and parameters.  To extract 
dense cores, we used the {\it Herschel} maps at 160 $\mu$m, 250 $\mu$m, 350 
$\mu$m, and 500 $\mu$m, regridded to pixels in common, as inputs.  These four
wavelengths bracket the peak of continuum emission from cold (T = 10-20 K) dust
in dense cores.  We also included the high-resolution column density map (see
panel (a) of Figures \ref{fig:fig1}-\ref{fig:fig5}) as an additional ``wavelength"
input to ensure identified objects consist of locally high column density.
In addition, for the ``detection stage" of this extraction, the native 160
$\mu$m maps were substituted with ones corrected for anisotropic temperature
gradients.  These corrected maps were made by converting the native 160 $\mu$m
maps to pseudo-column density maps using color temperatures determined from the
ratios of the native intensities at 160 $\mu$m and 250 $\mu$m, at the latter's
resolution. To extract YSOs or protostars, we used only the {\it Herschel} maps
at 70 $\mu$m.  YSOs and protostars will heat their surrounding dust to temperatures 
higher than those of cold cores, making them stand out obviously at 70 $\mu$m as
point-like objects \citep[e.g., see][]{Dunham08}.

For both sets of extractions, source fluxes and sizes are measured from the
native {\it Herschel} maps of all five wavelengths, appropriately deblended,
background subtracted, and aperture corrected.  Source lists were constructed 
from identified sources deemed reliable following separate standard criteria
for each extraction.  These criteria are listed in Appendix B.  Finally, to
reduce potential contamination by background galaxies, source positions were 
cross-checked with the NASA/IPAC Extragalactic Database (NED) and the SIMBAD 
database.  {\it Herschel} sources found within 6$\arcsec$, i.e., approximately
half the $13\farcs5$ resolution of {\it Herschel} at 160 $\mu$m, of a known 
background galaxy were flagged in the final catalogues.  These latter sources
are not included as cores in the analysis below.

Finally, each remaining source was visually checked.  For inclusion in the
final catalogues, a source had to be visible as a peak at its location in
the images of at least two {\it Herschel} wavelengths and the high-resolution 
column density map.  In total, 115 objects were excluded from the final
catalogues as a result of such checks.

Table \ref{tab:tab4} lists the numbers of dense cores obtained from each
Cepheus cloud.  In total, we identify 832 dense cores from all five clouds
observed, and each cloud has $\sim$150-200 dense cores.  Appendix C describes
the online material available for this paper, including a catalogue of the
observed properties of all Cepheus dense cores extracted from the various
maps, such as positions, fluxes, and sizes.  Table C1 lists the information
provided in this catalogue.  The online material also includes thumbnail
images of each extracted core at each {\it Herschel} wavelength and in the 
high-resolution column density map.  Figures C1 and C2 show example thumbnail
images of two cores, HGBS\_J204306.0+675009 (a robust prestellar core; see \S
3.3 below) and HGBS\_J203906.3+680216 (a protostellar core; see \S 3.3 below).

\startlongtable
\begin{deluxetable*}{cccccc}
\tablecaption{Numbers of Dense Cores (including Starless Cores, Prestellar Cores, and 
Protostellar Cores) in the Cepheus Flare clouds\label{tab:tab4}}
\tablehead{
& & \colhead{Unbound} & \colhead{Candidate} & \colhead{Robust}\\
& \colhead{Dense} & \colhead{Starless} & \colhead{Prestellar} & \colhead{Prestellar} & \colhead{Protostellar}\\ 
\colhead{Field} & \colhead{Cores} & \colhead{Cores} & \colhead{Cores} & \colhead{Cores} & \colhead{Cores}
}
\startdata
L1157 & 153 & 86 & 61 & 40 & 6 \\
L1172 & 156 & 98 & 52 & 31 & 6 \\
L1228 & 205 & 132 & 71 & 40 & 2 \\
L1241 & 131 & 98 & 33 & 14  & 0 \\
L1251 & 187 & 90 & 86 & 53 & 11 \\
\hline
Total & 832 & 504 & 303 & 178 & 25 \\
\enddata
\tablecomments{The locations of unbound starless cores, candidate prestellar cores,
and protostellar cores in the Cepheus clouds are shown in green, red, and
yellow, respectively, in Figures \ref{fig:fig1}-\ref{fig:fig5}.  Robust prestellar cores
are a subset of candidate prestellar cores.}
\end{deluxetable*}

We employed a secondary automated object identification algorithm to provide
an independent assessment of the sources extracted from the {\it Herschel}\/
images of the Cepheus clouds via {\it getsources}.  For these assessments, we
chose to use the {\it Cardiff Sourcefinding AlgoRithm} \citep[{\it CSAR};][]{Kirk13},
an algorithm that identifies sources in single-wavelength images by following
intensities down from maxima in the images until neighboring sources or a 
noise threshold is met.  The {\it CSAR} algorithm effectively functions as a 
conservative variant of the widely used {\it Clumpfind} algorithm of 
\cite{Williams94} in two dimensions.  We used {\it CSAR} on the high-resolution
column density map only (i.e., Figures \ref{fig:fig1}-\ref{fig:fig5}).
The catalogue produced by {\it CSAR} was checked against that created by {\it
getsources}, and sources found to be in common (i.e., with peaks located within 
6\arcsec\ of each other) are highlighted as such in the {\it getsources} catalogues
with a flag (see Appendix C and online material).  The percentages of objects
identified by both algorithms are 54\%, 46\%, 40\%, 60\%, and 49\% for L1157,
L1172, L1228, L1241, and L1251, respectively, values broadly consistent with
those obtained in other Gould Belt Clouds studied with {\it Herschel}, e.g.,
45\% in Aquila \citep{Konyves15}.  In general, {\it getsources} can identify
fainter objects than {\it CSAR} because it incorporates information from
multiple wavelengths for its assessments.

\subsection{Dense Core Masses and Sizes}

For each dense core extracted from the {\it Herschel} maps, an SED was constructed
using integrated fluxes corrected for immediate background emission (e.g., from a
host filament).  As with the determinations of column densities and temperatures
across each cloud, the dense core SEDs were fit with a modified blackbody model, 
assuming the same dependencies on $\kappa_{\nu}$ and $\beta$ (see $\S$3.1) to 
determine masses ($M_{\rm core}$) and line-of-sight averaged temperatures
($T_{\rm dust}$), following {\it Herschel} GBS standard procedures \citep{Konyves15}.  If a
core has more than three bands in which it is significant (i.e., Sig$_{\rm \lambda}
> 5$, where Sig$_{\rm \lambda}$ is the monochromatic significance determined by {\it 
getsources}) and if the 350 $\mu$m flux is higher than the 500 $\mu$m flux, two SED
fits were made.  In the first fit, the 70 $\mu$m flux was included, the errors used
were the ``detection error" = total core flux/Sig$_{\rm \lambda}$, and the weights
of the fitting were 1/(detection error)$^{2}$. In the second fit, the 70 $\mu$m flux
was neglected, the errors used were the ``measurement error" determined by {\it 
getsources}, and the weights of the fitting were 1/(measurement error)$^{2}$.  If
the mass estimate between both runs varied by less than a factor of two, we used the
mass and temperature from the second fit.  If the mass estimates differed by more
than a factor of two, the mass was calculated from the flux of the longest wavelength 
of significant flux (i.e., Sig$_{\rm \lambda} > 5$), assuming a temperature
corresponding to the median core temperature from those cores that passed the fitting 
test described above.  Approximately 45\% of cores had mass estimates from the different 
fits that differed by less than a factor of 2.  The median dust temperatures for cores 
were 11.3~K, 13.6~K, 11.9~K, 11.8~K, and 11.4~K for L1157, L1172, L1228, L1241, and 
L1251, respectively.

The observed size of each core was determined as the geometrical average of
the FWHMs of its major and minor axes in the high-resolution (18$\farcs$2) column
density map of its host Cepheus cloud ($FWHM^{\rm a}_{\rm N_{\rm H_{\rm 2}}}$
and $FWHM^{\rm b}_{\rm N_{\rm H_{\rm 2}}}$, respectively.  This angular size was 
converted to a physical size ($R_{\rm core}^{\rm obs}$) assuming the $\sim$360~pc 
distance to the Cepheus Flare clouds (see $\S$1).  A deconvolved radius was also determined via $R_{\rm core}^{\rm decon}$ = 
(${R_{\rm core}^{\rm obs}}^{2}$ - $\overline{HPBW_{N_{H_{2}}}}^{2}$)$^{1/2}$
where  $\overline{HPBW_{N_{H_{2}}}}$ is the physical size of the 18$\farcs$2 beam,
i.e., 0.032~pc at the $\sim$360~pc distance of the core.

We use the determined mass and size of each core to obtain estimates of their peak
column densities ($N_{\rm H_{\rm 2}}^{\rm peak}$), determined from the peak flux
densities at the 36$\farcs$3 resolution of the 500 $\mu$m data (see Appendix C 
and online material).  In addition, we determine the average column densities of
each core before and after deconvolution ($N_{\rm H_{\rm 2}}^{\rm ave}$ and $N_{\rm
H_{\rm 2}}^{\rm ave,d}$, respectively).  Next, we determine ``peak" (i.e., from
the peak column density value; $n_{\rm H_{\rm 2}}^{\rm peak}$) and average volume 
densities before and after deconvolution ($n_{\rm H_{\rm 2}}^{\rm ave}$ and $n_{\rm 
H_{\rm 2}}^{\rm ave,d}$, respectively.  The ``peak'' volume density was determined
using the peak column density assuming a Gaussian spherical distribution where 
$n_{\circ}$ = $N_{\circ}$~/~($\sqrt{2\pi}\sigma$) and $\sigma$ is the standard
deviation of the Gaussian distribution.

The online material accompanying this paper includes a catalogue of derived
physical properties of all Cepheus cores, as described in Appendix C.  Table
C2 lists the quantities found in this catalogue, including mass, size, column
density and volume density estimates.

\begin{figure*}
\plotone{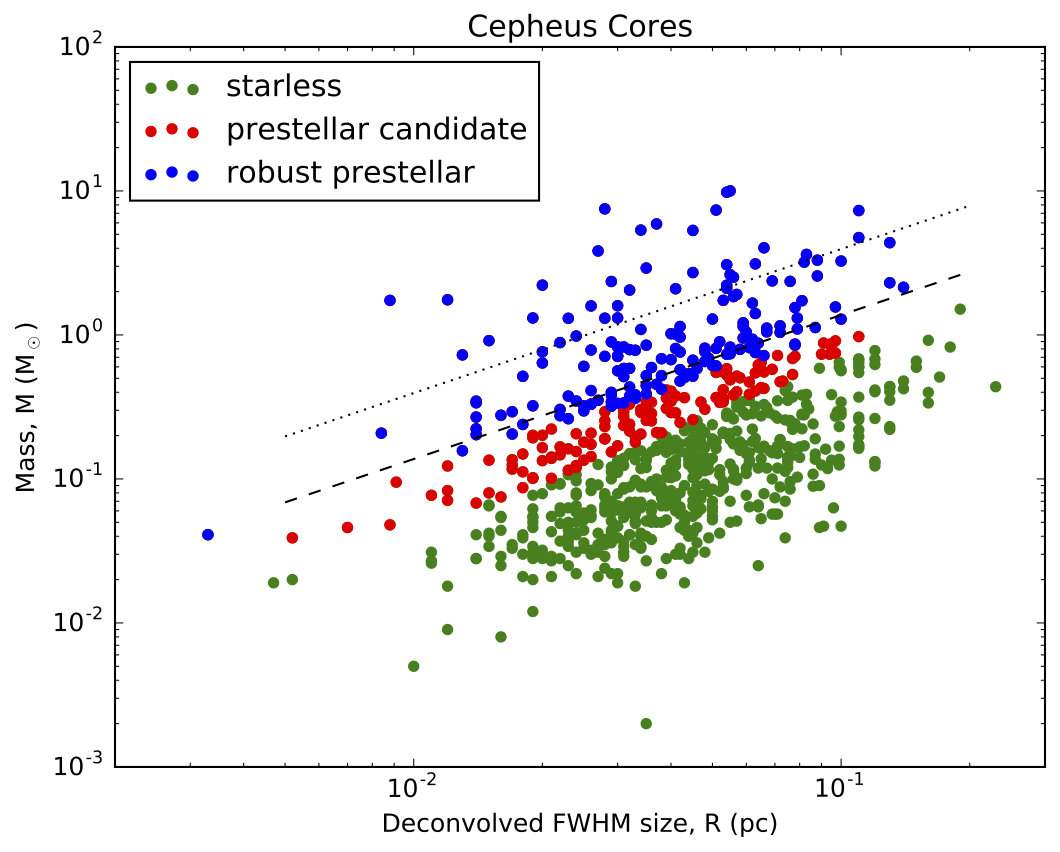}
\caption{Mass ($M_{\rm core}$) vs. deconvolved FWHM size ($R_{\rm
core}^{\rm decon}$) for cores extracted from all five Cepheus clouds.
Green, red, and blue dots denote starless cores, candidate prestellar
cores, and robust prestellar cores, respectively.  Note that robust
prestellar cores are a subset of candidate prestellar cores. The
critical Bonnor-Ebert mass for isothermal cores with $T=7$ K and
$T=20$ K are shown as black dashed and dotted lines,
respectively.}
\label{fig:fig8}
\end{figure*}

Figure \ref{fig:fig8} shows the distribution of $M_{\rm core}$ vs.\
$R_{\rm core}^{\rm decon}$ for the core population of all five Cepheus
clouds combined.  The cores range in size over $\sim$0.01-0.2 pc and
range in mass over $\sim$0.02-10 M$_{\odot}$.  There are no significant
differences in the ranges of core mass and size between clouds.

Based on the derived mass and deconvolved sizes, we explore the dynamical 
stability of the Cepheus cores by comparing their physical properties to 
those of a critical Bonnor-Ebert (BE) sphere \citep{Ebert55,Bonnor56}.
For example, the mass of a critical BE sphere is defined as: 

$$ M_{\rm BE,crit} \approx 2.4 R_{\rm BE}c_{\rm S}^{2} / G, \eqno{(1)} $$

\noindent
where $R_{\rm BE}$ is the BE radius, $c_{\rm s}$ is the isothermal sound
speed, and $G$ is the gravitational constant.  Here we neglect any nonthermal 
contributions to the support of the core, e.g., from turbulence.  For each core,
we estimated $M_{\rm BE}$ assuming $R_{\rm BE}$ = $R_{\rm core}^{\rm decon}$
from the high-resolution column density map and assumed T = 10 K.  We thus
define the mass ratio $\alpha_{\rm BE}$ = $M_{\rm BE,crit}$ / $M_{\rm core}$.

We follow the guidance of \cite{Konyves15} and use the size-dependent
limiting BE mass ratio criterion $\alpha_{\rm BE}$ $\leq$ 5 $\times$
($HPBW_{\rm N_{\rm H_{\rm 2}}} / FWHM_{\rm N_{\rm H_{\rm 2}}}$)$^{0.4}$
to estimate the dynamical state of a core.  Those cores meeting this
criterion are deemed ``candidate prestellar cores" (i.e., potentially
bound by gravity) and those that do not are considered ``starless" (i.e.,
gravitationally unbound).  Figures \ref{fig:fig1}-\ref{fig:fig5} show
the locations of these types of cores in each Cepheus cloud.  By
restricting the criterion to $\alpha_{\rm BE}$ $\leq$ 2, we further
define a subset of candidate prestellar cores we call ``robust
prestellar cores."  Those dense cores deemed to be neither prestellar
nor protostellar are considered to be ``starless cores."  Table
\ref{tab:tab4} lists the numbers of starless cores, candidate prestellar
cores and robust prestellar cores found in each cloud.  In total, this
comparison reveals 504 starless cores, 303 candidate prestellar cores,
and 178 robust prestellar cores in the Cepheus clouds.  (Twenty-five
cores are identified as being protostellar; see \S3.4 below.)  Figure
\ref{fig:fig8} shows the mass vs.\ size distribution of each starless
or prestellar core population in the Cepheus clouds.  Figure \ref{fig:fig8}
also shows the expected mass ($M_{\rm core}$) vs.\ size ($R_{\rm core}^{\rm
decon}$) relationships for critical isothermal BE spheres at temperatures
of 7 K and 20 K.  All the robust prestellar cores have masses near to or
larger than the critical BE mass at a given size and temperature of 7 K.

\cite{Konyves15} conducted extensive completeness tests of the cores 
extracted from {\it Herschel}\/ maps of the Aquila Rift also via the 
{\it getsources} algorithm.  They estimated their sample was 90\% 
complete for cores of observed mass of 0.2 M$_{\odot}$, assuming an
Aquila distance of 260 pc.  (From other tests, they determined that
observed core masses underestimate the true core mass of dense cores
by 20-30\%.)  
Given the larger distance of the Cepheus clouds of $\sim$360 pc,  we
accordingly estimate our 90\% completeness as being 0.4 M$_{\odot}$.
Note, however, that Aquila has a lower Galactic latitude ($l$ $\sim$
2-3$\degr$ than the Cepheus clouds ($l$ $\sim$ 13-20$\degr$), and thus
likely includes more background emission from Galactic cirrus.  Hence,
we consider our adopted value for Cepheus to be a conservative estimate
of our observations' true completeness.  Indeed, our completeness
estimate of 0.4~M$_{\odot}$ is largely consistent with those determined
for other Gould Belt clouds via modeling similar to that performed by
\cite{Konyves15}, after factoring in distance differences 
\citep[e.g.,][Pezzuto et al. 2020, in press]{Benedettini18,Ladjelate20}.

\subsection{Protostellar Cores}

Our extractions of sources at 70 $\mu$m yields 64 sources across all five
Cepheus clouds.  The numbers of 70 $\mu$m sources detected in each cloud
are 8, 12, 12, 2, and 30 in L1157, L1172, L1228, L1241, and L1251,
respectively.

To determine the populations of protostellar cores in the Cepheus clouds,
we compare the sources detected at 70 $\mu$m to the dense cores detected at
160-500 $\mu$m.  We classify as protostellar those dense cores where a 70
$\mu$m source is located within its FWHM ellipse.  The catalogues provided 
in the online material list the observed and derived physical properties
of the protostellar cores found in the Cepheus clouds studied here.  Only
25 protostellar cores are identified over all five clouds and Figures 
\ref{fig:fig1}-\ref{fig:fig5} show the locations of the protostellar cores 
in each cloud.  Table \ref{tab:tab4} lists the numbers of protostellar
cores found in each cloud.  L1241 has no protostellar cores identified in 
its midst, consistent with the finding of \cite{Kirk09} from {\it Spitzer} 
data that L1241 is without YSOs.  With 11 protostellar cores, L1251 has the 
largest number of such cores in the Cepheus clouds, nearly half the 
identified population, suggesting it is currently the most active 
star-forming cloud of the five studied here.  See $\S$4.2 for discussion
of the relative star-forming activity of these clouds.

Given the focus of this paper on the Cepheus clouds' prestellar core
population, we do not discuss further the populations of ``naked" 70
$\mu$m sources or protostellar cores.

\subsection{Filamentary Substructure}

As with all other clouds observed as part of the HGBS, the {\it Herschel}
data reveal that the five Cepheus clouds have extensive substructure, much
of it filamentary in morphology (see Figures \ref{fig:fig1}-\ref{fig:fig5}).
In particular, L1157, L1228, and L1251 appear to be dominated by filaments.
L1172 and L1241 also exhibit filaments but the former is dominated by a high
column density clump (NGC 7023) and filaments in the latter appear to be
more diffuse.

\begin{figure*}
\plotone{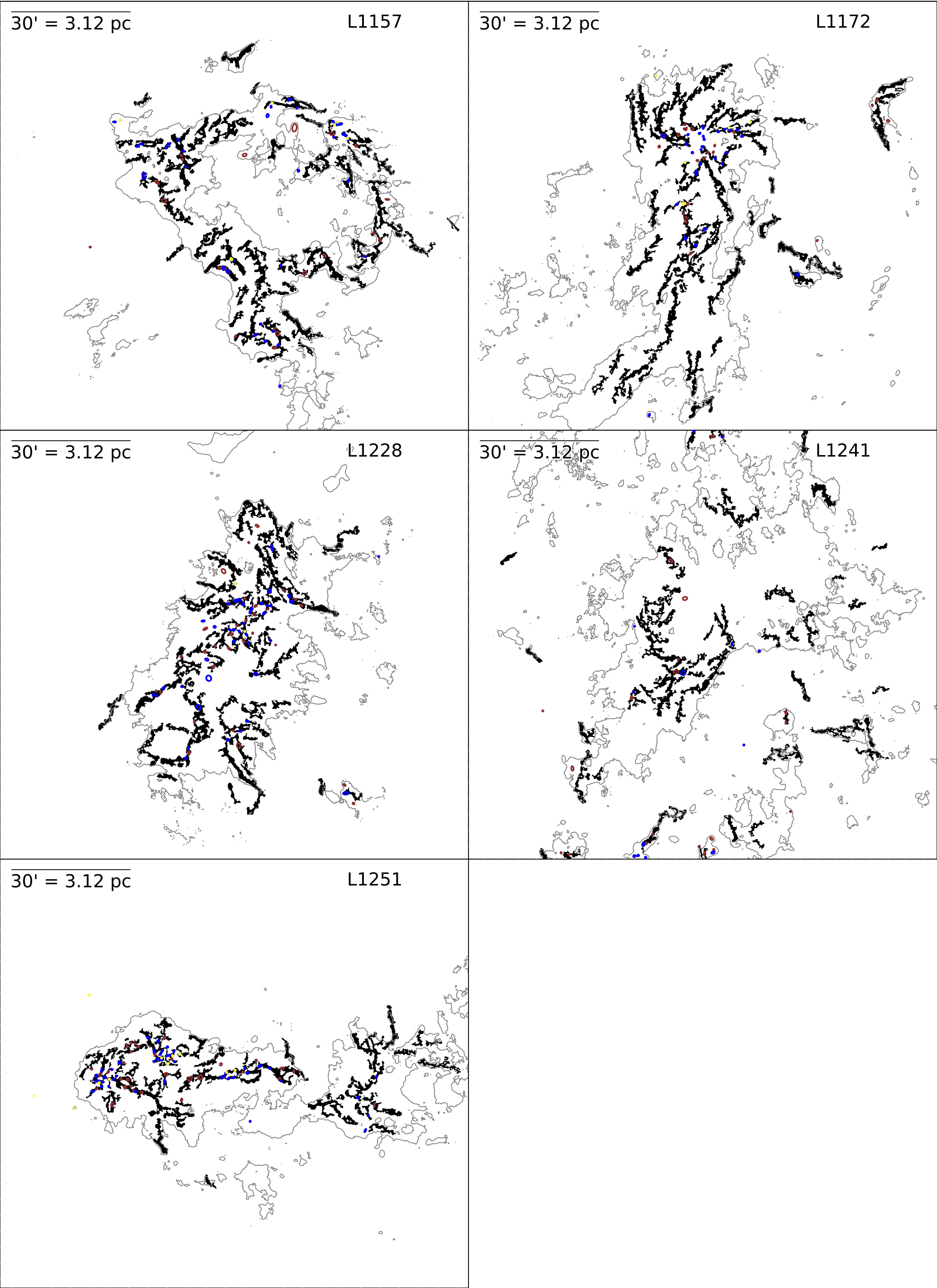}
\caption{Areas of filaments identified from the H$_{2}$ column density map
derived from {\it Herschel} data (dark pixels).  Spatial scales up to 0.125
pc (i.e., 72$^{\prime\prime}$ at 358 pc) are shown.  In each field, red, blue
and yellow ellipses indicate the locations and sizes of candidate prestellar
cores, robust prestellar cores, and protostellar cores, respectively.  For
context, the grey contour denotes the 500 $\mu$m intensity level of 94.5 mJy 
beam$^{-1}$.}
\label{fig:fig9}
\end{figure*}

To quantify the locations of filamentary structure in the Cepheus clouds,
we used the ``getfilaments" option with standard parameters during our core 
extractions with {\it getsources} \citep[see][]{Men'shchikov13}. This option 
provides as output the locations of long contiguous features in input maps 
that are identified using standard extraction parameters as part of the
multi-scale processing of those maps by {\it getsources}.  Figure
\ref{fig:fig9} shows the locations of filaments identified in each Cepheus
cloud from their respective {\it Herschel}-derived H$_{2}$ column density
maps up to spatial scales of 0.125 pc (i.e., 72$^{\prime\prime}$ at the 358
pc distance to the Cepheus clouds), along with the positions and sizes of
associated cores.  

\begin{figure*}
\plotone{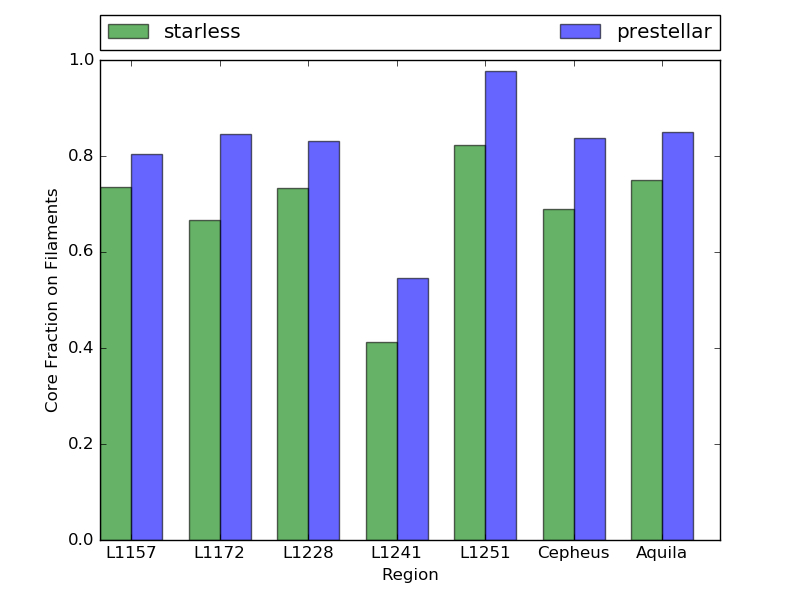}
\caption{Histogram of percentages of cores found to
be coincident with filamentary structure in the 
Cepheus clouds.  Green indicates starless cores and
blue indicates candidate prestellar cores.}
\label{fig:fig10}
\end{figure*}

Figure \ref{fig:fig9} illustrates how cores in Cepheus are predominantly
located on filaments.  Figure \ref{fig:fig10} shows a histogram of the 
fractions of cores by type that are coincident with filaments in the Cepheus 
clouds.  For this comparison, a core was considered to be ``on filament" if
its central pixel overlapped with the area of an identified filament, i.e..,
the locations identified in Figure \ref{fig:fig9}.  Some variation between
clouds is seen with L1241 having cores with the least association with filaments
($\sim$40-50\%) and L1251 having almost all its cores  ($\sim$80-100\%)
associated with filaments.  Excluding L1241, $\sim$75\% of starless cores
and $\sim$80\% of candidate prestellar cores are coincident with filaments
in Cepheus.  The percentages of robust prestellar cores on filaments are
similar to those of candidate prestellar cores on filaments.  These percentages are
roughly equivalent to those obtained for other HGBS clouds, e.g., Aquila
\citep[see Figure 14 of][]{Konyves15}.

\section{Discussion} \label{sec:disc}

\subsection{Star Formation in Low Column Density Environments}

Though the Cepheus clouds show extensive filamentary substructure and their cores are largely associated with that substructure, the clouds have relatively low column densities.  For example, the Cepheus clouds have median H$_{2}$ column densities of $\sim$6-10 $\times$ 10$^{20}$ cm$^{-2}$ (see Table \ref{tab:tab2}), corresponding to extinction levels of $A_{V}$ $\approx$ 0.6-1.0 \citep{Bohlin78}.  For this discussion, we do not subtract foreground/background column densities determined from emission by dust that is arguably unrelated to the cloud.  From the histograms of column density in each map shown in Figure \ref{fig:fig7}a, we surmise that the foreground/background column densities for these clouds are likely on the order of $\sim$1-2 $\times$ 10$^{20}$ cm$^{-2}$ or so.  Since the filament and cores in which we are interested in each cloud typically have column densities an order of magnitude than this value or more, we do not correct for systematic increases in their total column densities here.
    
To put the column densities of the Cepheus clouds into perspective, we note that there has been considerable discussion in recent years about a threshold column density for core formation in filaments of $A_{V}$ $\approx$ 7-8.  The physics behind this threshold may be understood from the isothermal infinite cylinder model of \cite{Ostriker64}, which becomes critically stable when its line mass (mass per unit length) $M_{\rm line}$ = 2$c_{\rm s}^{2}/G$ $\approx$ 16 M$_{\odot}$ pc$^{-1}$ at 10 K.  Assuming a common filament width of 0.1 pc \citep[see][]{Arzoumanian11,Arzoumanian19}, this nominal line mass occurs at $A_{V}$ $\approx$ 7-8 in filaments \citep{Andre14}.  Indeed, such a threshold appeared evident in the Aquila Rift cloud, where 75\% of prestellar cores are associated with filaments with line masses $\geq$ 16 M$_{\odot}$ pc$^{-1}$ \citep{Konyves15}.  More recent works than \cite{Ostriker64}, however, have posited that ``transcritical'' filaments, i.e., with line masses within a factor of $\sim$2 of this critical value (e.g., 8-32~M$_{\odot}$~pc$^{-1}$ at 10 K), are actually those susceptible to fragmentation, leading to less of a sharp threshold and more of a smooth transition for core formation with column density in filaments, as has been observed \citep{Inutsuka97,Fischera12,Arzoumanian19,Konyves20}.

Only 0.02-0.9\% of pixels in all five Cepheus clouds have column densities $>$8 $\times$ 10$^{21}$ cm$^{-2}$, i.e., $A_{V}$ $>$ 8.   Nevertheless, modest star formation is indeed occurring in the Cepheus clouds.  Such activity is evident by the 303 candidate prestellar cores and 25 protostellar cores identified in the {\it Herschel} data alone (Table \ref{tab:tab4}).  Moreover, \cite{Kirk09} using {\it Spitzer} data find 93 YSO candidates, mostly Class II objects, in L1172, L1228, and L1251.  Notably, they find only one YSO candidate in L1241.  As mentioned earlier, L1157 was not included in their study.

In terms of core formation activity, the Cepheus clouds appear similar to other low column density environments observed by {\it Herschel}, e.g., Lupus I, III, and IV \citep{Benedettini15,Benedettini18}.  In these particular clouds, \cite{Benedettini15} do find column density PDFs of similar morphology to those of the Cepheus clouds that also peak at $0.5-1.0$ $\times$ 10$^{21}$ cm$^{-2}$ (see \S3.1). More recently, \cite{Benedettini18} identify $\sim$30\% fewer numbers of cores in Lupus than what we find over all five Cepheus clouds, e.g., 532 dense cores of which 102 are candidate prestellar cores.  For comparison, the Lupus clouds are $\sim$160 pc from the Sun \citep{Dzib18}, less than half the distance to the Cepheus clouds.  As with Cepheus, \cite{Benedettini18} find that almost all Lupus prestellar cores are associated with filaments, though only a third of the Lupus starless cores are so associated.  They also find that 90\% of Lupus prestellar cores are located in backgrounds of $A_{V}$ $\geq$ 2, an extinction level much lower than that seen in other Gould Belt clouds \citep[e.g., Aquila; see][]{Konyves15}.  

\cite{Benedettini15} find the median column densities of filaments in the Lupus I, III, and IV clouds to be 1.2-1.9 $\times$ 10$^{21}$ cm$^{-2}$.  Relatedly, they find the median line mass of filaments in the Lupus clouds to be $\sim$3 M$_{\odot}$ pc$^{-1}$ (see their Figure 15), somewhat lower than the transcritical range of 8-32 M$_{\odot}$ pc$^{-1}$ for cylinder stability at 10~K \citep{Arzoumanian19}.  Clearly, filaments remain extremely relevant to core (and star) formation, even in instances like Lupus where their median column densities are lower than that range.  Given the large degree of correspondence of cores with filaments in Lupus but lower column densities, \cite{Benedettini15} and \cite{Benedettini18} suggest that the condition for overdensity needed for filament fragmentation, i.e., within the transcritical range of column densities, may be only reached locally in Lupus filaments.  Indeed, observed filaments are not constant column density structures, and can exhibit significant variations along their lengths \citep[e.g., see][]{Roy15}.  Hence, a low average line mass for an entire filament is not a good parameter for determining whether or not stars will form in that filament.

\begin{figure*}
\plottwo{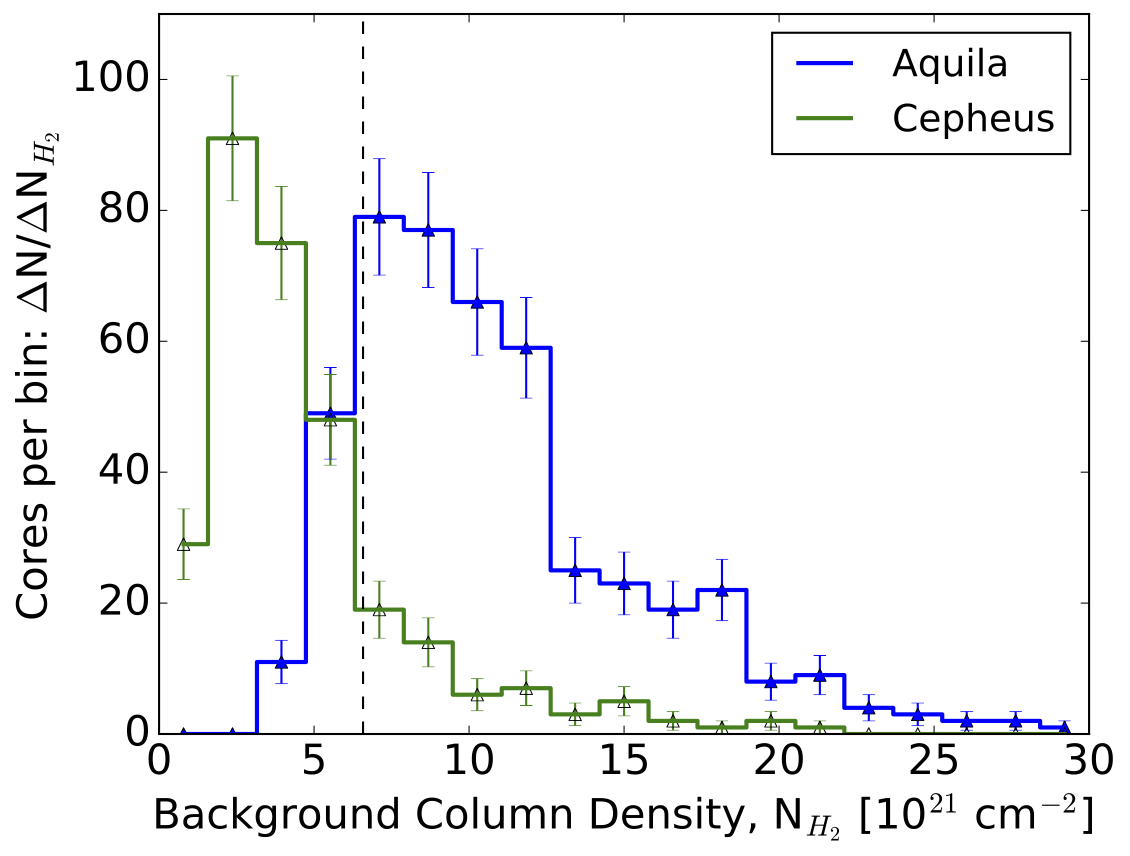}{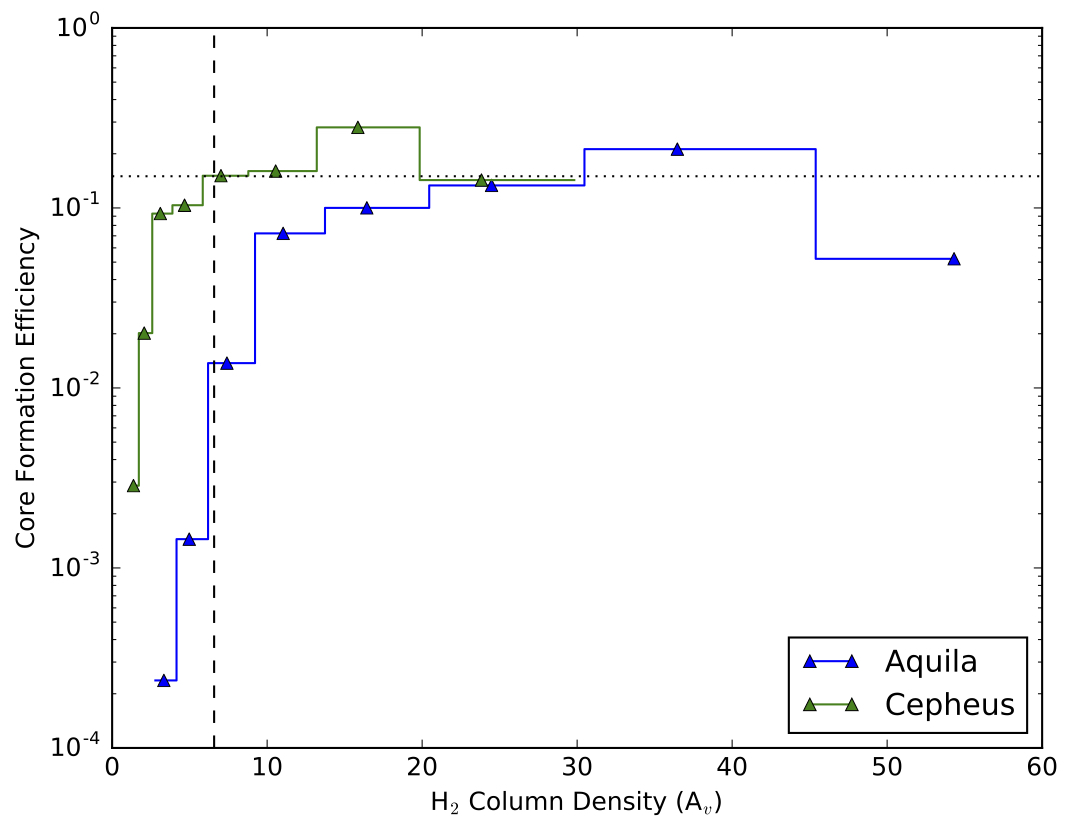}
\caption{{\it Left panel}: Histogram of numbers of candidate prestellar cores vs. background H$_{2}$ column densities of cores in the Cepheus clouds (green) and the Aquila Rift (blue).  Error bars are from Poisson statistics.  The dashed line indicates a background column 
density of $A_{V}$ $\approx$ 7; see \citealt{Konyves15}. {\it Right panel}: Observed differential core formation efficiencies (CFE$_{\rm obs}$) as a function of background column density expressed in units of $A_{V}$ for the Cepheus clouds (green) and the Aquila Rift (blue).} 
\label{fig:fig11}
\end{figure*}

Table \ref{tab:tab6} lists the median and mean column densities of filaments in the 
Cepheus clouds (see Figure 9).  These values were computed from the column density map 
using the definitions of filaments shown in Figure \ref{fig:fig9}, and reflect all mass 
traced within those filaments, including cores.  The median filament column densities
are 1.8-2.7 $\times$ 10$^{21}$ cm$^{-2}$, about a factor of $\sim$1.5 higher than the 
range of those in the Lupus clouds.  Adopting an average filamentary width of 0.1 pc 
\citep{Arzoumanian11,Arzoumanian19}, the median Cepheus filament column densities 
translate into median line masses of 4-6 M$_{\odot}$ pc$^{-1}$.  The mean column 
densities of the Cepheus filaments, however, are 2.2-4.4 $\times$ 10$^{21}$ cm$^{-2}$, 
which translate into mean line masses of 5-10 M$_{\odot}$ pc$^{-1}$, i.e., 25-66\%
higher than the median line masses, indicating that the line masses of filaments in
these clouds are skewed somewhat above the median, unlike in Lupus.  This range of line 
masses is just below to within the range of transcritical masses at 10 K, i.e., $M_{\rm 
line}$ = 8-32 M$_{\odot}$ pc$^{-1}$.

In Figure \ref{fig:fig11} ({\it left}), we show the distribution of background column densities for candidate prestellar cores in the Cepheus clouds.  The distribution clearly peaks at a column density of $\sim$2-4 $\times$ 10$^{21}$ cm$^{-2}$, similar to the mean column densities of the filaments and much lower than the threshold $A_{V} \approx 7$ column density.  Indeed, $\sim$80\% of candidate prestellar cores in Cepheus are found at background column densities of $A_{V}$ $\leq$ 7.  For comparison, Figure \ref{fig:fig11} ({\it left}) also shows the distribution of background column densities for candidate prestellar cores in the Aquila Rift, where only 20\% are found at $A_{V}$ $\leq$ 7 \citep{Konyves15}.  Note that a significant tail at the higher ends of the distributions are also seen, i.e., up to $\sim$20 $\times$ 10$^{21}$ cm$^{-2}$ in Cepheus.  If we assume these background column densities are indicative of the original filament environments in which the respective cores formed, then by number $\sim$56\% of Cepheus candidate prestellar cores formed in filaments with line masses below the transcritical range and $\sim$44\% from filaments with line masses within that range.  The original filament line masses of course would be larger if the mass we see now in cores was initially distributed more widely within their host filaments.  With that possibility in mind, the split is likely on the order of 50-50.

For further comparison, Figure \ref{fig:fig11} ({\it right}) shows the observed differential core formation efficiencies ($CFE_{\rm obs}$) of both the Cepheus clouds and Aquila, where $CFE_{\rm obs}$($A_{V}$) = $\Delta M_{\rm cores}$($A_{V}$) / $\Delta M_{\rm cloud}$($A_{V}$). Here, $\Delta M_{\rm cores}$($A_{V}$) is the mass of the prestellar cores within a given bin of background $A_{V}$ and $\Delta M_{\rm cloud}$($A_{V}$) is the cloud mass estimated from the column density map in the same $A_{V}$ bin.  In Aquila, the $CFE_{\rm obs}$ rises from very low values at low $A_{V}$ and levels off at $\sim$15\% at high $A_{V}$, with a transition around $A_{V}$ $\approx$ 7 \citep{Konyves15}.  In Cepheus, however, the $CFE_{\rm obs}$ rises much more quickly with $A_{V}$, and levels off a similar value of $\sim$15\% at lower $A_{V}$, with a transition at $A_{V}$ $\approx$ 3. 

\startlongtable
\begin{deluxetable*}{ccc}
\tablecaption{Median and Mean H$_{2}$ Column Densities of Filaments in Cepheus Flare Clouds\label{tab:tab6}}
\tablehead{
& \colhead{Median} & \colhead{Mean} \\
\colhead{Field} & \colhead{Column Density} & \colhead{Column Density} \\
& (cm$^{-2}$) & (cm$^{-2}$) 
}
\startdata
L1157 & 1.9E+21 & 2.5E+21 \\
L1172 & 1.8E+21 & 2.4E+21 \\
L1228 & 2.3E+21 & 3.0E+21 \\
L1241 & 2.0E+21 & 2.2E+21 \\
L1251 & 2.7E+21 & 4.4E+21 \\
\enddata
\end{deluxetable*}

Given the column densities of filaments in the Cepheus clouds, core formation within them is likely proceeding in a manner that bridges the behaviors identified earlier in Lupus and Aquila.  Clearly, the strong association between filaments and prestellar cores throughout Cepheus implicates the role of filaments in the core formation process, as elsewhere.  In Cepheus filaments with line masses below the transcritical range though, i.e., $M_{\rm line}$ $<$ 8 M$_{\odot}$ pc$^{-1}$, cores are likely forming more sporadically and only where conditions have allowed localized filament fragmentation, as seen in Lupus.  In cases with line masses within (but at the low end of) the transcritical range, i.e., $M_{\rm line}$ $>$ 8 M$_{\odot}$ pc$^{-1}$, however, cores are likely forming more {\it en masse} due to widespread filamentary fragmentation, as seen in Aquila.  Following Figure \ref{fig:fig11} ({\it left}), roughly half of the cores in Cepheus may have formed in the former way and half in the latter.

\begin{figure*}
\plotone{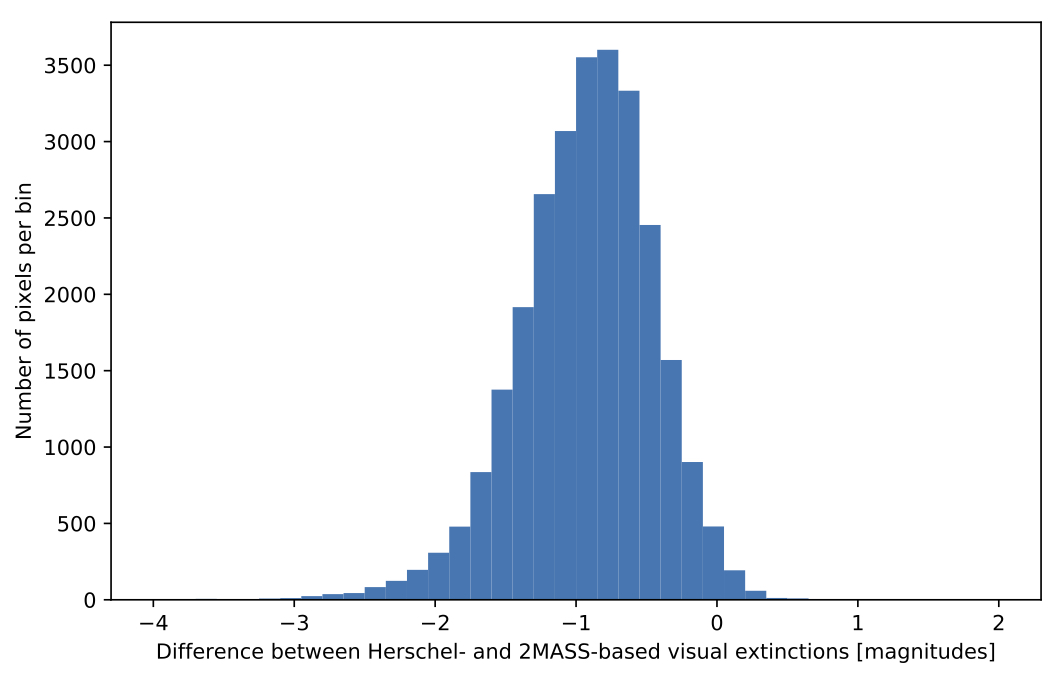}
\caption{Histogram showing the numbers of pixels in all five Cepheus clouds with differences between extinctions based on {\it Herschel} data and those based on 2MASS data.  The {\it Herschel}-based column densities were first smoothed to the 4$^{\prime}$ FWHM resolution of the 2MASS extinction maps, regridded onto the same 2$^{\prime}$ $\times$ 2$^{\prime}$ pixels of the 2MASS extinction maps, and converted to A$_{V}$ using the conversion factor of \cite{Bohlin78}.}
\label{fig:fig12}
\end{figure*}

Note that for our column density determinations, we use dust opacity values that are standard for the HGBS, i.e., $\kappa_\nu$ = 0.144 cm$^{-2}$ at 250 $\mu$m, which assumes a dust-to-gas ratio of 100 and a power-law dependence with wavelength of index $\beta$ = 2 (see \S 3.1).  Earlier comparisons of HGBS column densities with extinction maps \citep[][Pezzuto et al. 2020, in press]{Benedettini15,Konyves20}, however, have indicated such {\it Herschel}-based column densities may underestimate the true column densities at lower extinctions, e.g., $A_{V} < 4$.  To explore this possibility for Cepheus, we used near-infrared data from 2MASS \citep{Strutskie06} to determine extinction maps of each cloud at 4$^{\prime}$ FWHM resolution and compared these extinctions to those expected from {\it Herschel} data.  Specifically, Figure \ref{fig:fig12} contains a histogram showing the difference between the extinctions from the two datasets for $A_{V} < 4$.  For this histogram, the 18$^{\prime\prime}$ FHWM resolution {\it Herschel}-based column densities were smoothed to the 4$^{\prime}$ FWHM resolution of the 2MASS extinction maps, regridded to the same 2$^{\prime}$ $\times$ 2$^{\prime}$ pixels of the 2MASS extinction maps, and converted to extinctions using the common $A_{V}$-to-H$_{2}$ conversion factor of N(H$_{2}$) = $A_{V}$ $\times$ 0.94 $\times$ 10$^{21}$ from \cite{Bohlin78}.  The comparison clearly shows the {\it Herschel} data consistently produce lower extinctions than those expected from 2MASS data by $\sim$0-2 magnitudes, with a peak difference at $\sim$1 magnitude.  Though the intrinsic resolutions of the datasets differ significantly, the {\it Herschel}-based column density estimates we use as the basis of our discussion here may be systematically lower by $\sim$1 $\times$ 10$^{21}$ cm$^{-2}$.  Accordingly, the line masses of Cepheus filaments may be systematically larger by $\sim$2 M$_{\odot}$ pc$^{-1}$, somewhat increasing the fraction of such filaments in the transcritical line mass range.  See \cite{Benedettini15} and \cite{Konyves20} for discussions of possible sources of the disparity between extinctions and {\it Herschel}-derived column densities.


Even with a modest increase in local column densities, star formation is occurring in Cepheus in relatively low column density environments.  Cores have likely formed in relatively low numbers in Cepheus due to its significant number of lower column density filaments, relative to clouds widely containing filaments at higher (i.e., critical) column densities.  Namely, with widespread fragmentation being less available in lower column density filaments, relatively fewer cores would be expected to be produced per filament.  Though the range of mean column density in the Cepheus clouds is narrow, it is notable that the numbers of candidate and robust prestellar cores in each cloud roughly track monotonically (though less than linearly) with the mean column density of the filaments in each cloud, with the lowest numbers of prestellar cores found in L1241 and the highest in L1251 (see Tables \ref{tab:tab4} and \ref{tab:tab6}).  In addition, the numbers of Cepheus candidate prestellar cores show that $\sim$half at present have formed in lower column density filaments (see Figure \ref{fig:fig11} ({\it left})).  The low number of prestellar cores likely has had a concomitant effect on the productivity of star formation in Cepheus, and likely accounts for these clouds' relatively modest protostellar yields.  Indeed, the relative unavailability of transcritical filaments widely in lower column density clouds may simply explain why such clouds do not form as many cores and stars as clouds with higher column densities do.  

\begin{figure*}
\plotone{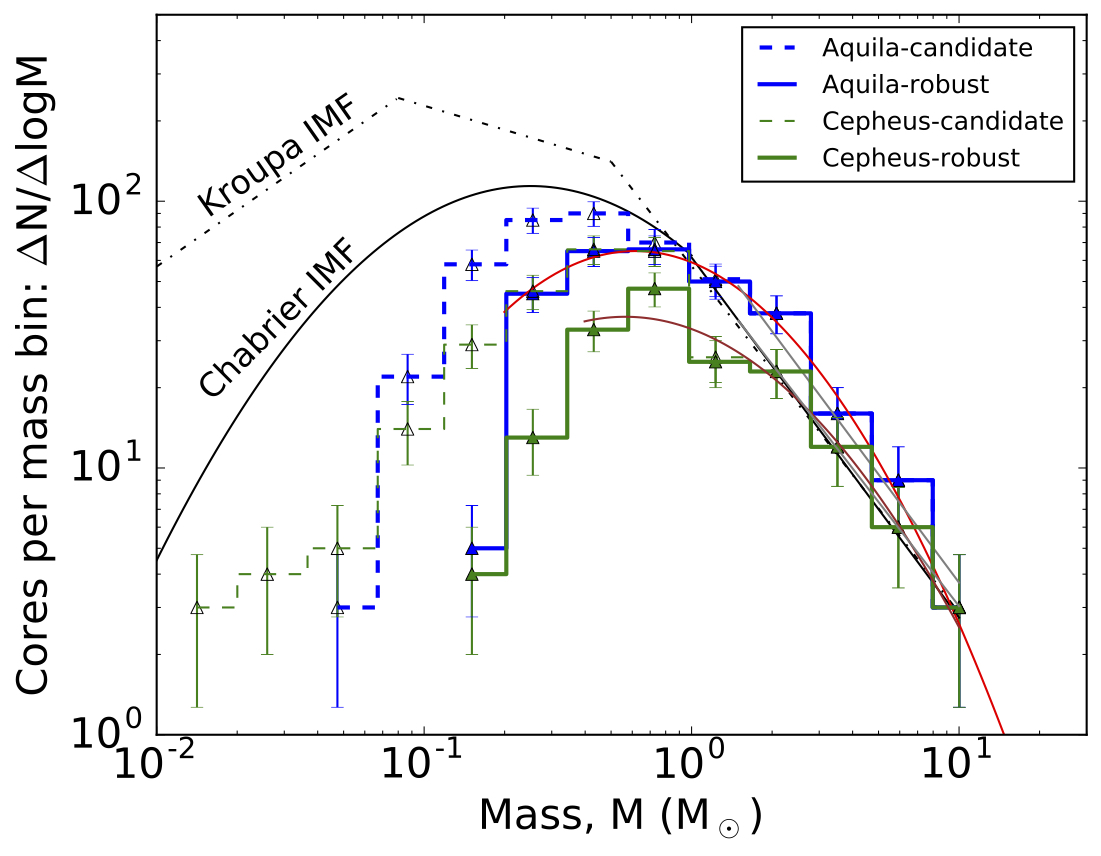}
\caption{Mass functions of candidate prestellar (dashed histograms) and robust prestellar (solid histograms) cores in the Cepheus clouds (green) and the Aquila Rift (blue). The numbers of candidate and robust prestellar cores in each sample are 303 and 187 (Cepheus) and 446 and 292 (Aquila), respectively.  Error bars are from Poisson statistics. The brown and red lines indicate lognormal fits to the Cepheus and Aquila robust prestellar mass functions, excluding the lowest mass bins which are likely incomplete. The lognormal fits have peaks at 0.56 M$_{\odot}$ and 0.62 M$_{\odot}$  and standard deviations of 0.54 and 0.48, respectively. For comparison, the black solid and dash-dotted lines shows the stellar initial mass functions (IMFs) of Chabrier (2005) and Kroupa (2001), respectively, after scaling by a factor of 10$^{3}$.}
\label{fig:fig13}
\end{figure*}

\subsection{Core Mass Functions}

Figure \ref{fig:fig13} shows the number distributions of candidate
and robust prestellar core mass in Cepheus.  The two core mass functions
(CMFs) are virtually the same at the high mass end, i.e., $M_{obs}$ $\geq$ 
1~M$_{\odot}$.  Both CMFs peak at $\sim$0.7~M$_{\odot}$, but the robust
prestellar CMF falls off more quickly to lower masses than the candidate
prestellar CMF does, likely because it is more difficult for lower mass
cores to satisfy the requirement that a robust prestellar core have
$\alpha_{BE} \leq 2$ (see \S3.3).  Indeed, the cores of the robust
prestellar CMF are arguably those most likely to collapse imminently to
form a new generation of stars.  Notably, we find no robust prestellar
cores in Cepheus of mass below 0.1 M$_{\odot}$, but a few candidate
prestellar cores of mass $<$ 0.1 M$_{\odot}$ are seen.  These scarcities 
of low-mass prestellar cores are likely related to our expectation that 
our source extractions are 90\% complete to cores of mass $\approx$
0.4 M$_{\odot}$.  Note, however, that the peaks of the prestellar CMFs
are well above the 90\% completeness limit, indicating that they have not
been artificially induced due to incompleteness.  As found in other
clouds studied in the HGBS, the robust prestellar CMF in Cepheus shows an
overall lognormal shape.  Indeed, a lognormal fit to that CMF, performed
with Levenberg-Marquardt least-squares minimization over bins at and above
the 0.4 M$_{\odot}$ mass limit of 90\% completeness, has a peak at 0.56
M$_{\odot}$ $\pm$ 0.21 M$_{\odot}$ and a width $\sigma$ = 0.54 $\pm$ 0.06
dex.  We note that these errors are likely lower limits since we do
not consider uncertainties in the masses of individual cores when fitting
these CMFs.

Figure \ref{fig:fig13} also shows the candidate and robust prestellar
CMFs for the Aquila Rift, as found by \cite{Konyves15} using a
distance\footnote{Based on {\it GAIA} data, \cite{Ortiz-Leon18} have
recently suggested an Aquila distance of 436~pc $\pm$ 9~pc, which would
shift the Aquila CMFs upward in mass by a factor of $\sim$2.8 but not
change significantly their prestellar core memberships or lognormal shapes.
We have retained the earlier distance for this discussion as it is unclear
how much of Aquila corresponds to 436~pc (see Palmeirim et al.~2020, in
preparation).  We note that the ultimate conclusions of this paper do not
depend on the distance to Aquila.} of 260 pc.  The robust prestellar CMFs
of Cepheus and Aquila show two important similarities and one difference.
The first similarity is in the widths of lognormal fits to the two robust 
prestellar CMFs, e.g., 0.54 $\pm$ 0.06 dex for Cepheus vs.\ 0.48 $\pm$ 0.02
dex for Aquila.  The second similarity is in the peaks of these CMFs, e.g.,
0.56 M$_{\odot}$ $\pm$ 0.21 M$_{\odot}$ for Cepheus vs.\ 0.62 M$_{\odot}$
$\pm$ 0.04 M$_{\odot}$ for Aquila.  The difference, however, is between the
heights of the clouds' CMFs, which can be simply attributed to the fact
that Cepheus has $\sim$60\% as many robust prestellar cores than Aquila,
e.g., 178 vs.\ 292, respectively.  Nevertheless, this comparison reveals
that the lognormal fits to the robust prestellar CMFs in both clouds are
very similar.  Note the 90\% completeness core mass limit for Cepheus
(0.4 M$_{\odot}$) is roughly twice that of Aquila (0.2 M$_{\odot}$).


For further comparison, Figure \ref{fig:fig13} also shows the stellar initial
mass functions (IMFs) determined by \cite{Kroupa01} for single stars and by 
\cite{Chabrier05} for multiple systems.  These IMFs have been each scaled by
a factor of 10$^{3}$ so they can be plotted alongside the CMFs.  Both IMFs are
also lognormal in character, though they peak at much lower masses than the
Cepheus and Aquila CMFs.  The 0.54 $\pm$ 0.06 dex width of Cepheus' robust 
prestellar CMF is remarkably similar to the 0.55 dex width of the Chabrier IMF
of Galactic disk stellar systems \citep{Chabrier05}, but is narrower than the 
Kroupa IMF of single stars.  Indeed, comparing the peaks of the Cepheus 
robust prestellar CMF and the Chabrier IMF, an efficiency factor of $\epsilon$ 
$\approx$ 0.3-0.4 is implicated, similar to the $\epsilon$ $\approx$ 0.4
estimated from the Aquila CMF by \cite{Konyves15}.  (Of course, this picture
assumes the efficiency factor is constant over a wide range of mass.)  At
their higher mass ends, the slopes of both the Cepheus and Aquila CMFs are also 
consistent with each other within uncertainties, i.e., -1.31 $\pm$ 0.01 and -1.35 
$\pm$ 0.14, respectively.  In turn, these values are themselves consistent with the 
-1.35 slope of the high-mass end of the IMF determined by \cite{Salpeter55}. 


\startlongtable
\begin{deluxetable*}{ccc}
\tablecaption{Results of Lognormal Fits to Individual Cepheus Cloud
Robust Prestellar CMFs at M $\geq$ 0.4 M$_{\odot}$\label{tab:tab5}}
\tablehead{
& \colhead{Lognormal} & \colhead{Lognormal}\\
& \colhead{Peak Mass} & \colhead {Width}\\ 
\colhead{Field} & \colhead{(M$_{\odot}$)} & \colhead{(dex)}
}
\startdata
L1157 & 0.87 $\pm$ 0.21 & 0.55 $\pm$ 0.08\\
L1172 & 0.62 $\pm$ 0.35 & 0.58 $\pm$ 0.16\\
L1228 & 0.41 $\pm$ 0.65 & 0.50 $\pm$ 0.30 \\
L1241 &  ... & ...\\
L1251 & 0.75 $\pm$ 0.19 & 0.41 $\pm$ 0.09\\
\hline
Cepheus & 0.56 $\pm$ 0.21 & 0.54 $\pm$ 0.06\\
\enddata
\end{deluxetable*}

\begin{figure*}
\plotone{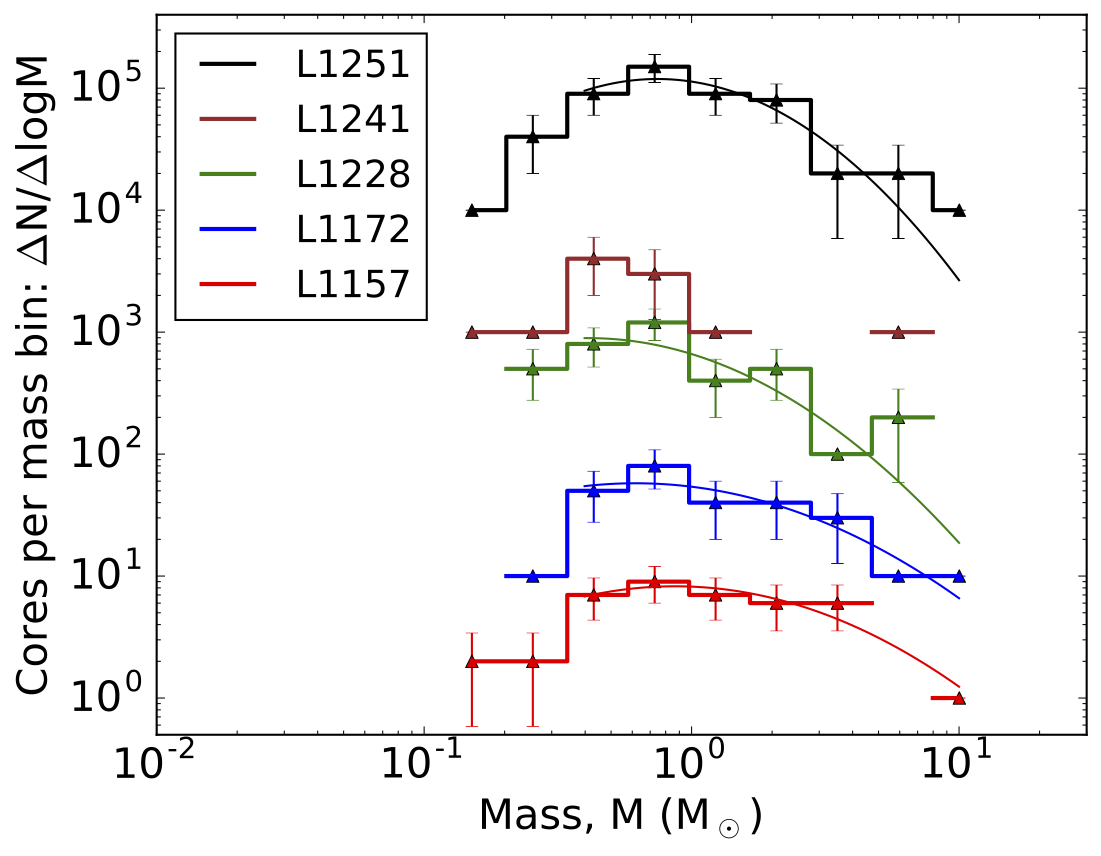}
\caption{Mass functions of robust prestellar cores in each Cepheus cloud:
L1157 (red), L1172 (blue), L1228 (green), L1241 (brown), and L1251 (black).
Error bars are from Poisson statistics.  To allow easier comparison of the
cloud CMFs, the values for the latter four clouds have been artificially
shifted up by multiplying each successively by an increasing order of
magnitude.  Lognormal fits to L1157, L1172, L1228, and L1251 at M $\geq$ 0.4
M$_{\odot}$ are also shown.  Note that the L1241 mass function was
unable to be fit by a lognormal.} 
\label{fig:fig14}
\end{figure*}

Figure \ref{fig:fig13} shows the CMF compiled from all five Cepheus clouds,
including 178 robust prestellar cores.  To explore deeper into the makeup
of the lognormal CMF, it is instructive to examine such CMFs drawn from the
individual Cepheus clouds.  Accordingly, Figure \ref{fig:fig14} shows the 
robust prestellar CMFs from each Cepheus cloud, assembled from populations
of 14 (L1241) to 53 (L1251) prestellar cores each (see Table \ref{tab:tab4}).
(Note that the CMFs in Figure 13 have been shifted vertically by multiples
of orders of magnitude so their shapes can be more easily compared.)
Some variation between the shapes of the individual cloud CMFs is seen. The
CMFs of L1172 and L1251 appear the most lognormal-like, but those of L1228
and L1241 look remarkably flat.  With five central bins of similar height,
the peak of L1157's CMF is hard to define and so this CMF seems somewhere
in between a lognormal and flat distribution.  Of course, the smaller
numbers of cores per bin in each cloud's CMFs make it hard to tell by eye
if the distributions  differ significantly. 

To provide a quantitative sense of the morphologies of the robust 
prestellar CMFs of each cloud, we fit lognormals to each CMF at and
above 0.4 M$_{\odot}$, the 90\% completeness core mass limit.  Table
\ref{tab:tab5} shows the results of the peak mass and width of the
lognormal fit to each cloud's robust prestellar CMF with uncertainties.
The results for the combined robust prestellar CMF for all five Cepheus
clouds are also listed.   A lognormal fit is not possible for L1241,
especially given the 0.4 M$_{\odot}$ lower limit restricting the sample
available to fit.  Also, the lognormal fit for L1228 is rather poor, as
evidenced by the large uncertainties in peak mass and width listed in
Table \ref{tab:tab5}.  Meanwhile, the peak masses of the robust
prestellar CMFs from the three clouds with reasonable lognormal fits
show some possible variation but are largely consistent within errors
with 0.6 M$_{\odot}$, a little higher but within errors of the peak
mass of the combined CMF.  (The peak mass of the combined CMF is
slightly lower than the peak masses of the CMFs of L1157, L1172, and
L1251 due to L1228 and L1241 adding mostly lower mass cores to the
ensemble.)  Furthermore, the widths of the three CMFs vary between
0.41 dex and 0.58 dex, but are largely consistent within 1~$\sigma$
of the 0.54 dex width of the combined CMF and the 0.55 dex width
of the Chabrier system IMF.  

A visual comparison of the individual column density and temperature
maps of the Cepheus clouds (see Figures \ref{fig:fig1}-\ref{fig:fig5}, 
\ref{fig:fig6}) shows apparent differences between them, with their characters
ranging from more diffuse (L1241) to highly filamentary (L1251) to
cluster dominated (L1172).  To place these differences in context,
recall that the Cepheus clouds are  distributed quite widely
on the sky.  Though they share a common name, the Cepheus clouds are
more of a loose association rather than a complex.  Hence, it is likely
that each is at its own stage and ability in producing stars, one related
to its total mass, its fraction of dense gas, its immediate physical
environment, and time.   With these differences, however, it is
perhaps not surprising that the individual cloud CMFs shown in Figure 
\ref{fig:fig14} differ in appearance, though the small sample sizes make
it difficult to be sure.  Nevertheless, the Cepheus clouds are broadly
similar in terms of their column density and temperature distributions
(see Figure \ref{fig:fig7}), which may explain the broad similarity of
the lognormal fits to the CMFs of at least three of the clouds.  The
combination of all five cloud CMFs, however, into a single CMF that is
similar in peak mass and width to those seen elsewhere (e.g., in Aquila)
is a remarkable demonstration of how the lognormal core mass function
arises out of a wide range of initial conditions.

The common peak and width of the robust prestellar CMFs of various nearby
clouds may speak to commonalities in how their cores (and stars) formed.
Namely, filaments are seen to be central to developing prestellar cores
in clouds.  Such filaments likely have a turbulent origin, as bulk motions
within clouds drive gas together to form sheets and filaments.  Indeed, 
sheets likely fragment very easily into filaments; see \cite{Andre14}.  
Given
their turbulent origins, the filaments themselves likely retain density 
perturbations along their lengths consistent with that turbulence.  For
example, \cite{Roy15} noted a potential link between the density fluctuations 
seeded by turbulence in filaments and the CMF.  Hence, the similar peaks
and widths of robust prestellar CMFs seen in different clouds could be itself
an expression of the common influence of turbulence forming filaments and
seeding their density fluctuations.  
As a result, the 
system IMF arises from molecular clouds where multiple lognormal populations
of robust prestellar cores inefficiently produce stars.

More star-forming clouds than just Cepheus and Aquila need to be examined
to see if the morphological differences and similarities of CMFs noted here
are found elsewhere,  e.g., see Fiorellino et al. (2020, submitted).
Furthermore, it is important to retain the perspective that the Cepheus clouds
(and even Aquila), with relatively modest star formation activity, are relatively 
minor contributors to the total amount of star formation in the Galaxy.
Indeed, the IMF would be set by the much larger engines
of star formation in the Galaxy, i.e., Giant Molecular Clouds.  Therefore,
future detailed examinations of the CMFs in those clouds are also necessary.

\section{Summary and Conclusions} \label{sec:sum}

In this paper, we examined the SPIRE and PACS observations of five Cepheus
Flare clouds, L1157, L1172, L1228, L1241, and L1251, that were obtained as
part of the {\it Herschel} Gould Belt Survey key project.  We summarize 
our main findings below.

1. The Cepheus clouds are relatively low mass clouds of the Gould Belt.
Given the column densities obtained from the {\it Herschel} data, we estimate
their masses to be 800 M$_{\odot}$ (L1157) to 2300 M$_{\odot}$ (L1241).  Like
other Gould Belt clouds observed by {\it Herschel}, the Cepheus clouds exhibit
significant filamentary substructure.  The character of the substructure does
appear to vary from cloud to cloud, from relatively diffuse (L1241) to more
pronounced (L1251) to cluster dominated (L1172).

2. The column density probability density functions (PDFs) obtained from
the {\it Herschel} observations are generally similar, peaking around 1
$\times$ 10$^{21}$ cm$^{-2}$ with a powerlaw-like decline to higher column
densities.  The column density PDF of L1241, however, is noticeably narrower
than the others.  The temperature PDFs of all five clouds show similar peaks
around 14 K.  Those of L1157, L1228, and L1251 have similar widths while
those of L1241 and L1172 are narrower and wider, respectively.  These
differences are likely due to the absence and presence of significant
internal heating sources (i.e,. protostars) in these latter clouds,
respectively

3. Using the {\it getsources} automated source identification algorithm,
we identify 832 dense cores in the {\it Herschel} 160-500 $\mu$m data.  Of
the dense cores, 504 are classified as starless (i.e., gravitationally
unbound) and 303 are classified as prestellar core candidates from their
locations in a mass vs.\ size diagram.  A subset of 178 cores from the
latter are considered ``robust'' prestellar cores.  The remaining 25 dense
cores were found to be coincident with sources extracted independently
from the 70 $\mu$m data alone, and are classified as protostellar cores.

4. The {\it getsources} algorithm also identified filamentary structure
in each Cepheus cloud.  With some variation between clouds, $\sim$75\% of 
starless cores and $\sim$80\% of prestellar cores are found to be coincident
with filaments.  L1251 has the highest percentages of cores coinicident with
filaments (80-100\%) while L1241 has the lowest (40-60\%).

5. The distribution of the background column densities 
of the population of prestellar cores in the Cepheus 
clouds peaks at 2-4 $\times$ 10$^{21}$ cm$^{-2}$,
but has a small tail leading up to background column densities 
of $\sim$20 $\times$ 10$^{21}$ cm$^{-2}$.  Approximately half of Cepheus'
candidate prestellar cores appear to have formed in filaments with line masses
within but at the lower end of the ``transcritical" range at T = 10 K, i.e.,
$M_{\rm line}$ = 8-32 M${\odot}$ pc$^{-1}$ and half in filaments with line
masses lower than that range.  (Further investigation of the opacity values
to use at lower extinctions is needed.)  In the former case, greater numbers
of cores are expected from widespread filament fragmentation while fewer are
expected in the latter due to fragmentation only occurring where localized
conditions warrant.  As a result, Cepheus is forming fewer cores than higher
column density clouds like Aquila.  

6. The mass function of robust prestellar cores (CMF) in all five Cepheus
clouds combined is lognormal in shape, with a peak mass of 0.56 M$_{\odot}$
and a width of 0.54 dex.  In comparison, the Aquila Rift CMF has a lognormal
shape as well, with a similar peak mass of 0.62 M$_{\odot}$ and a similar 
width of 0.48 dex.  The Cepheus CMF is consistent with the system IMF of 
\cite{Chabrier05}, assuming a mass-independent efficiency factor $\epsilon$
= 0.3-0.4.

7. The robust prestellar CMFs of L1157, L1172, and L1251 can be
also fit by lognormals with peak masses consistent within errors with
$\sim$0.6 M$_{\odot}$ and widths broadly consistent with the 0.55 dex
width of the system IMF of \cite{Chabrier05}.  The flatter CMFs of L1228
and L1241, however, were unable to be reliably fit with lognormals.

Though filamentary substructure is ubiquitous in the Cepheus clouds, this
substructure has mean column densities largely below or at the low end of
the range of ``transcritical" values associated with the radial cylindrical
fragmentation mechanism enabling core formation in filaments.  Core formation
and evolution in Cepheus appears to bridge that observed in lower and higher
column density clouds, such as Lupus and Aquila, respectively.  As a result,
both localized fragmentation where conditions permit and more widespread core 
fragmentation in transcritical filaments are both occurring, producing cores
of seemingly equal number.  The CMFs of the individual Cepheus clouds reflect
their current core formation potential but in aggregate they reflect a more
generalized distribution of prestellar core origins by encompassing a range of 
environments.  Indeed, the common width of the aggregate Cepheus CMF and the
Aquila CMF of $\sim$0.5 dex, hints at a common origin, perhaps due to seeding
of the fluctuations that evolve into cores by turbulence.   Assuming a
mass-independent factor for inefficiently converting core mass into stars, the
system IMF may originate from CMFs of similar width from numerous clouds.

\acknowledgments
We thank the input of an anonymous referee whose comments greatly improved
this paper.  J.D.F. and J.K. acknowledge the financial support of a Discovery
Grant from the NSERC of Canada.  N.S. and S.B. acknowledge support by the
Agence National de Recherche (ANR/France) and the Deutsche Forschungsgemeinschaft 
(DFG/Germany) through the project ``GENESIS" (ANR-16-CE92-0035-01/DFG1591/2-1).
SPIRE has been developed by a consortium of institutes led by Cardiff Univ. 
(UK) and including: Univ. Lethbridge (Canada); NAOC (China); CEA, LAM (France); 
IFSI, Univ. Padua (Italy); IAC (Spain); Stockholm Observatory (Sweden); Imperial 
College London, RAL, UCL- MSSL, UKATC, Univ. Sussex (UK); and Caltech, JPL, NHSC, 
Univ. Colorado (USA). This development has been supported by national funding 
agencies: CSA (Canada); NAOC (China); CEA, CNES, CNRS (France); ASI (Italy); 
MCINN (Spain); SNSB (Sweden); STFC, UKSA (UK); and NASA (USA). PACS has been 
developed by a consortium of institutes led by MPE (Germany) and including UVIE 
(Austria); KUL, CSL, IMEC (Belgium); CEA, OAMP (France); MPIA (Germany); IFSI, 
OAP/AOT, OAA/CAISMI, LENS, SISSA (Italy); IAC (Spain). This development has been 
supported by the funding agencies BMVIT (Austria), ESA-PRODEX (Belgium), CEA/CNES 
(France), DLR (Germany), ASI (Italy), and CICT/MCT (Spain).  This work received 
support from a Discovery grant from the National Science and Engineering Council 
of Canada. This research has made use of the SIMBAD database, operated at CDS, 
Strasbourg (France), and of the NASA/IPAC Extragalactic Database (NED), operated 
by the Jet Propulsion Laboratory, California Institute of Technology, under 
contract with the National Aeronautics and Space Administration.  PACS has
been developed by a consortium of institutes led by MPE (Germany) and including
UVIE (Austria); KUL, CSL, IMEC (Belgium); CEA, OAMP (France); MPIA (Germany);
IFSI, OAP/AOT, OAA/CAISMI, LENS, SISSA (Italy); IAC (Spain).  This development
has been supported by the funding agencies BMVIT (Austria), ESA-PRODEX (Belgium),
CEA/CNES (France), DLR (Germany), ASI (Italy), and CICT/MCT (Spain)

%

\vspace{5mm}
\facility{{\it Herschel Space Observatory}}


\software{{\it getsources} \citep{Men'shchikov12},
		  HIPE \citep{Ott11},  
          Scanamorphos \citep{Roussel13} 
          }



\appendix

\section{{\it Herschel}\/ Observations of Cepheus Clouds}

\begin{figure*}
\gridline{\fig{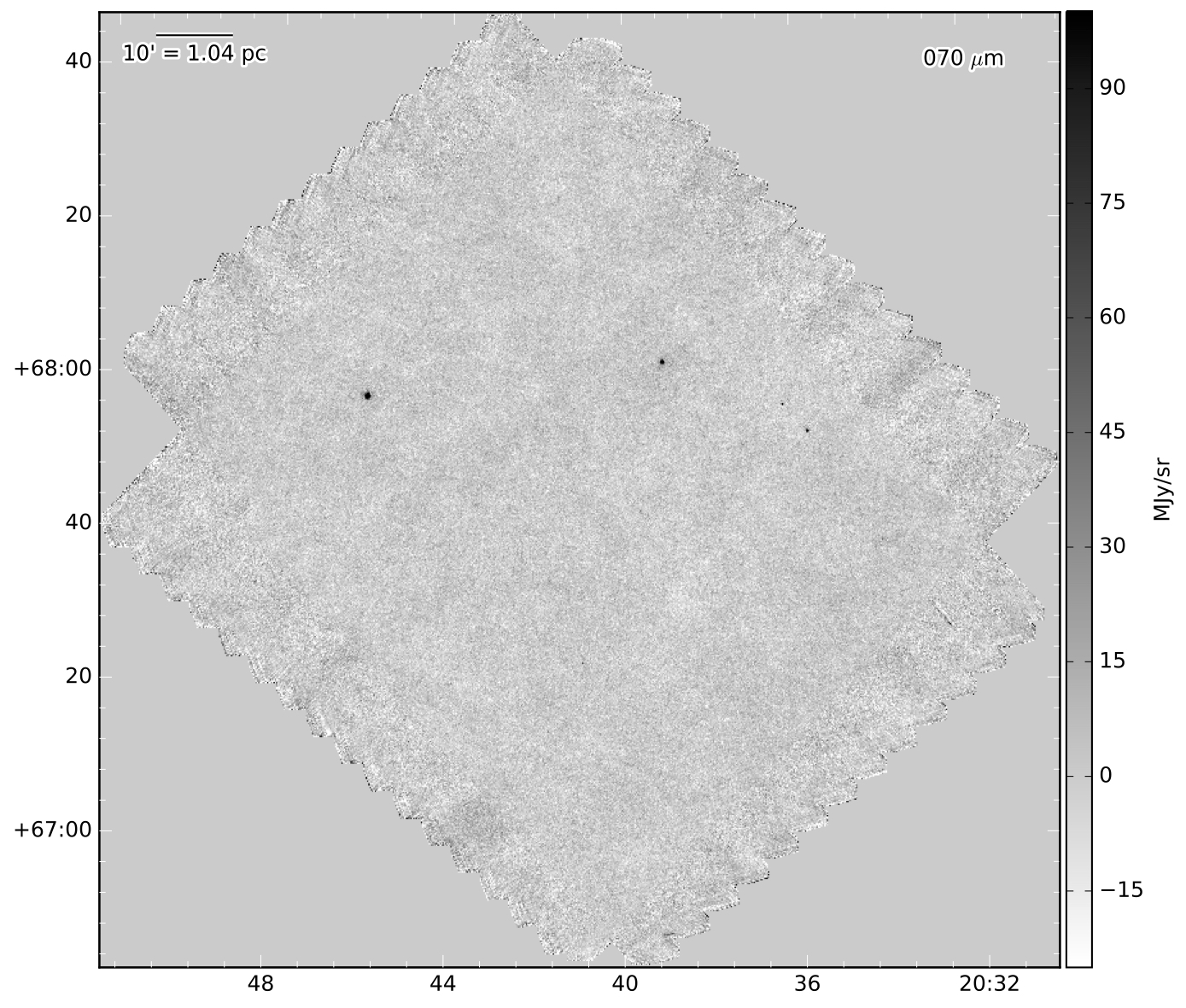}{0.4\textwidth}{(a)}
          \fig{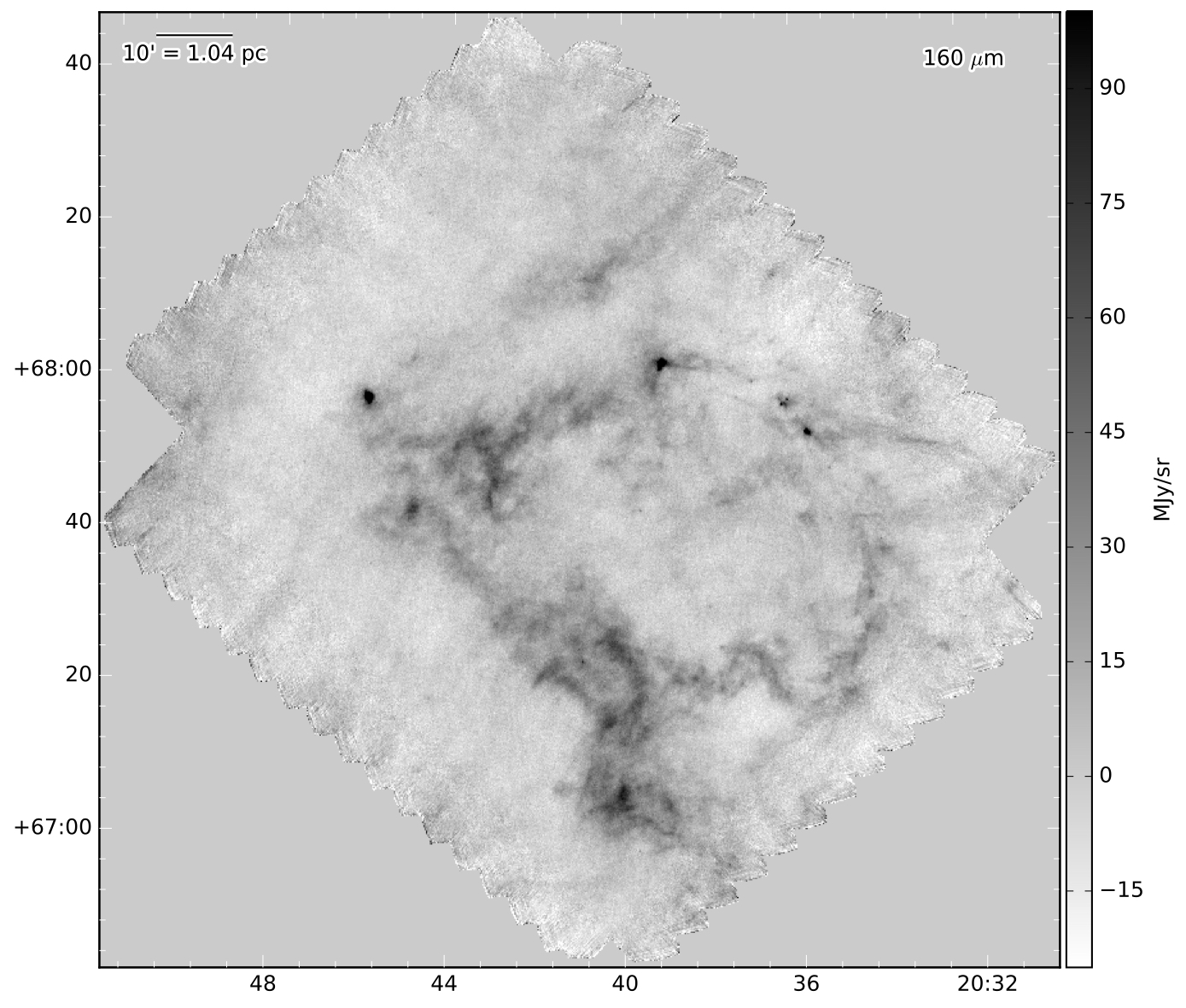}{0.4\textwidth}{(b)}
          }
\gridline{\fig{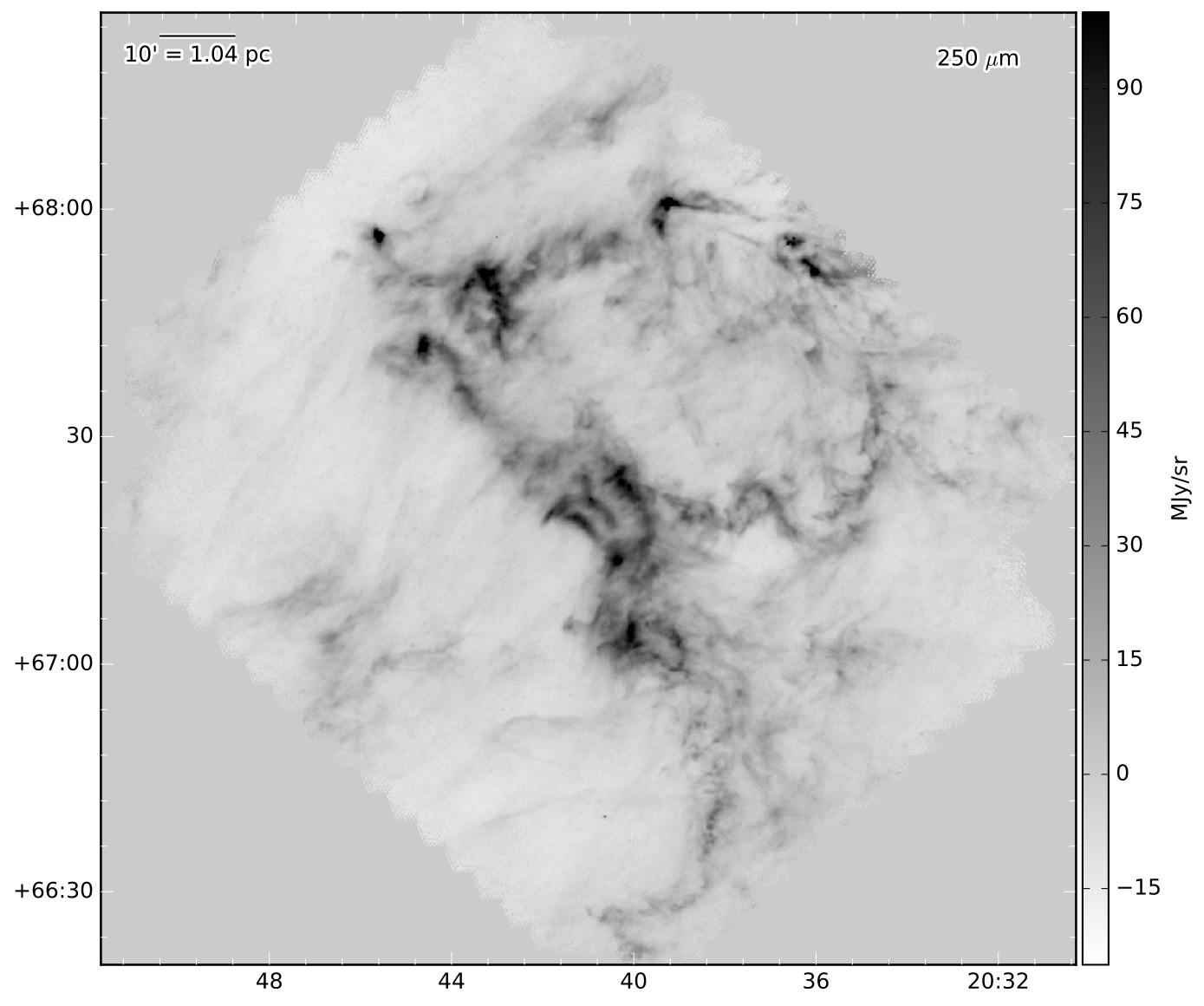}{0.4\textwidth}{(c)}
          \fig{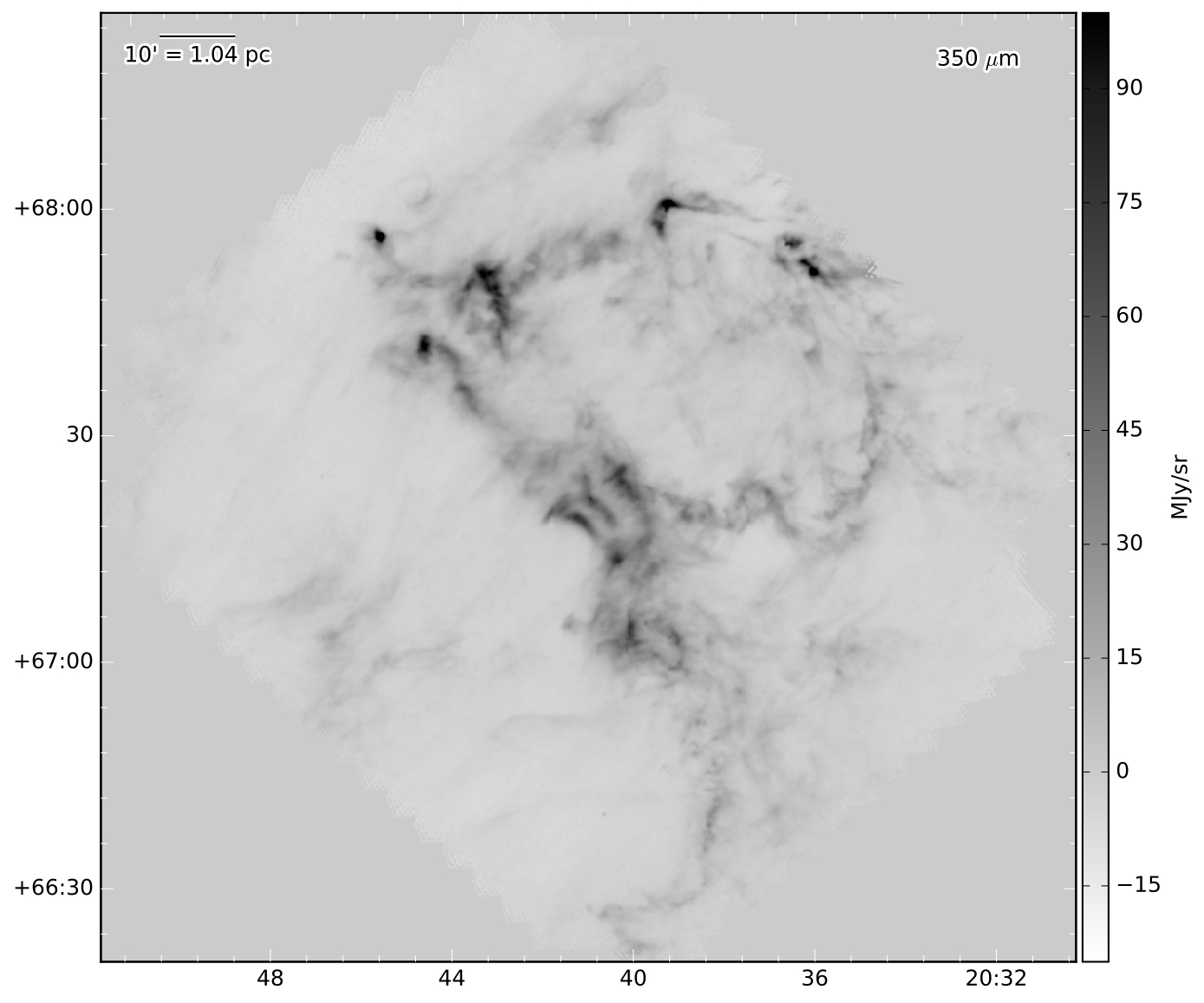}{0.4\textwidth}{(d)}
          }
\gridline{\fig{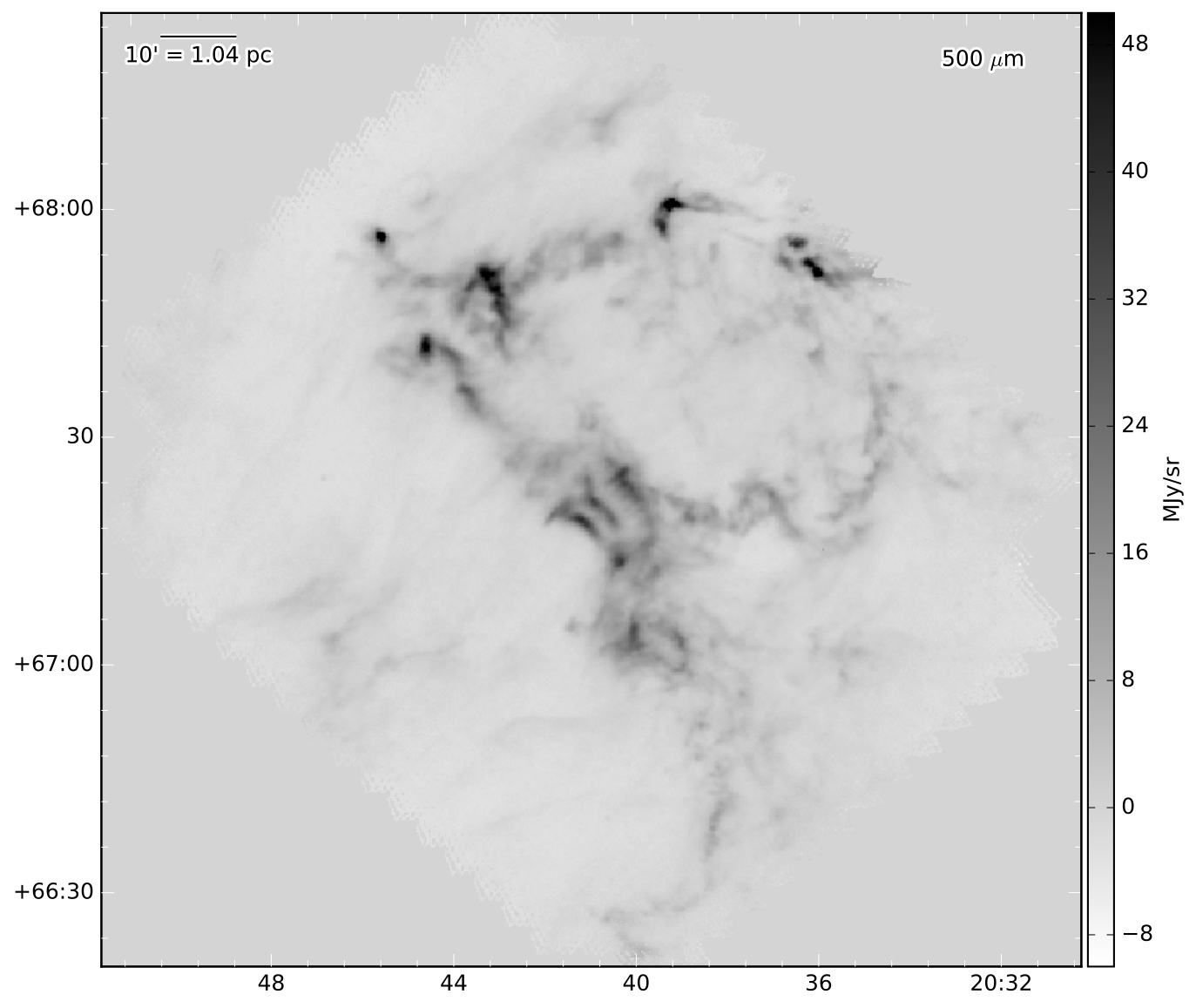}{0.4\textwidth}{(e)}
          }
\caption{{\it Herschel}\/ observations of L1157 at
(a) 70 $\mu$m; $-$25~MJy~sr$^{-1}$ to 100~MJy~sr$^{-1}$, (b) 160 $\mu$m; $-$25~MJy~sr$^{-1}$ to 100~MJy~sr$^{-1}$, (c) 160 $\mu$m; $-$25~MJy~sr$^{-1}$ to 100~MJy~sr$^{-1}$, (d) 350 $\mu$m; $-$25~MJy~sr$^{-1}$ to 100~MJy~sr$^{-1}$, and
(e) 500 $\mu$m; $-$10~MJy~sr$^{-1}$ to 50~MJy~sr$^{-1}$.
\label{fig:a1}}
\end{figure*}

\begin{figure*}
\gridline{\fig{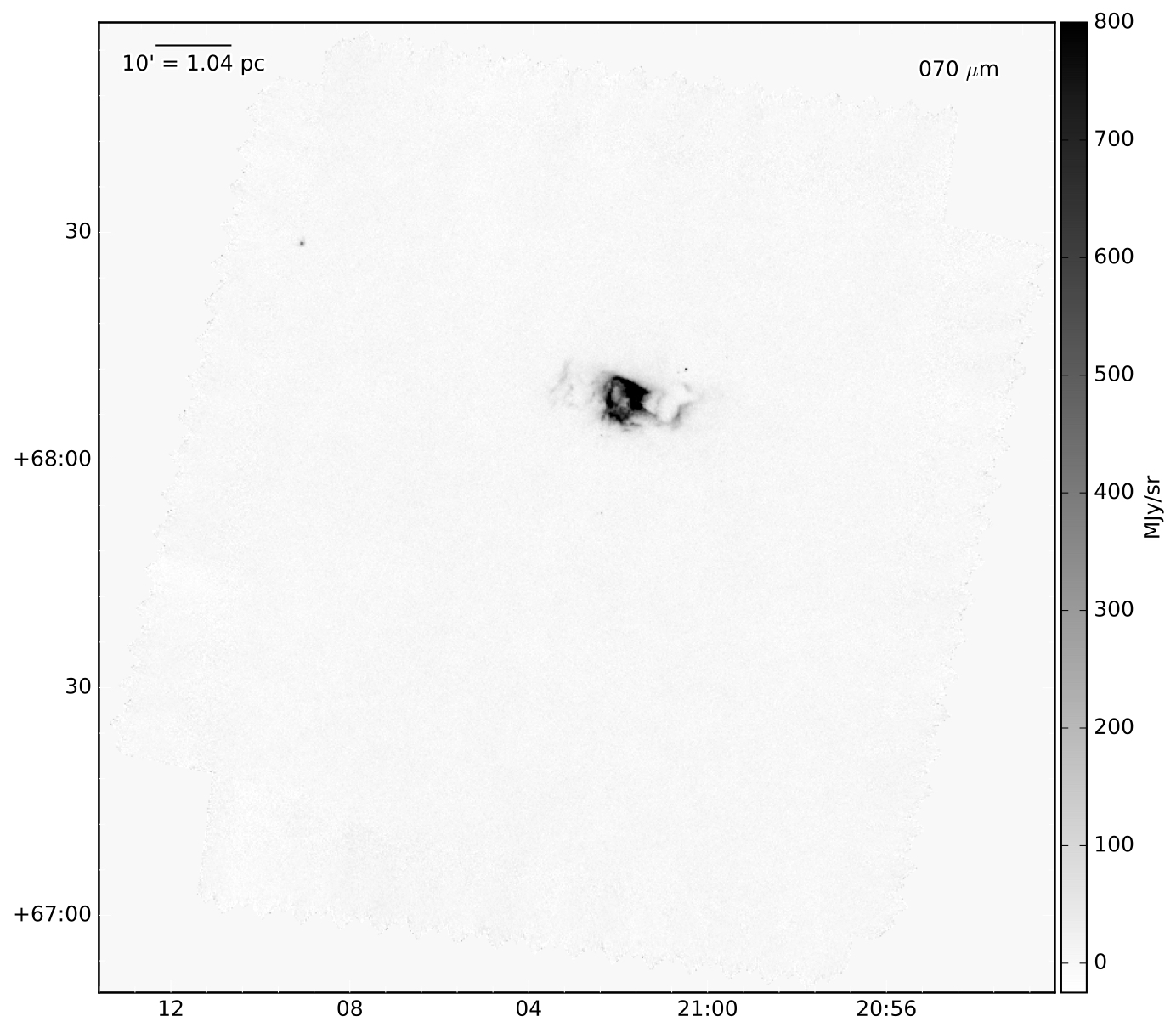}{0.4\textwidth}{(a)}
          \fig{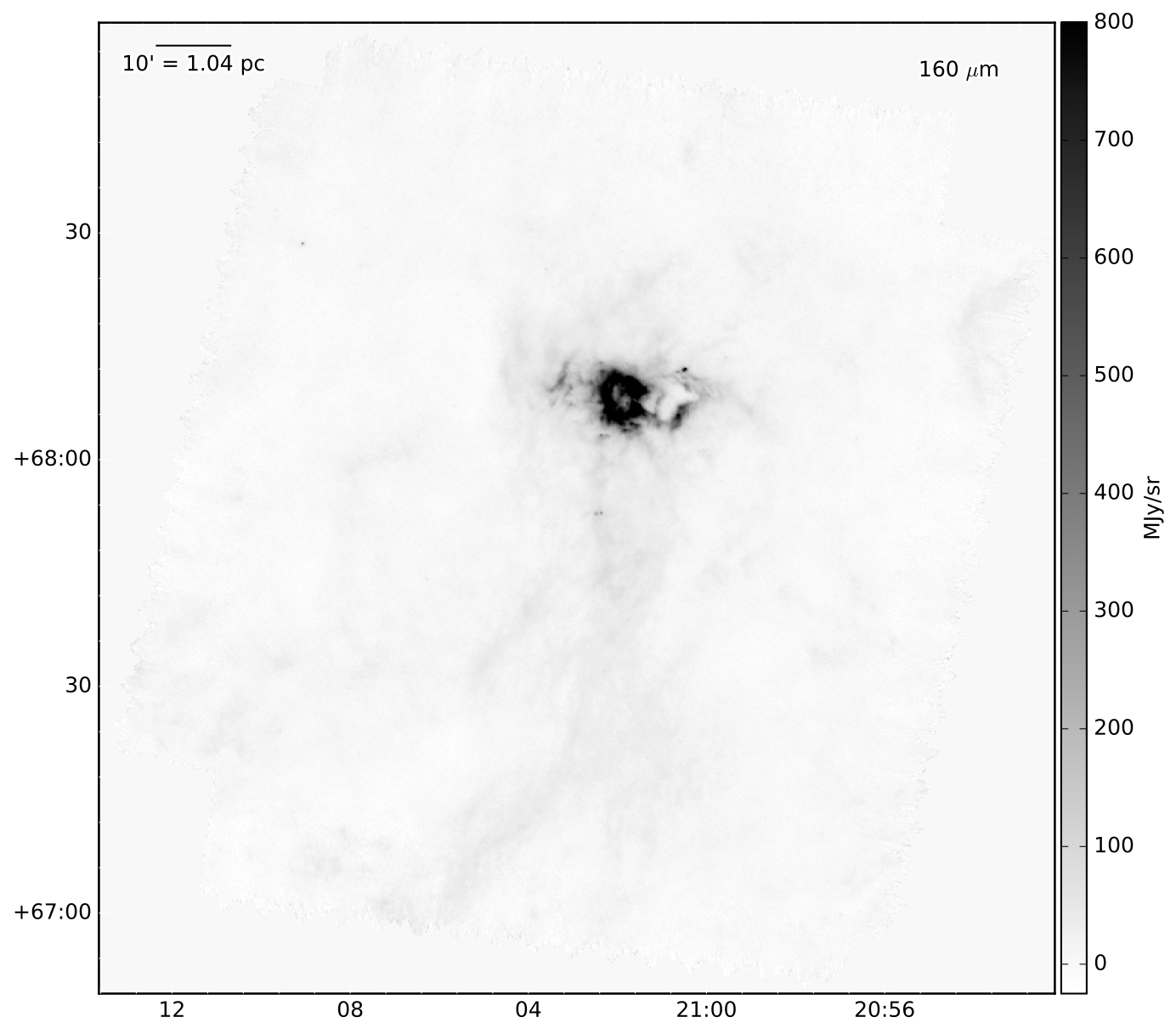}{0.4\textwidth}{(b)}
          }
\gridline{\fig{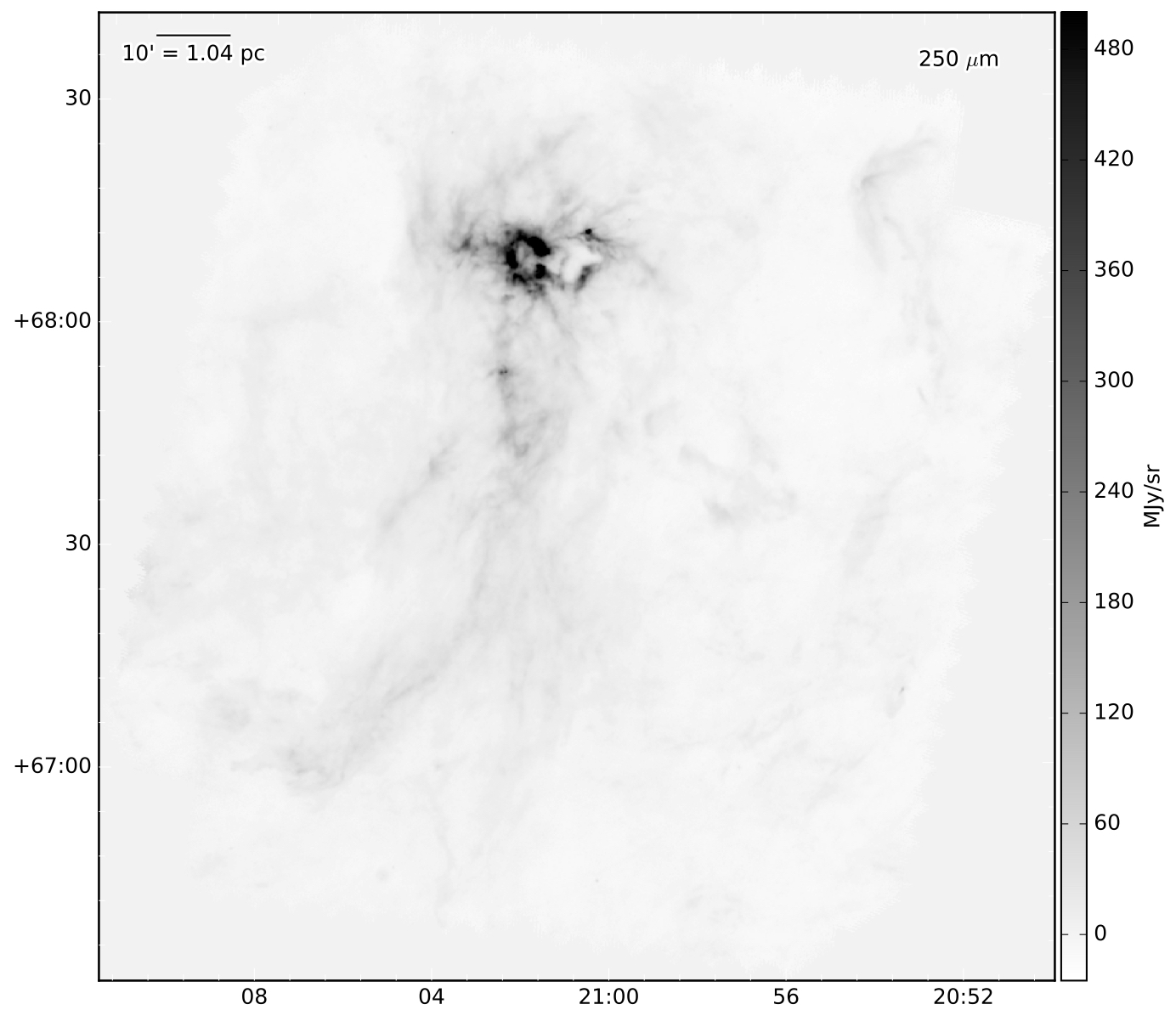}{0.4\textwidth}{(c)}
          \fig{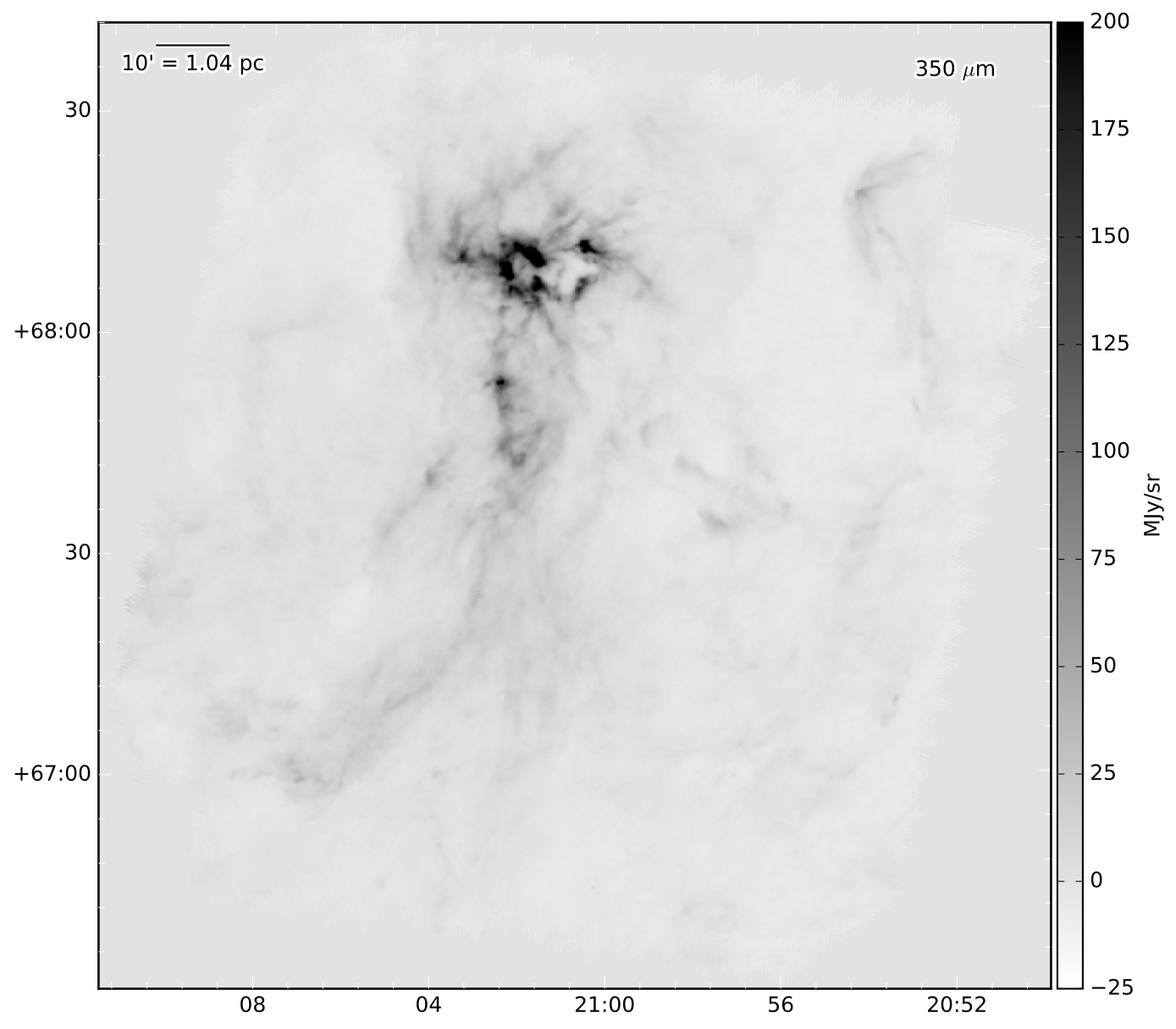}{0.4\textwidth}{(d)}
          }
\gridline{\fig{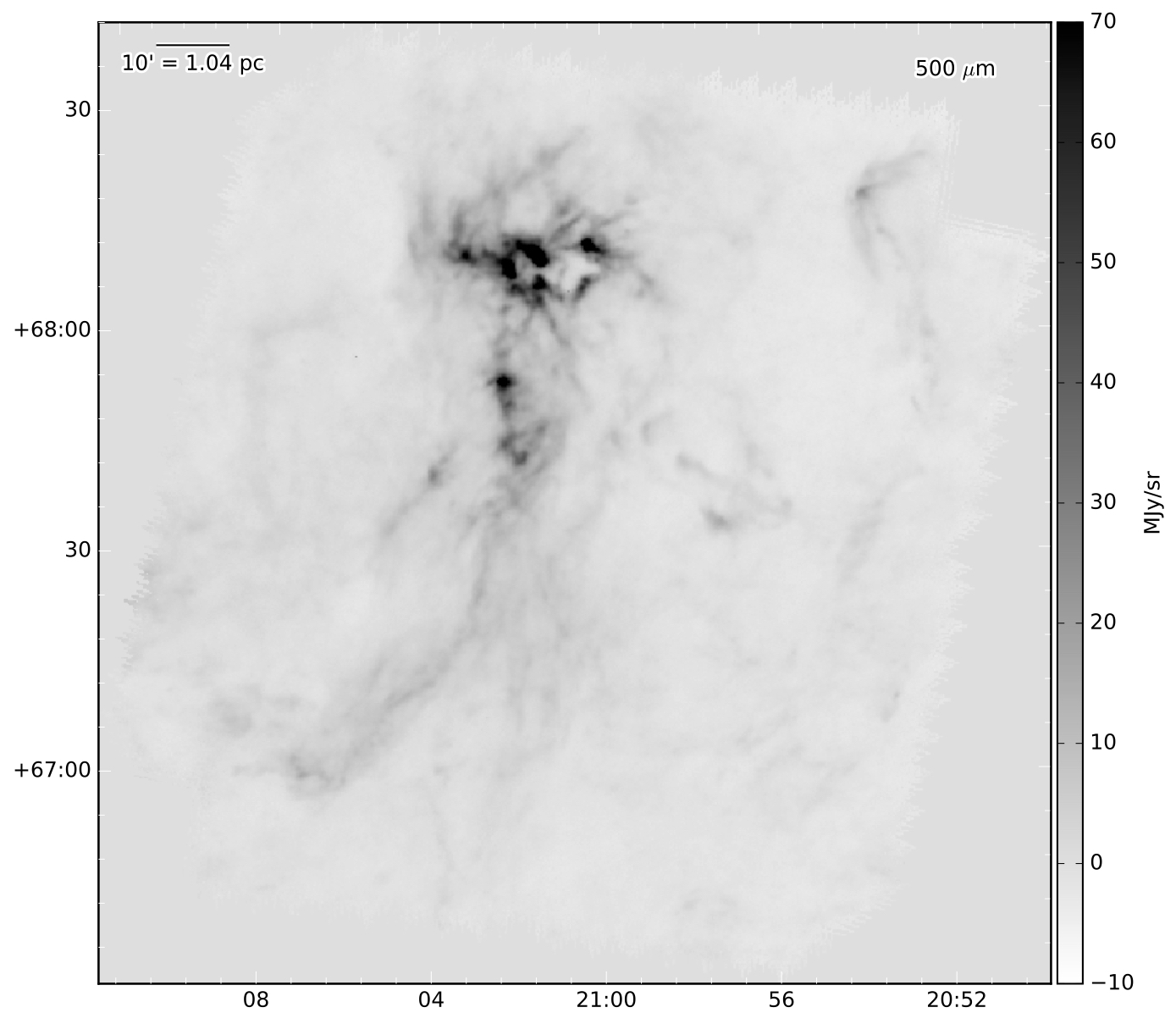}{0.4\textwidth}{(e)}
          }
\caption{{\it Herschel}\/ observations of L1172 at
(a) 70 $\mu$m; $-$25~MJy~sr$^{-1}$ to 100~MJy~sr$^{-1}$, (b) 160 $\mu$m; $-$25~MJy~sr$^{-1}$ to 100~MJy~sr$^{-1}$, (c) 160 $\mu$m; $-$25~MJy~sr$^{-1}$ to 100~MJy~sr$^{-1}$, (d) 350 $\mu$m; $-$25~MJy~sr$^{-1}$ to 100~MJy~sr$^{-1}$, and
(e) 500 $\mu$m; $-$10~MJy~sr$^{-1}$ to 50~MJy~sr$^{-1}$.
\label{fig:a2}}
\end{figure*}

\begin{figure*}
\gridline{\fig{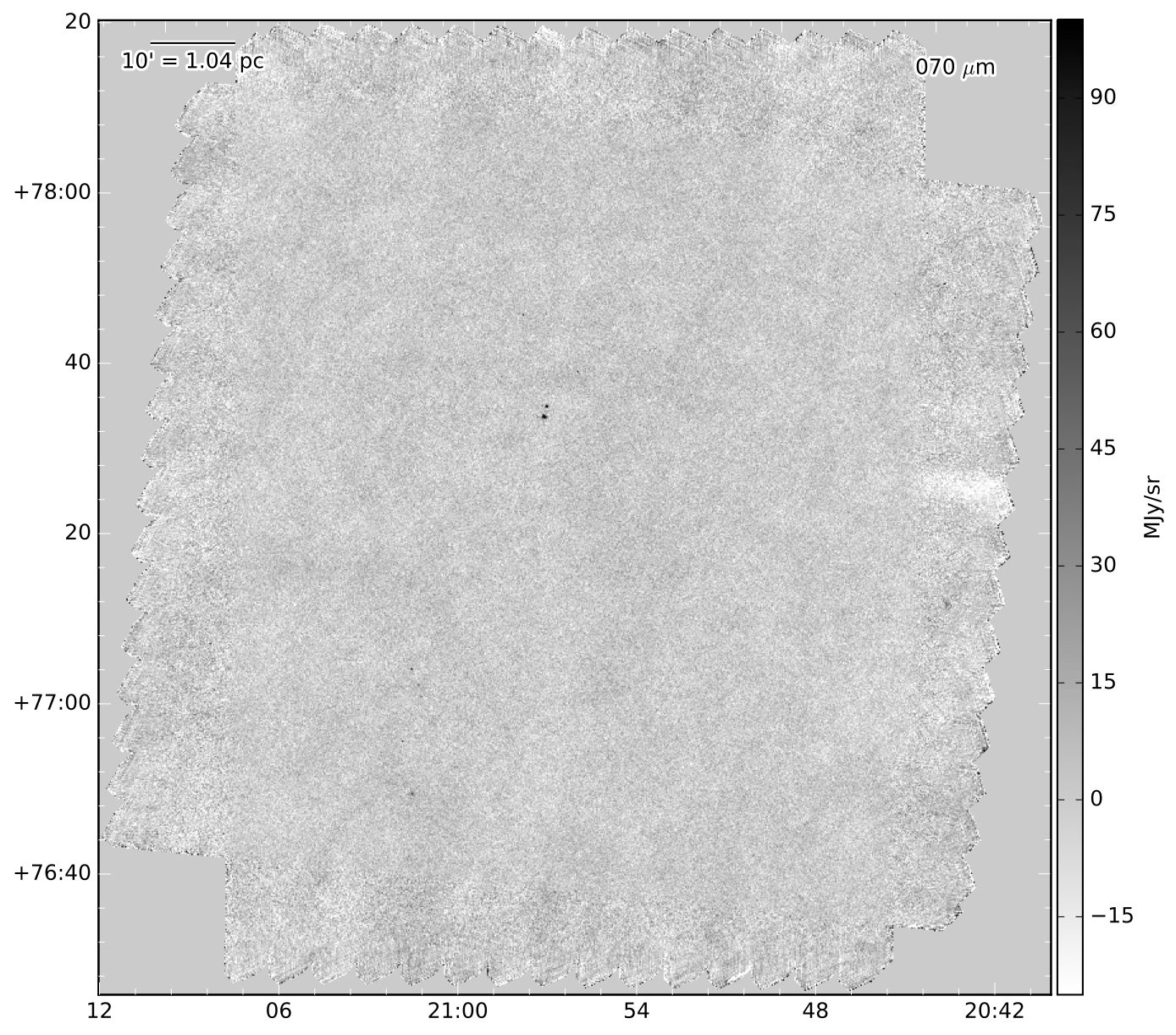}{0.4\textwidth}{(a)}
          \fig{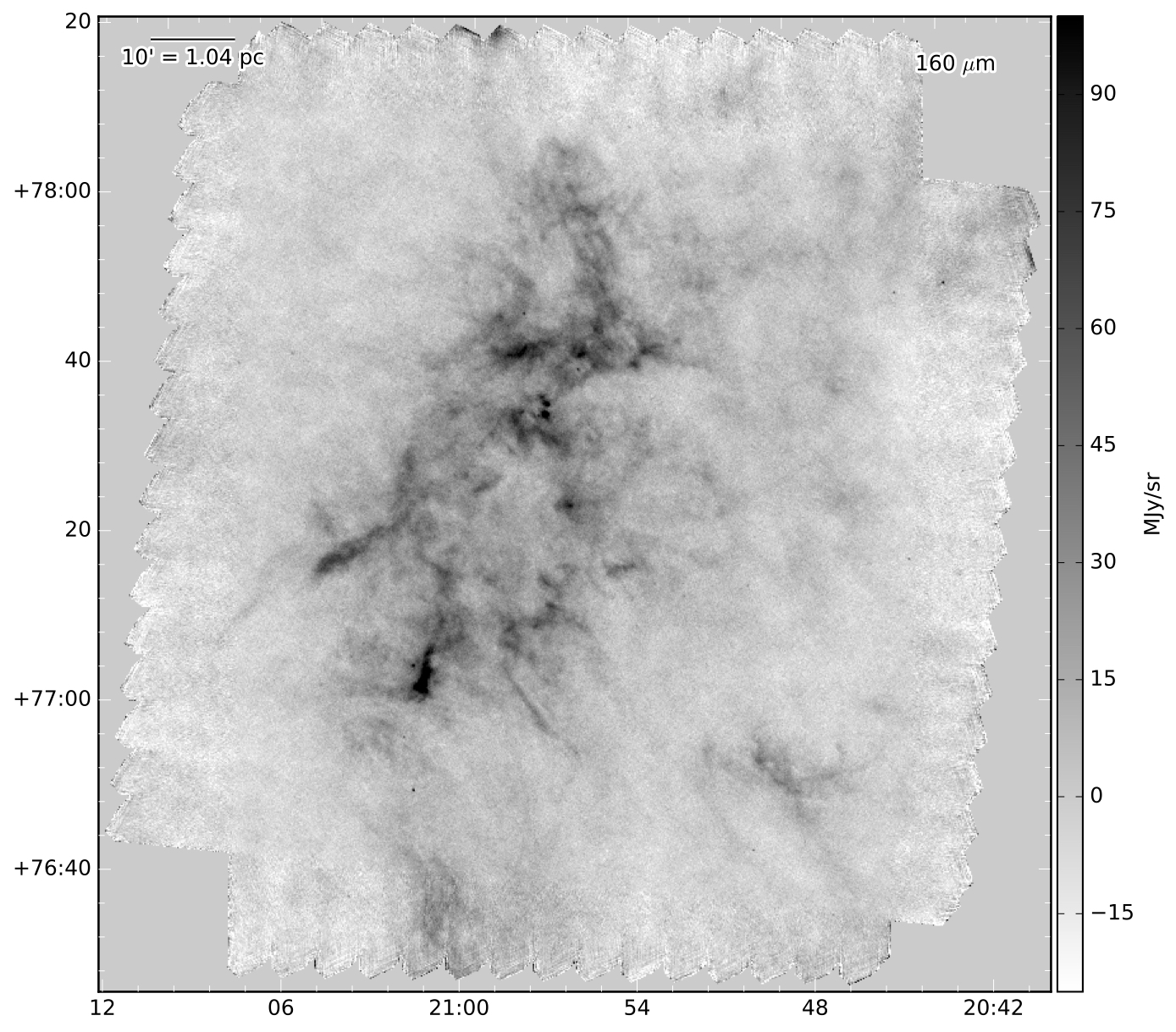}{0.4\textwidth}{(b)}
          }
\gridline{\fig{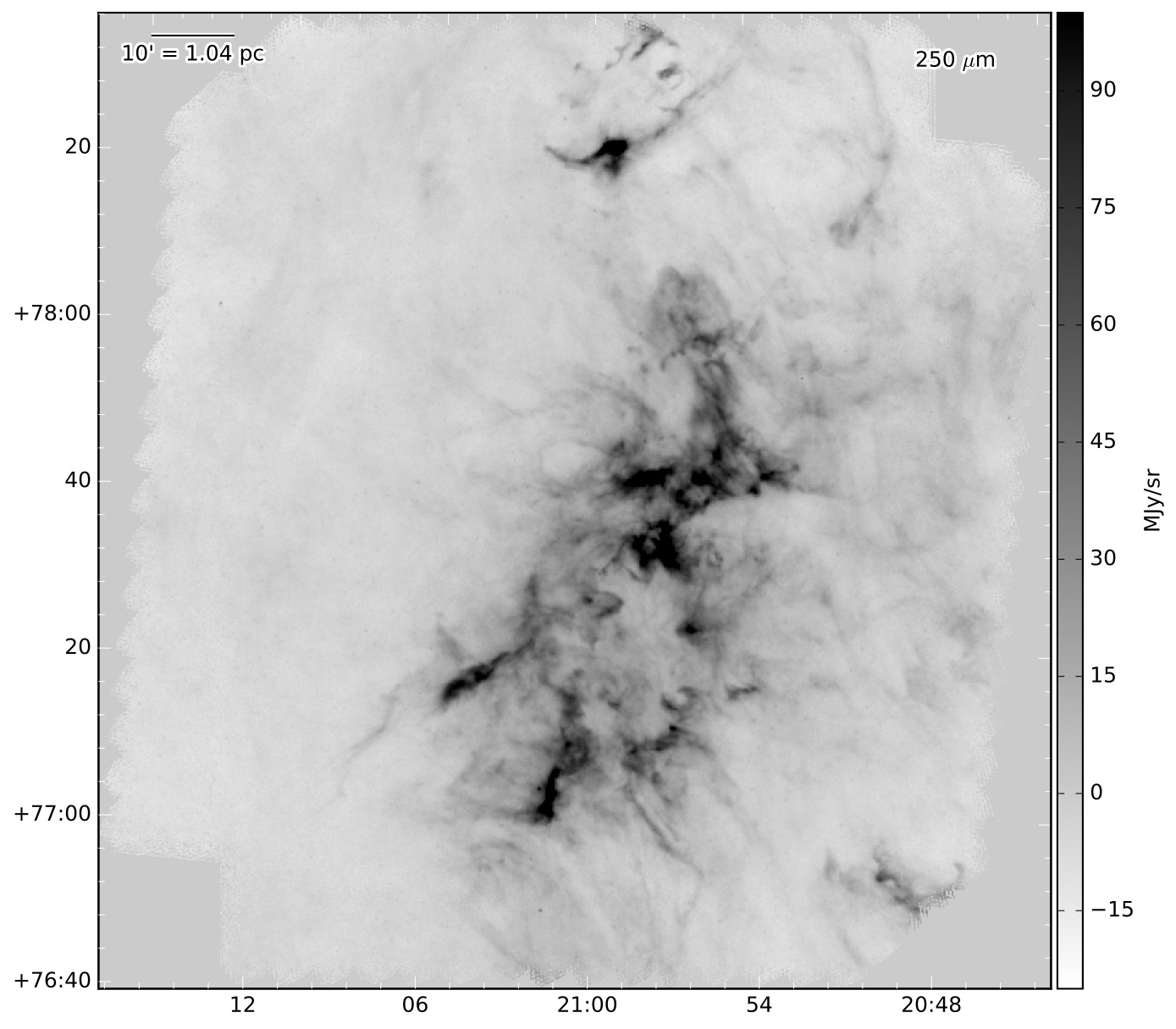}{0.4\textwidth}{(c)}
          \fig{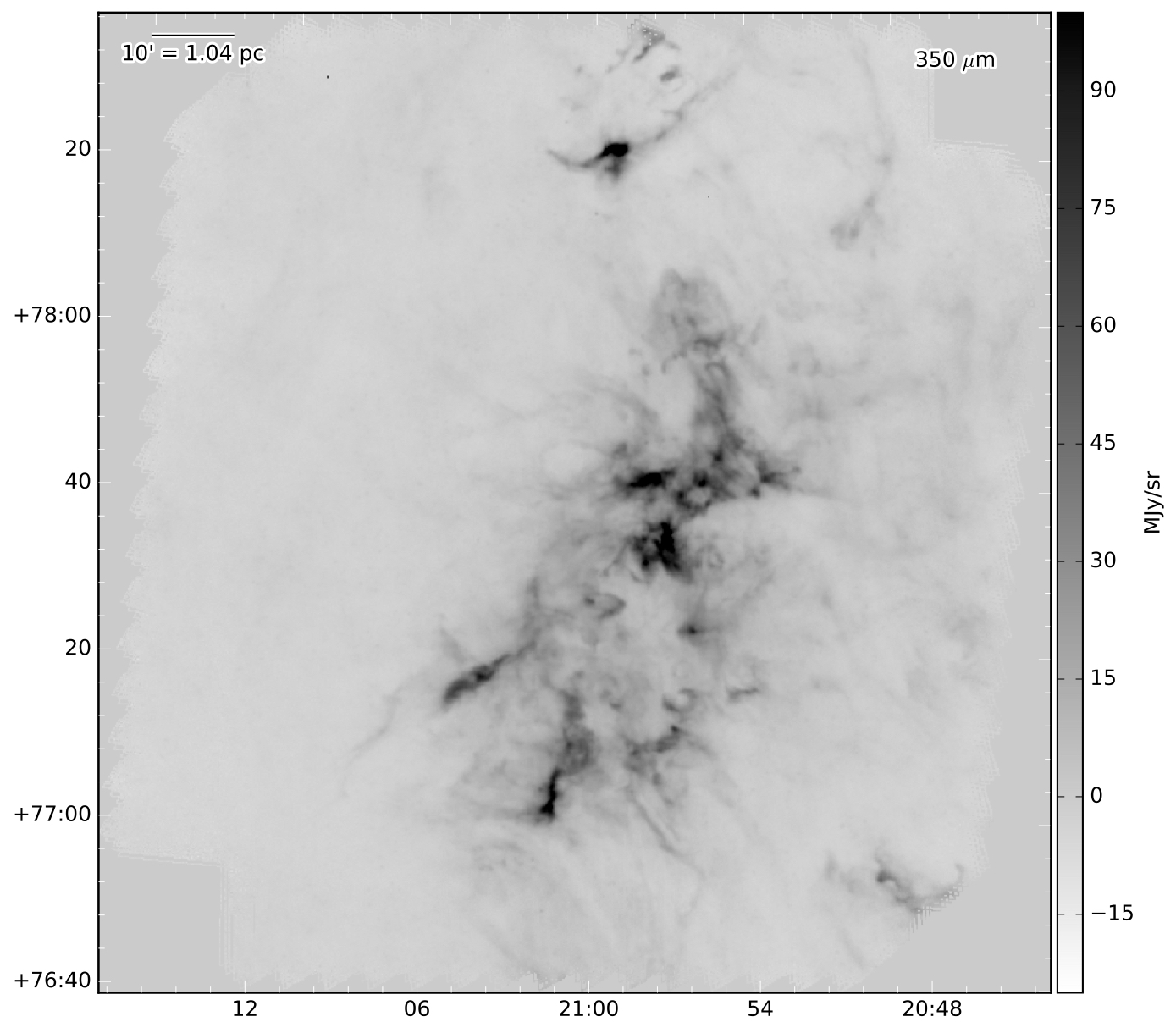}{0.4\textwidth}{(d)}
          }
\gridline{\fig{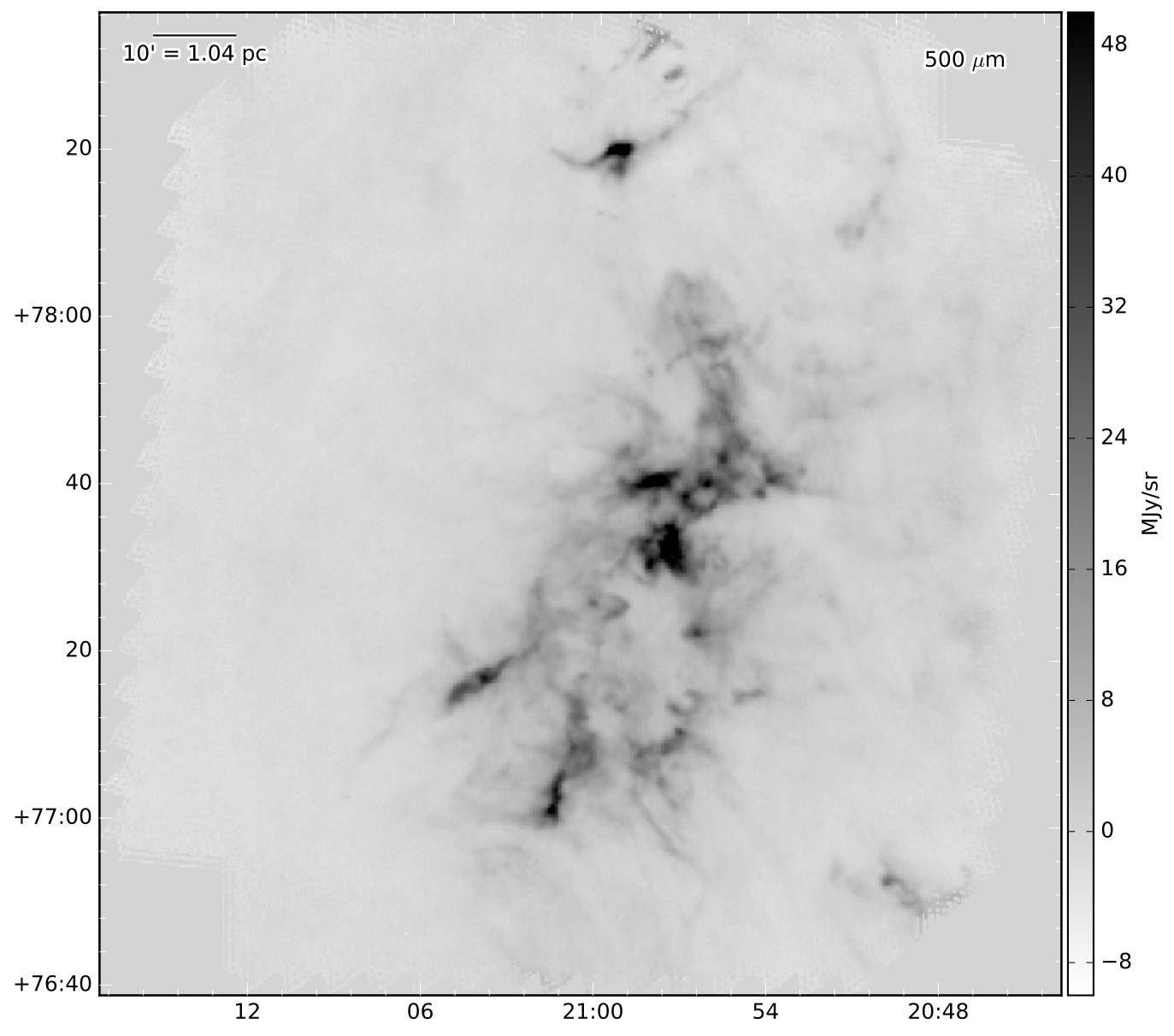}{0.4\textwidth}{(e)}
          }
\caption{{\it Herschel}\/ observations of L1228 at
(a) 70 $\mu$m; $-$25~MJy~sr$^{-1}$ to 100~MJy~sr$^{-1}$, (b) 160 $\mu$m; $-$25~MJy~sr$^{-1}$ to 100~MJy~sr$^{-1}$, (c) 160 $\mu$m; $-$25~MJy~sr$^{-1}$ to 100~MJy~sr$^{-1}$, (d) 350 $\mu$m; $-$25~MJy~sr$^{-1}$ to 100~MJy~sr$^{-1}$, and
(e) 500 $\mu$m; $-$10~MJy~sr$^{-1}$ to 50~MJy~sr$^{-1}$.
\label{fig:a3}}
\end{figure*}

\begin{figure*}
\gridline{\fig{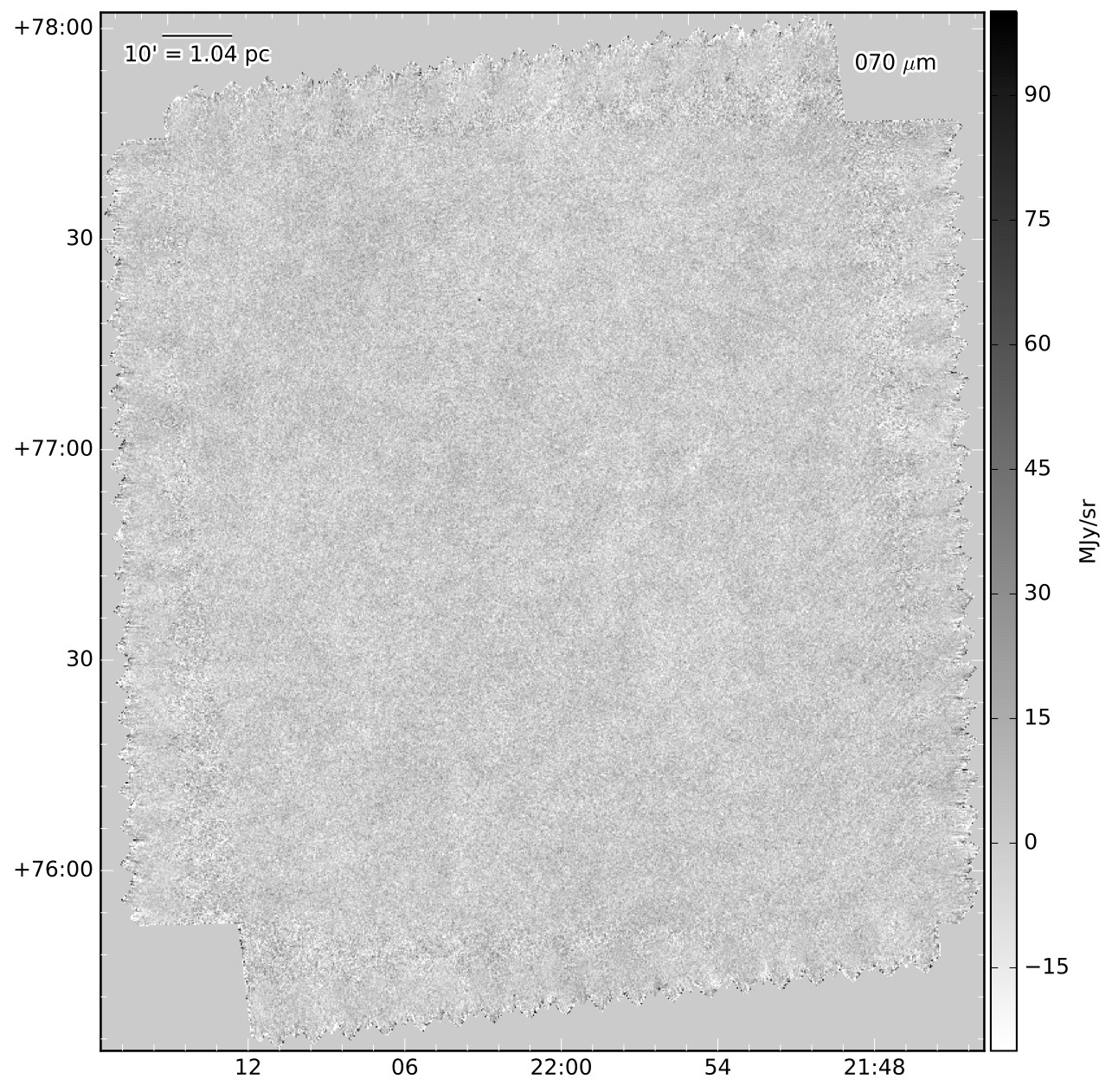}{0.35\textwidth}{(a)}
          \fig{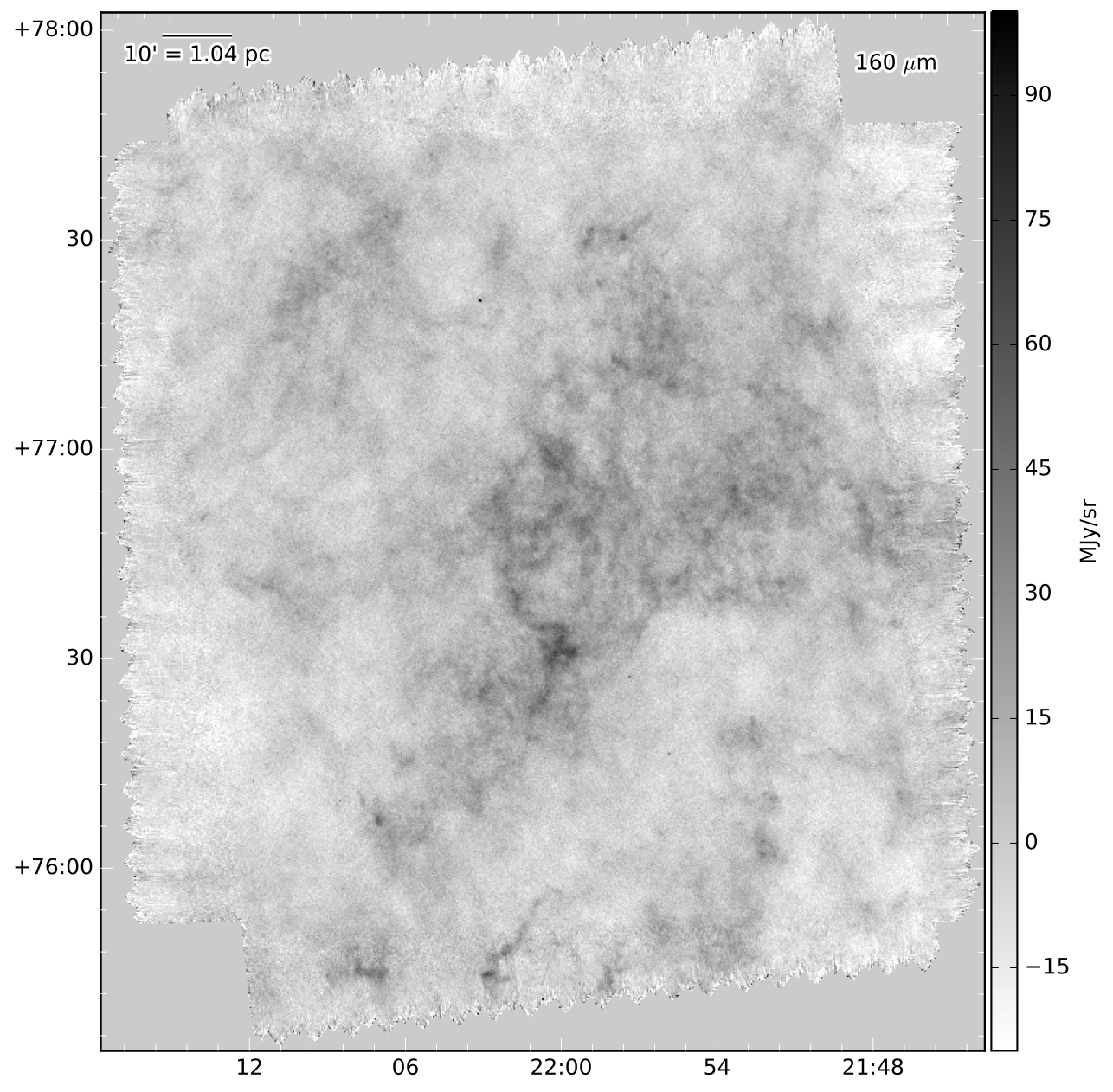}{0.35\textwidth}{(b)}
          }
\gridline{\fig{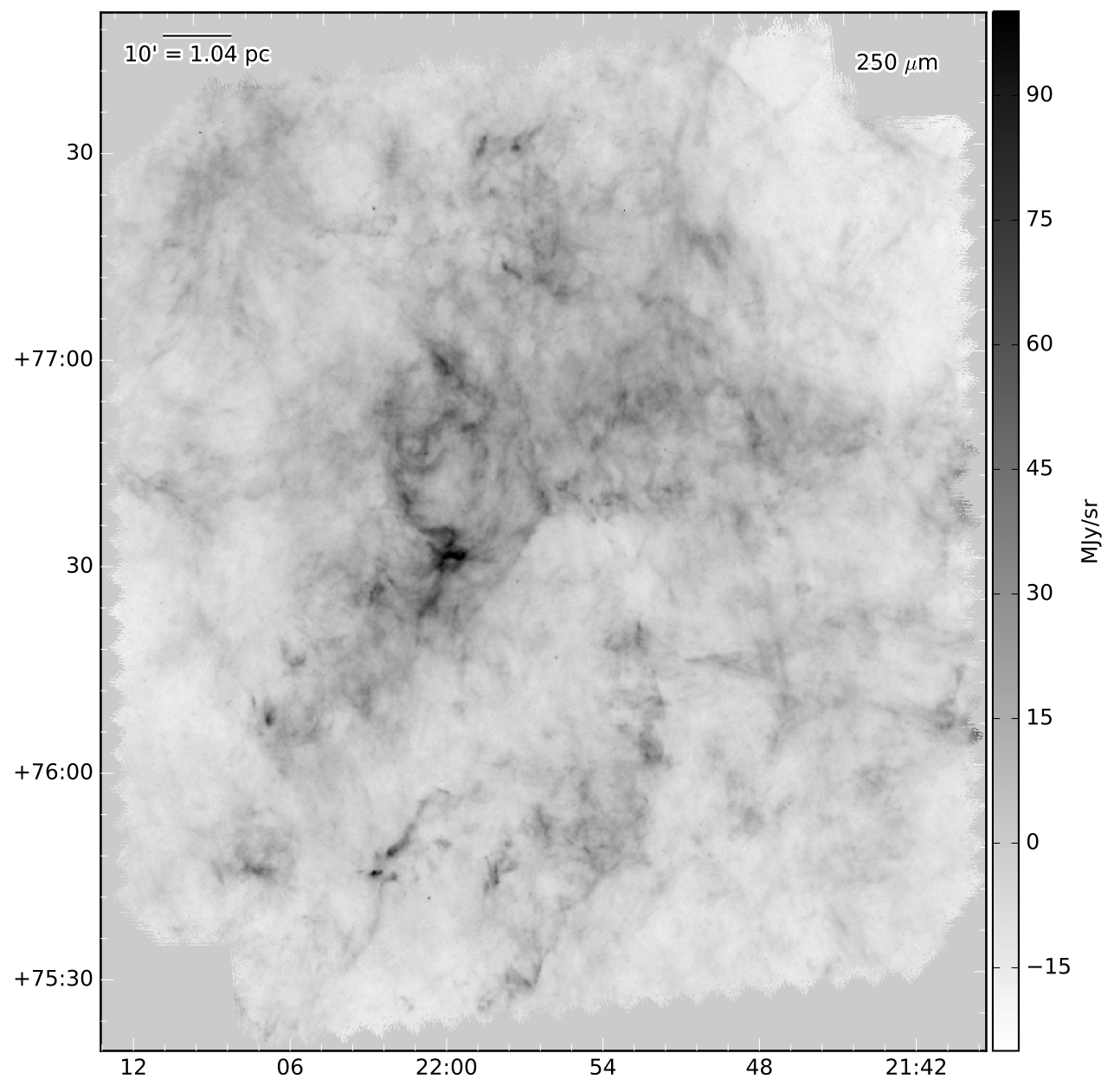}{0.35\textwidth}{(c)}
          \fig{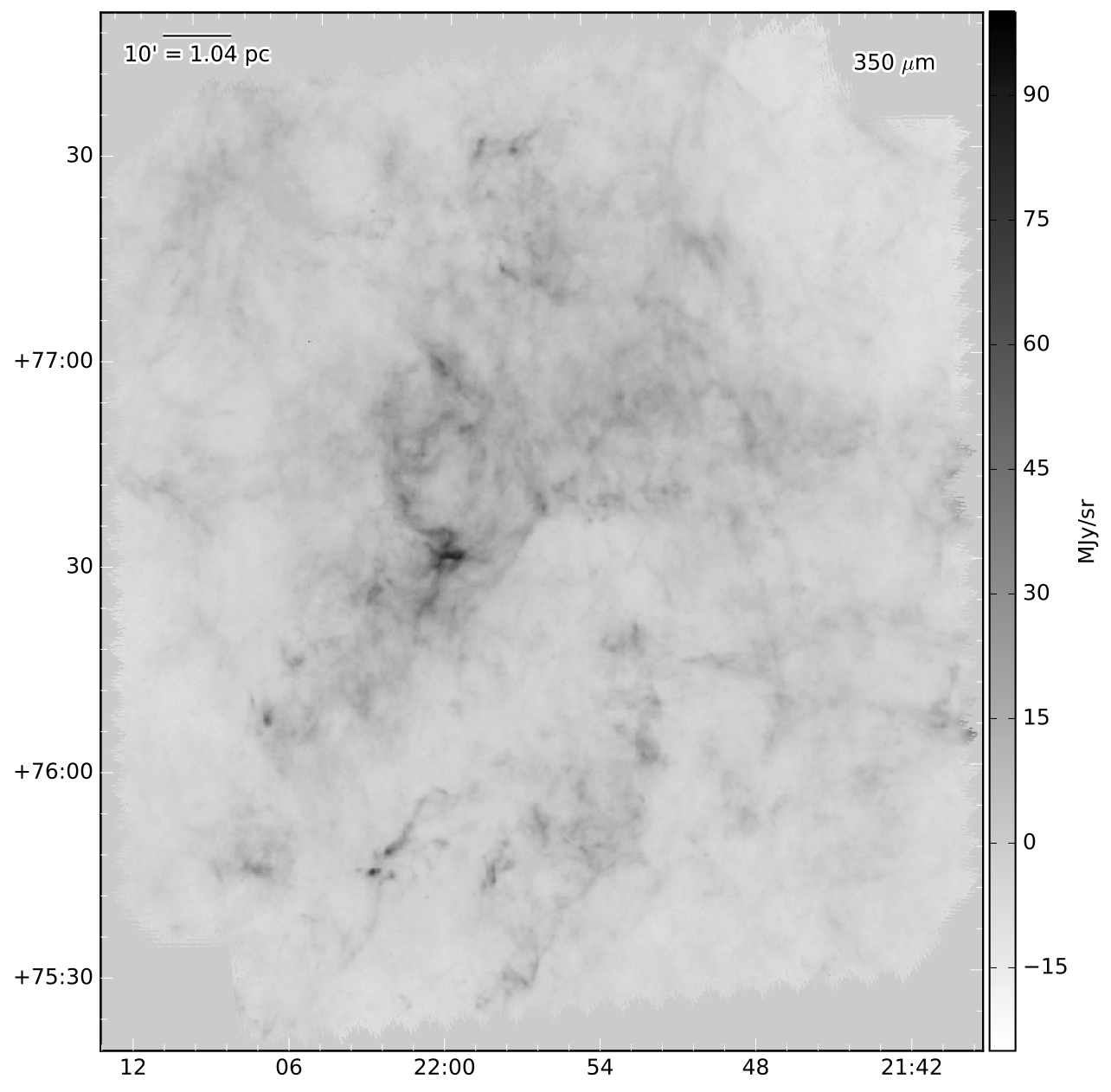}{0.35\textwidth}{(d)}
          }
\gridline{\fig{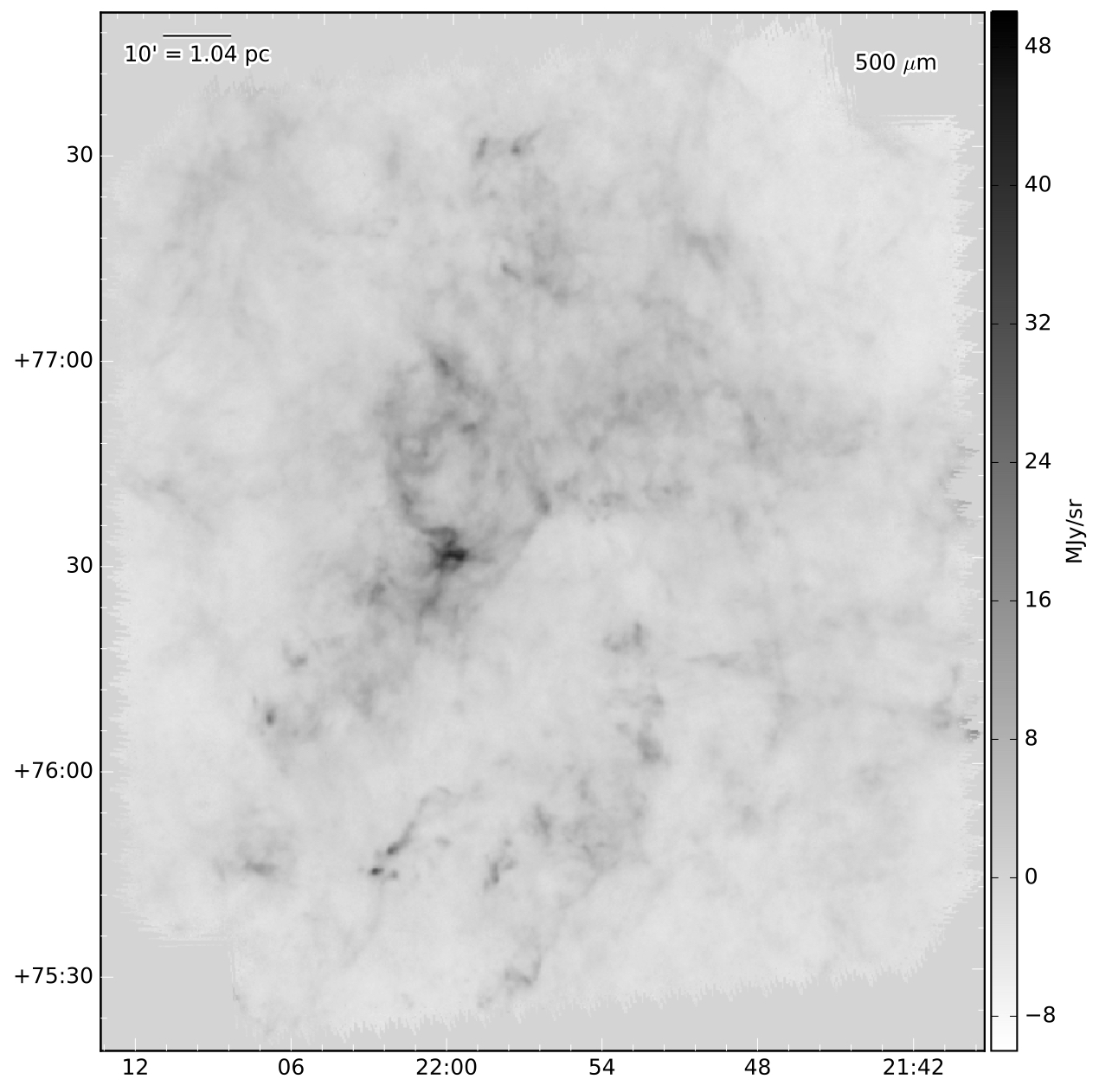}{0.35\textwidth}{(e)}
          }
\caption{{\it Herschel}\/ observations of L1241 at
(a) 70 $\mu$m; $-$25~MJy~sr$^{-1}$ to 100~MJy~sr$^{-1}$, (b) 160 $\mu$m; $-$25~MJy~sr$^{-1}$ to 100~MJy~sr$^{-1}$, (c) 160 $\mu$m; $-$25~MJy~sr$^{-1}$ to 100~MJy~sr$^{-1}$, (d) 350 $\mu$m; $-$25~MJy~sr$^{-1}$ to 100~MJy~sr$^{-1}$, and
(e) 500 $\mu$m; $-$10~MJy~sr$^{-1}$ to 50~MJy~sr$^{-1}$.
\label{fig:a4}}
\end{figure*}

\begin{figure*}
\gridline{\fig{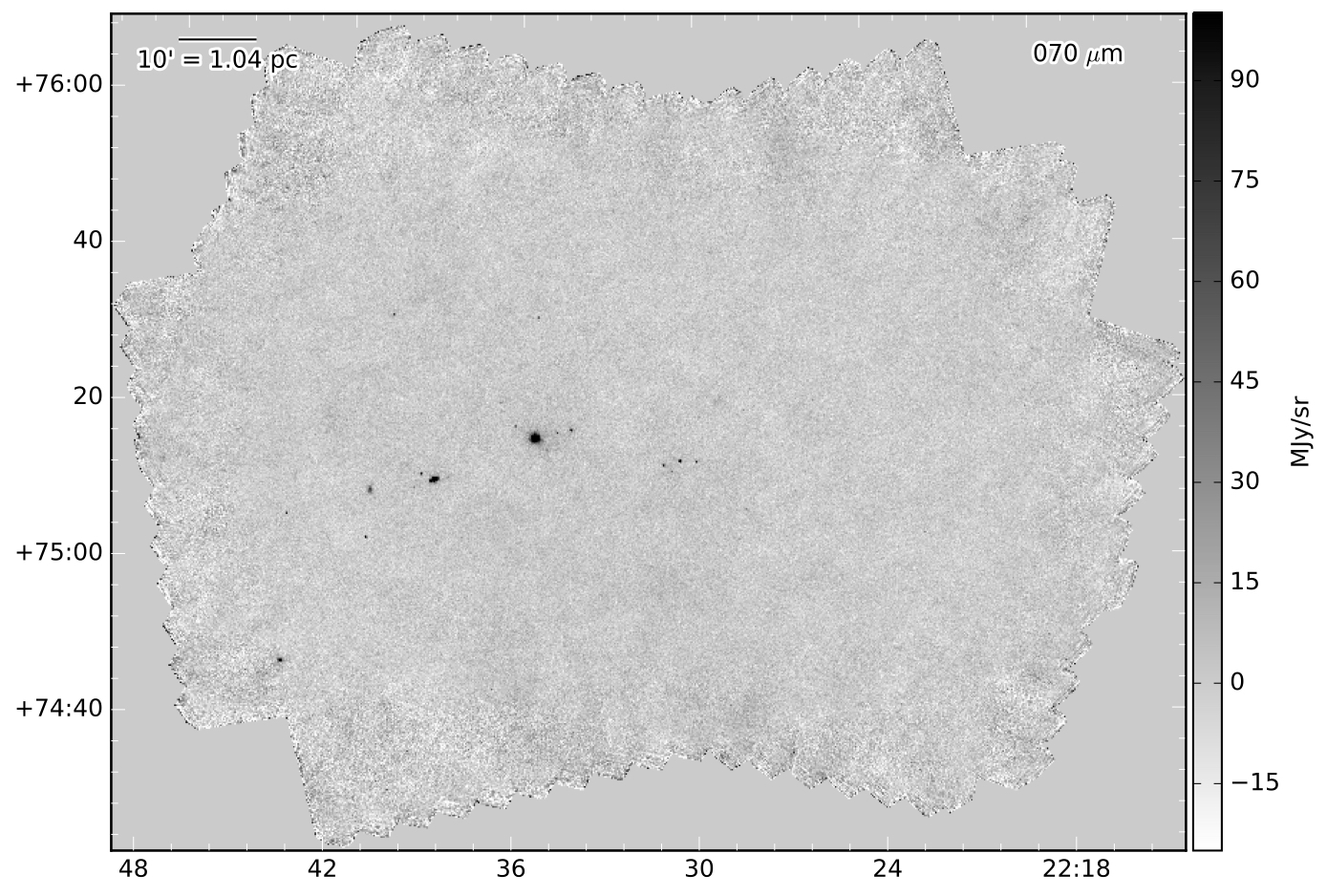}{0.4\textwidth}{(a)}
          \fig{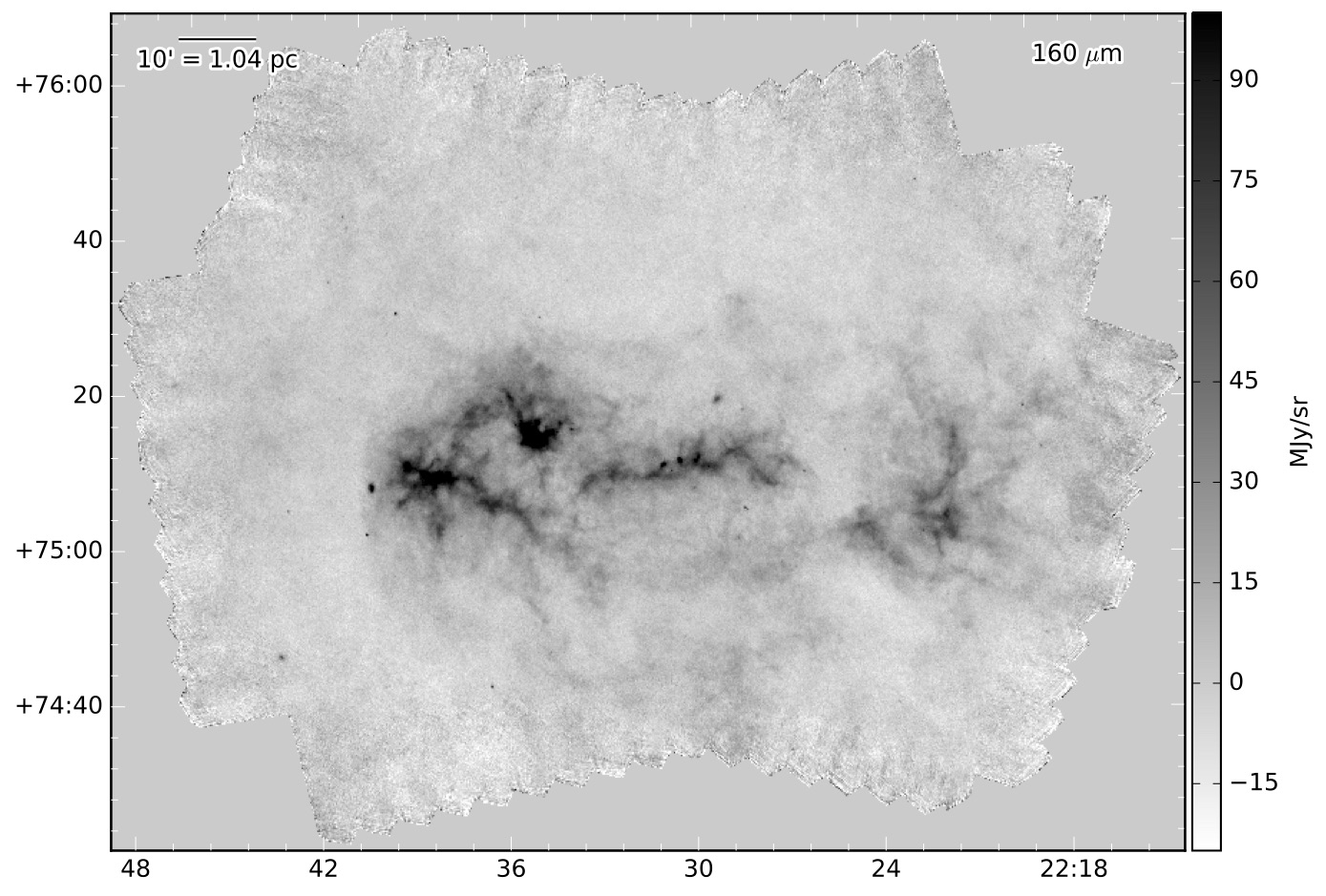}{0.4\textwidth}{(b)}
          }
\gridline{\fig{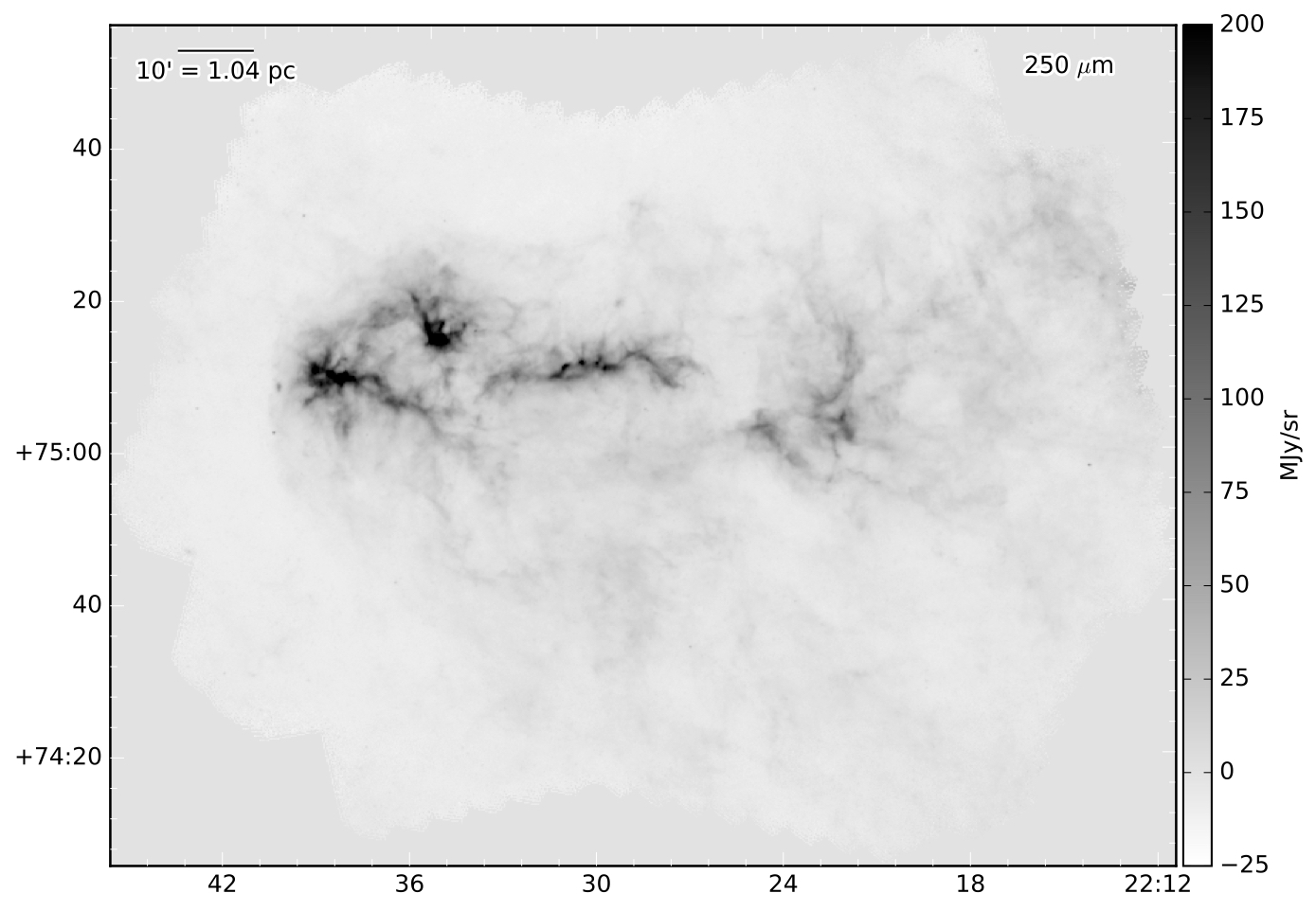}{0.4\textwidth}{(c)}
          \fig{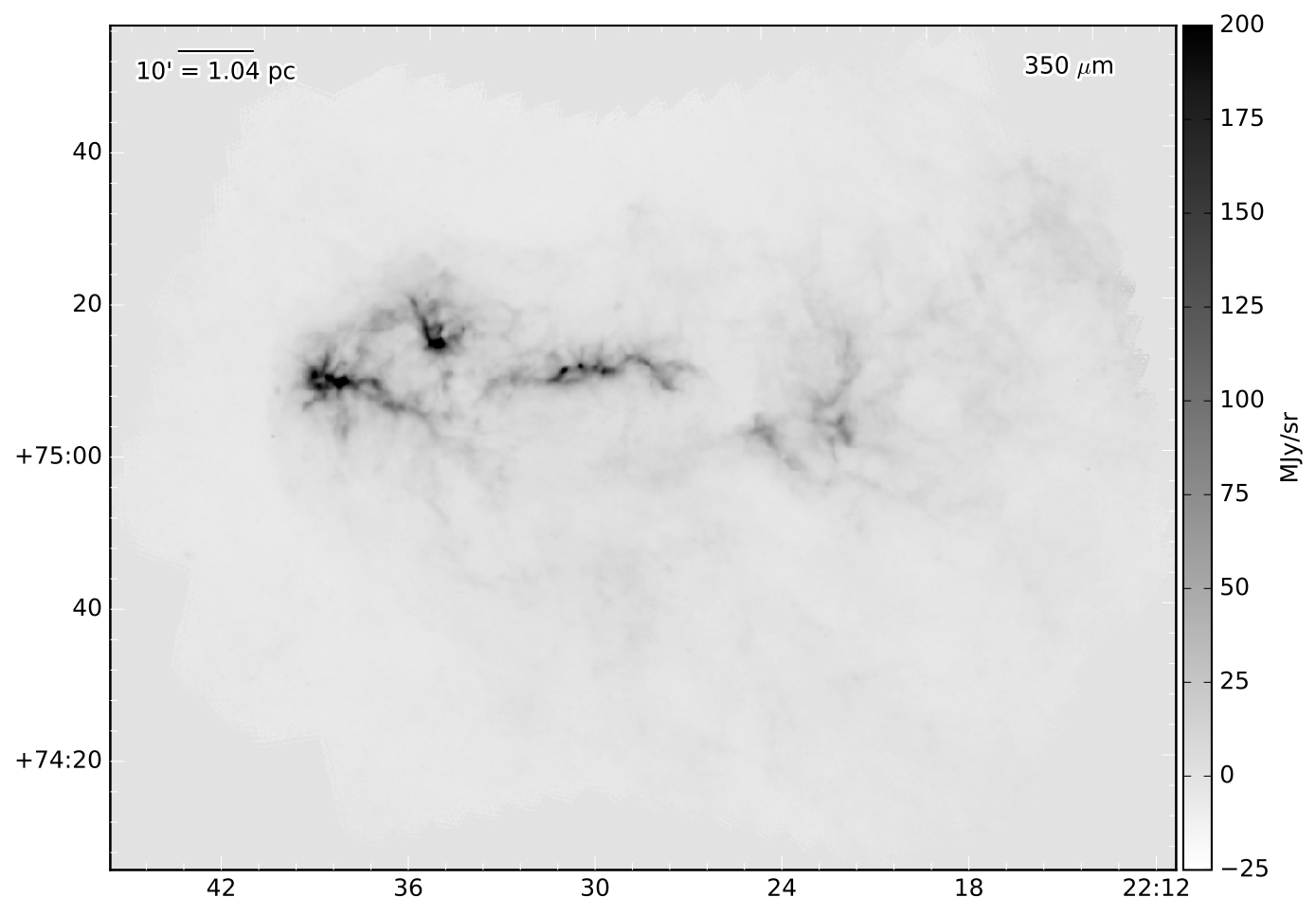}{0.4\textwidth}{(d)}
          }
\gridline{\fig{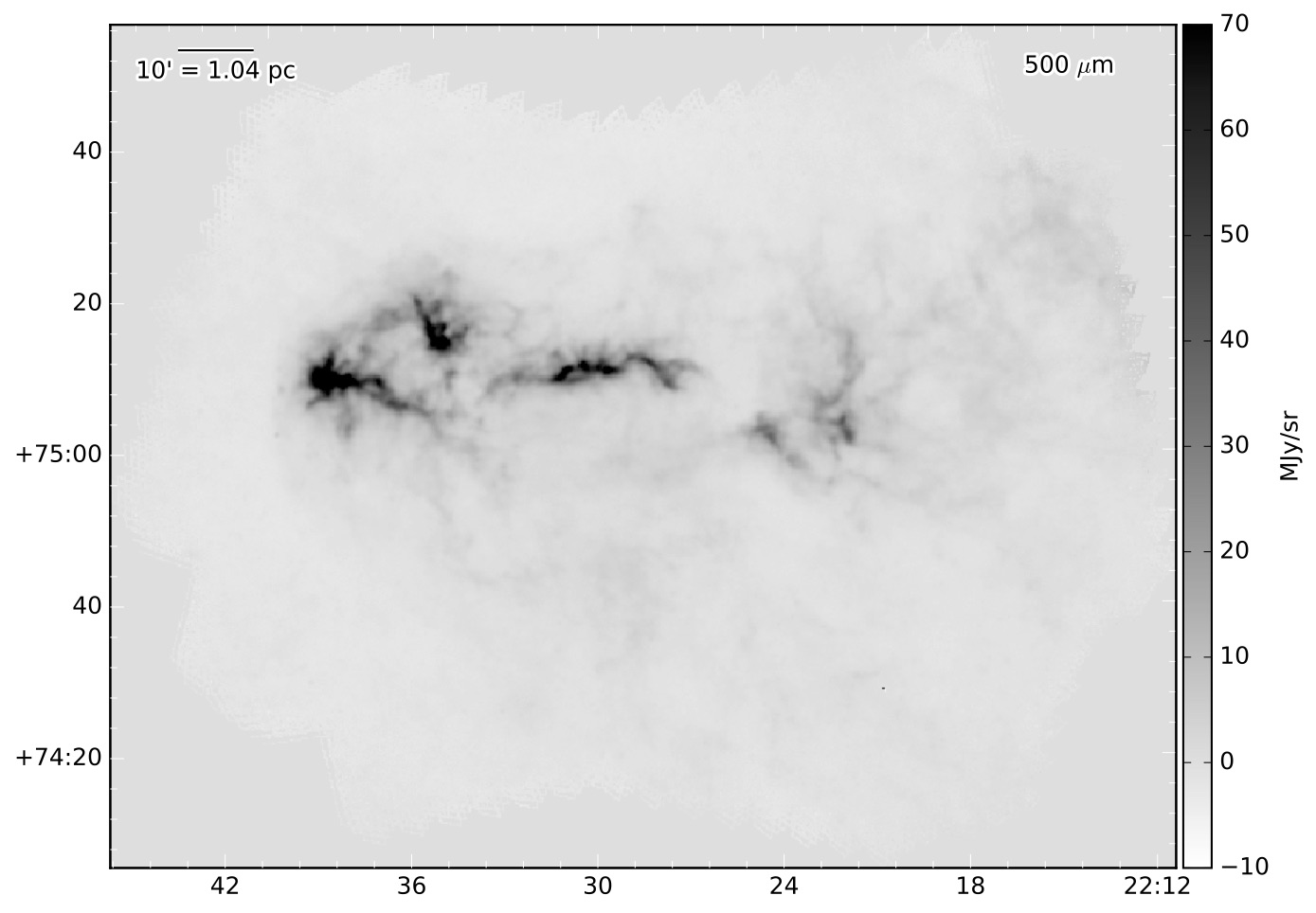}{0.4\textwidth}{(e)}
          }
\caption{{\it Herschel}\/ observations of L1251 at
(a) 70 $\mu$m; $-$25~MJy~sr$^{-1}$ to 100~MJy~sr$^{-1}$, (b) 160 $\mu$m; $-$25~MJy~sr$^{-1}$ to 100~MJy~sr$^{-1}$, (c) 160 $\mu$m; $-$25~MJy~sr$^{-1}$ to 100~MJy~sr$^{-1}$, (d) 350 $\mu$m; $-$25~MJy~sr$^{-1}$ to 100~MJy~sr$^{-1}$, and
(e) 500 $\mu$m; $-$10~MJy~sr$^{-1}$ to 50~MJy~sr$^{-1}$.
\label{fig:a5}}
\end{figure*}

In Figures \ref{fig:a1} to \ref{fig:a5}, we provide the {\it 
Herschel}\/ images of L1157, L1172, L1228, L1241, and L1251 at
70 $\mu$m, 160 $\mu$m, 250 $\mu$m, 350 $\mu$m, and 500 $\mu$m,
respectively, at their native resolutions and without the
respective {\it Planck} offsets added. 

\section{Reliability Criteria for {\it getsources}\/ Extractions}

In this Appendix, we list the criteria applied to the list of sources 
detected by the {\it getsources} algorithm to select reliable extractions.
These criteria are the same as those applied by other HGBS teams to 
extract reliable sources from {\it Herschel} data of other nearby molecular
clouds \citep[e.g.,][]{Konyves15}.

The criteria applied to the list of dense cores are:

\begin{itemize}
\item{Column density detection significance greater than 5 in the high-resolution
column density map (Sig$_{\rm N_{\rm H_{\rm 2}}} > 5$)};
\item{Global detection significance over all wavelengths greater than 10;}
\item{Global goodness $>$ 1, where goodness is an output quality parameter of
{\it getsources}, combining global SNR and source reliability;}
\item{Column density measurement SNR $>$ 1 in the high-resolution column density
map;}
\item{Monochromatic detection significance greater than 5 (Sig$_{\rm \lambda} > 5$) in at least two bands
between 160 $\mu$m and 500 $\mu$m; and}
\item{Flux measurement with SNR $>$ 1 in at least one band between 160 $\mu$m and
500 $\mu$m for which the monochromatic detection significance is simultaneously
greater than 5.}
\end{itemize}

The criteria applied to the list of YSOs/protostars are:

\begin{itemize}
\item{Monochromatic detection significance greater than 5 in the 70 $\mu$m 
band (Sig$_{\rm 070} > 5$);}
\item{Positive peak and integrated flux densities at 70 $\mu$m ($S_{\rm 070}^{\rm peak} > 0$ and $S_{\rm 070}^{\rm tot} > 0$);}
\item{Global goodness greater than or equal to 1;}
\item{Flux measurement with SNR $>$ 1.5 in the 70 $\mu$m band;}
\item{FWHM source size at 70 $\mu$m smaller than 1.6 times the 70 $\mu$m
beam size ($\overline{FWHM}_{070}$ $<$ 1.6 $\times$ 8$\farcs$4 = 13$\farcs$44); and}
\item{Estimated source elongation $<$ 1.30 at 70 $\mu$m, where source
elongation is defined as the ratio of the major and minor FWHM sizes ($FWHM_{\rm 070}^{\rm a}$ / $FWHM_{\rm 070}^{\rm b}$ $<$ 1.30).}
\end{itemize}

\section{Catalogues of Observed and Derived Physical Properties of Cepheus Dense Cores}

A catalogue of the observed properties of the Cepheus dense cores can be
found in online material.  Table C1 describes the entries in this catalogue
by column, with associated units and labels following the HGBS definitions
provided by \cite{Konyves15} or comparably used in the online material header.
For each dense core, the catalogue includes the host cloud identifier, i.e.,
its name (Col.\ 1), the object running number for that cloud (Col.\ 2), its
HGBS source name (3), and its J2000 position (4--9).  The HGBS source name
is defined with the prefix ``HGBS\_J'' directly followed by a tag given by
the sexagesimal coordinates of the J2000 position.  

For each of the five {\it Herschel} wavelengths, the catalogue includes the
detection significance (Cols.\ 10, 20, 30, 40, and 50, respectively). The
detection significance is given a special value of 0.0 if the core is not
visible in clean single scales.  In addition, the catalogue lists for each
wavelength the peak intensity and error (Cols.\ 11 $\pm$ 12, 21 $\pm$ 22, 31
$\pm$ 32, 41 $\pm$ 42, and 51 $\pm$ 52, respectively). Next, the catalogue
lists for each wavelength the contrast over the local background, i.e., the
ratio of the background-subtracted peak intensity to the local background
intensity (Cols.\ 13, 23, 33, 43, and 53, respectively).  The catalogue then
lists the peak flux density at 70 $\mu$m, 160 $\mu$m, 250 $\mu$m, and 350
$\mu$m in a 36$\farcs$3 beam, i.e., the resolution of the {\it Herschel} 
images at 500 $\mu$m (Cols.\ 14, 24, 34, and 44, respectively) and the total
integrated flux and error at each wavelength (Cols.\ 15 $\pm$ 16, 25 $\pm$ 26,
35 $\pm$ 36, 45 $\pm$ 46, and 54 $\pm$ 55).  The catalogue also lists for each
object its major and minor FWHM diameters (Cols.\ 17 \& 18, 27 \& 28, 37 \& 38,
47 \& 48, and 56 \& 57, respectively) and the position angle of its major axis
east of north (Cols.\ 19, 29, 39, 49, and 58, respectively), where a value
of ``-1" indicates that no size measurement was possible.  The catalogue also
provides for each object its respective detection significance in the
high-resolution column density map (Col.\ 59), the peak H$_{2}$ column
density at 18$\farcs$2 resolution, i.e., the resolution of the {\it Herschel}
images at 250 $\mu$m (60), the column density contrast over the background (61),
the peak H$_{2}$ column density in a 36$\farcs$3 beam, i.e., the resolution
of the {\it Herschel} images at 500 $\mu$m (62), the column density of the local
background as determined by {\it getsources} (63), the major and minor FWHM
diameters and the position angle of the major axis east of north in the
high-resolution column density map (64--66), and the number of {\it Herschel}
bands at which the core has been significantly identified, i.e., Sig$_{\lambda}$
$>$ 5 (67).  

In addition, the catalogue lists (Col.\ 68) a flag indicating if the core
was also identified by CSAR, i.e., ``2'' if the core has a counterpart also
found by CSAR within 6$^{\prime\prime}$, ``1'' if no close CSAR counterpart
exists but the peak position of a CSAR source is found within the FHWM contour
of the core in the high-resolution column density map, or ``0" otherwise.  
Furthermore, the catalogue lists (Col.\ 69) the core type, i.e., either
starless, (candidate) prestellar, or protostellar.  The catalogue also lists
the closest SIMBAD, NED, and {\it Spitzer} counterparts \citep[see][]{Kirk09},
if any, within 6$^{\prime\prime}$ of the {\it Herschel} peak position (Cols.
70, 71, and 72, respectively).  

\begin{figure}
\plotone{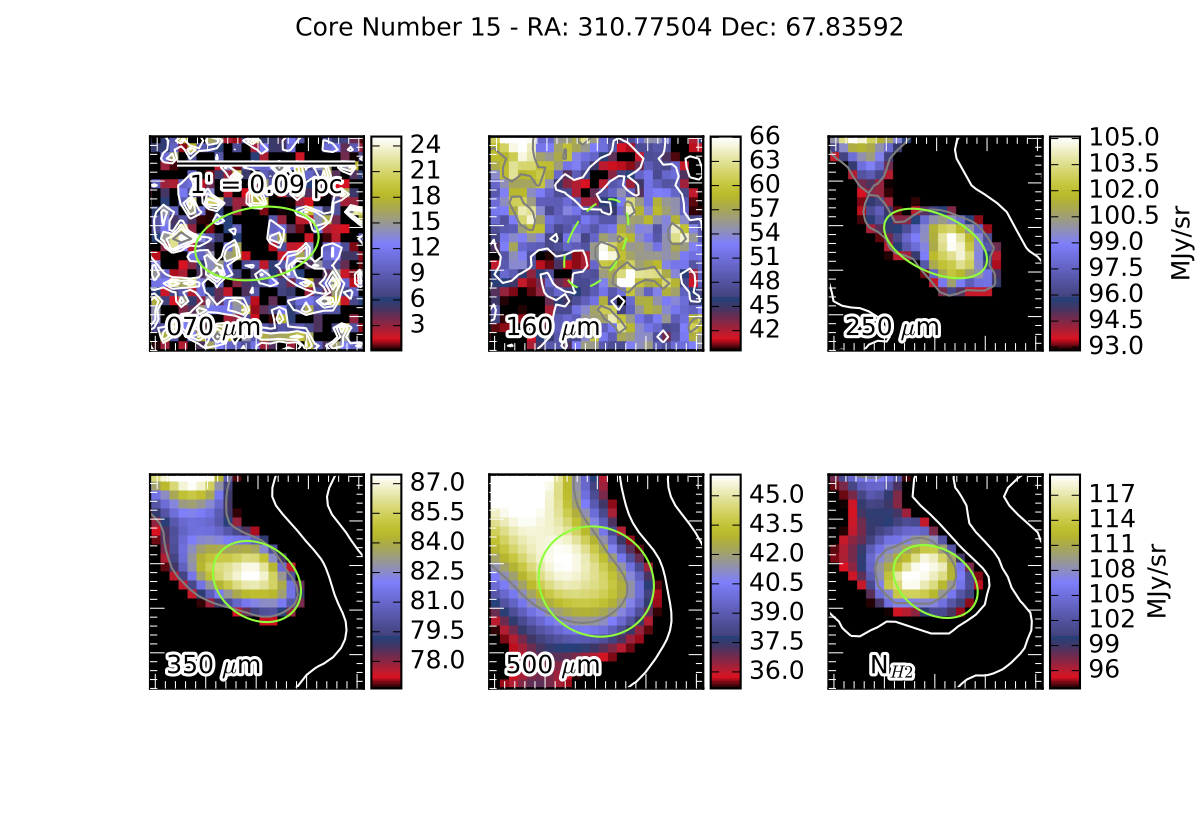}
\caption{Example thumbnail images of a dense core (L1157-15) at 70 $\mu$m
(upper left), 160 $\mu$m (upper center), 250 $\mu$m (upper right), 350 $\mu$m
(lower left), 500 $\mu$m (lower center) and in the high-resolution column
density map (lower right).  Green ellipses show the extents of the
object at each wavelength determined by {\it getsources}.  Based on its mass
and size, this object is estimated to be a robust prestellar core.} 
\label{fig:c1}
\end{figure}

\begin{figure}
\plotone{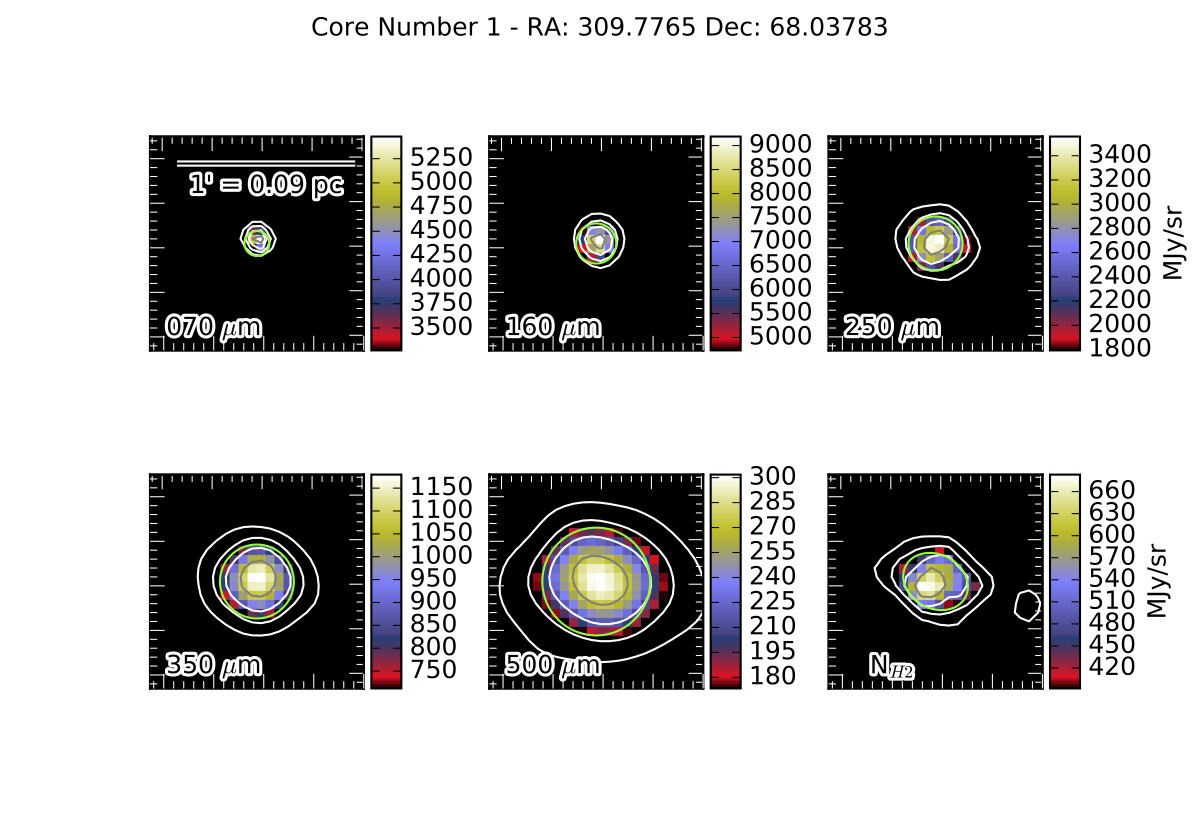}
\caption{Example thumbnail images of a protostellar core (L1157-1) at 70 $\mu$m 
(upper left), 160 $\mu$m (upper center), 250 $\mu$m (upper right), 350 $\mu$m 
(lower left), 500 $\mu$m (lower center) and in the high-resolution column
density map (lower right).  Green ellipses show the extent of the objects 
determined by {\it getsources}.} 
\label{fig:c2}
\end{figure}

Figures \ref{fig:c1} and \ref{fig:c2} show example thumbnail images of
emission at each wavelength and local column density for a robust 
prestellar core and a protostellar core, respectively.  The full suite
of thumbnail images for each dense core is also available as online
material.

A catalogue of the derived physical properties of each Cepheus dense core
can be found also in online material.  Table C2 describes the entries in this
catalogue by column, with associated units and labels again following the HGBS 
definitions provided by \cite{Konyves15} or comparably used in the online material 
header.  For each dense core, the catalogue includes its host cloud identifier
(Col.\ 1), its object running number (2), its HGBS source name (3), and its
J2000 coordinates (4--9).  In addition, the catalogue lists each core's
deconvolved and observed core radii, each obtained from the geometrical average
of the core's major and minor axis FWHMs, as measured in the high-resolution
column density map, after deconvolution from the 18\farcs2 HPBW of the map and
before deconvolution, respectively (10 \& 11).  As noted  by \cite{Konyves15},
these values provide estimates of the outer radius of the core when it can
be approximately described by a Gaussian distribution, as is the case for a
Bonnor-Ebert spheroid.  Next, the catalogue lists for each core its estimated 
mass and the associated 1 $\sigma$ uncertainty (12 \& 13) and its dust temperature
and the associated 1 $\sigma$ uncertainty (14 \& 15), following standard HGBS
practices.  (Note that the 1 $\sigma$ uncertainties include statistical errors
including calibration uncertainties but not uncertainties in dust opacity.)
Next, the catalogue lists for each core its peak column density at 36$\farcs$3 
resolution, as derived from a greybody SED fit to the core peak flux densities
measured at a common 36\farcs3 resolution at all wavelengths (16).  Furthermore,
the catalogue lists the average column density measured before and after
deconvolution, equal to 
($M_{\rm core}$ / $\pi R^{2}_{\rm core}$)(1/$\mu m_{\rm H}$) 
where $R_{\rm core}$ is the core radius either before or after deconvolution,
respectively, $\mu$ = 2.8, and $m_{\rm H}$ is the mass of the hydrogen
atom (17 \& 18).   The catalogue then includes the beam-averaged peak volume
density of each core at 36$\farcs$3 resolution (19), derived from the peak
column density and assuming a Gaussian spherical distribution, i.e., 
$n^{\rm peak}_{\rm H_{\rm 2}}$ = ($\sqrt{4~\rm{ln}2/\pi}$)($N^{\rm peak}_{\rm
H_{\rm 2}}/\overline{FWHM}_{\rm 500}$).  In
addition, the catalogue lists the average volume density measured before and
after deconvolution, equal to
($M_{\rm core}$/(4/3)$\pi R^{3}_{\rm core}$)(1/$\mu m_{\rm H}$), where as
before $R_{\rm core}$ is the core radius either before or after deconvolution,
respectively (20 \& 21).  Finally, the catalogue lists the Bonnor-Ebert mass
ratio $\alpha_{\rm BE}$ = $M_{\rm BE,crit}$/$M_{\rm core}$ (22), the core
type (either starless, (candidate) prestellar, or protostellar; 23), and comments
(24).  For the latter, if an greybody SED was unable to be fit to the core flux
densities, an entry of ``no\_SED\_fit" is given.

\clearpage
\startlongtable
\begin{deluxetable}{ccccc}
\tablenum{C1}
\tablecaption{Cepheus Dense Core Observed Properties Catalogue Entries}
\tablehead{
& & & & \colhead{Online} \\
& & & \colhead{HGBS} & \colhead{Material} \\
\colhead{Column} & \colhead{Unit} & \colhead{Description} & \colhead{Label} & \colhead{Label}
}
\startdata
1 & \nodata & Cloud identifier & \nodata & Cloud \\
2 & \nodata & Object running number & rNO & Seq \\
3 & \nodata & HGBS source name & Core name & Name \\
4 & h & Hour of Right Ascension (J2000) & RA$_{\rm 2000}$(h) & RAh \\
5 & m & Minute of Right Ascension (J2000) & RA$_{\rm 2000}$(m) & RAm\\
6 & s & Second of Right Ascension (J2000) & RA$_{\rm 2000}$(s) & RAs \\
7 & $^{\circ}$ & Degree of declination (J2000) & Dec$_{\rm 2000}$(d) & DEd \\
8 & $^{\prime}$ & Arcminute of declination (J2000) & Dec$_{\rm 2000}$(m) & DEm \\
9 & $^{\prime\prime}$ & Arcsecond of declination (J2000) & Dec$_{\rm 2000}$(s) & DEs \\
10 & \nodata & Detection significance at 70 $\mu$m & Sig$_{\rm 070}$ & Signi070 \\
11 & Jy beam$^{-1}$ & Peak intensity at 70 $\mu$m & $S^{\rm peak}_{\rm 070}$ & Sp070 \\
12 & Jy beam$^{-1}$ & 1 $\sigma$ uncertainty in peak intensity at 70 $\mu$m & $S^{\rm peak,err}_{\rm 070}$ & e\_Sp070 \\
13 & \nodata & Contrast over local background at 70 $\mu$m & $S^{\rm peak}_{\rm 070}$ / $S_{\rm bg}$ & Sp070/Sbg070 \\
14 & Jy beam$^{-1}$ & Peak flux density in a 36\farcs3 beam at 70 $\mu$m & $S^{\rm conv,500}_{\rm 070}$ & Sconv070 \\
15 & Jy & Total integrated flux at 70 $\mu$m & $S^{\rm tot}_{\rm 070}$ & Stot070 \\
16 & Jy & 1 $\sigma$ uncertainty in total integrated flux at 70 $\mu$m & $S^{\rm tot,err}_{\rm 070}$ & e\_Stot070 \\
17 & arcseconds & Major axis FWHM at 70 $\mu$m & $FWHM^{\rm a}_{\rm 070}$ & FWHMa070 \\
18 & arcseconds & Minor axis FWHM at 70 $\mu$m & $FWHM^{\rm b}_{\rm 070}$ & FWHMb070 \\
19 & degrees & Position angle of major axis at 70 $\mu$m & PA$_{\rm 070}$ & PA070 \\
20 & \nodata & Detection significance at 160 $\mu$m & Sig$_{\rm 160}$ & Signi160 \\
21 & Jy beam$^{-1}$ & Peak intensity at 160 $\mu$m & $S^{\rm peak}_{\rm 160}$ & Sp160 \\
22 & Jy beam$^{-1}$ & 1 $\sigma$ uncertainty in peak intensity at 160 $\mu$m & $S^{\rm peak,err}_{\rm 160}$ & e\_Sp160 \\
23 & \nodata & Contrast over local background at 160 $\mu$m & $S^{\rm peak}_{\rm 160}$ / $S_{\rm bg}$ & Sp160/Sbg160 \\
24 & Jy beam$^{-1}$ & Peak flux density in a 36\farcs3 beam at 160 $\mu$m & $S^{\rm conv,500}_{\rm 160}$ & Sconv160 \\
25 & Jy & Total integrated flux at 160 $\mu$m & $S^{\rm tot}_{\rm 160}$ & Stot160 \\
26 & Jy & 1 $\sigma$ uncertainty in total integrated flux at 160 $\mu$m & $S^{\rm tot,err}_{\rm 160}$ & e\_Stot160 \\
27 & arcseconds & Major axis FWHM at 160 $\mu$m & $FWHM^{\rm a}_{\rm 160}$ & FWHMa160 \\
28 & arcseconds & Minor axis FWHM at 160 $\mu$m & $FWHM^{\rm b}_{\rm 160}$ & FWHMb160 \\
29 & degrees & Position angle of major axis at 160 $\mu$m & PA$_{\rm 160}$ & PA160 \\
30 & \nodata & Detection significance at 250 $\mu$m & Sig$_{\rm 250}$ & Signi250 \\
31 & Jy beam$^{-1}$ & Peak intensity at 250 $\mu$m & $S^{\rm peak}_{\rm 250}$ & Sp250 \\
32 & Jy beam$^{-1}$ & 1 $\sigma$ uncertainty in peak intensity at 250 $\mu$m & $S^{\rm peak,err}_{\rm 250}$ & e\_Sp250 \\
33 & \nodata & Contrast over local background at 250 $\mu$m & $S^{\rm peak}_{\rm 250}$ / $S_{\rm bg}$ & Sp250/Sbg250 \\
34 & Jy beam$^{-1}$ & Peak flux density in a 36\farcs3 beam at 250 $\mu$m & $S^{\rm conv,500}_{\rm 250}$ & Sconv250 \\
35 & Jy & Total integrated flux at 250 $\mu$m & $S^{\rm tot}_{\rm 250}$ & Stot250 \\
36 & Jy & 1 $\sigma$ uncertainty in total integrated flux at 250 $\mu$m & $S^{\rm tot,err}_{\rm 250}$ & e\_Stot250 \\
37 & arcseconds & Major axis FWHM at 250 $\mu$m & $FWHM^{\rm a}_{\rm 250}$ & FWHMa250 \\
38 & arcseconds & Minor axis FWHM at 250 $\mu$m & $FWHM^{\rm b}_{\rm 250}$ & FWHMb250 \\
39 & degrees & Position angle of major axis at 250 $\mu$m & PA$_{\rm 250}$ & PA250 \\
40 & \nodata & Detection significance at 350 $\mu$m & Sig$_{\rm 350}$ & Signi350 \\
41 & Jy beam$^{-1}$ & Peak intensity at 350 $\mu$m & $S^{\rm peak}_{\rm 350}$ & Sp350 \\
42 & Jy beam$^{-1}$ & 1 $\sigma$ uncertainty in peak intensity at 350 $\mu$m & $S^{\rm peak,err}_{\rm 350}$ & e\_Sp350 \\
43 & \nodata & Contrast over local background at 350 $\mu$m & $S^{\rm peak}_{\rm 350}$ / $S_{\rm bg}$ & Sp350/Sbg350 \\
44 & Jy beam$^{-1}$ & Peak flux density in a 36\farcs3 beam at 350 $\mu$m & $S^{\rm conv,500}_{\rm 350}$ & Sconv350 \\
45 & Jy & Total integrated flux at 350 $\mu$m & $S^{\rm tot}_{\rm 350}$ & Stot350 \\
46 & Jy & 1 $\sigma$ uncertainty in total integrated flux at 350 $\mu$m & $S^{\rm tot,err}_{\rm 350}$ & e\_Stot350 \\
47 & arcseconds & Major axis FWHM at 350 $\mu$m & $FWHM^{\rm a}_{\rm 350}$ & FWHMa350 \\
48 & arcseconds & Minor axis FWHM at 350 $\mu$m & $FWHM^{\rm b}_{\rm 350}$ & FWHMb350 \\
49 & degrees & Position angle of major axis at 350 $\mu$m & PA$_{\rm 350}$ & PA350 \\
50 & \nodata & Detection significance at 500 $\mu$m & Sig$_{\rm 500}$ & Signi500 \\
51 & Jy beam$^{-1}$ & Peak intensity at 500 $\mu$m & $S^{\rm peak}_{\rm 500}$ & Sp500 \\
52 & Jy beam$^{-1}$ & 1 $\sigma$ uncertainty in peak intensity at 500 $\mu$m & $S^{\rm peak,err}_{\rm 500}$ & e\_Sp500 \\
53 & \nodata & Contrast over local background at 500 $\mu$m & $S^{\rm peak}_{\rm 500}$ / $S_{\rm bg}$ & Sp500/Sbg500 \\
54 & Jy & Total integrated flux at 500 $\mu$m & $S^{\rm tot}_{\rm 500}$ & Stot500 \\
55 & Jy & 1 $\sigma$ uncertainty in total integrated flux at 500 $\mu$m & $S^{\rm tot,err}_{\rm 500}$ & e\_Stot500 \\
56 & arcseconds & Major axis FWHM at 500 $\mu$m & $FWHM^{\rm a}_{\rm 500}$ & FWHMa500 \\
57 & arcseconds & Minor axis FWHM at 500 $\mu$m & $FWHM^{\rm b}_{\rm 500}$ & FWHMb500 \\
58 & degrees & Position angle of major axis at 500 $\mu$m & PA$_{\rm 500}$ & PA500 \\
59 & \nodata & Detection significance in high-resolution column density map & Sig$_{\rm N_{\rm H_{\rm 2}}}$ & SigniNH2 \\
60 & 10$^{21}$ cm$^{-2}$ & Peak H$_{2}$ column density at 18\farcs2 resolution & $N^{\rm peak}_{\rm H_{\rm 2}}$ & NpH2 \\
61 & \nodata & Contrast over local background in column density & $N^{\rm peak}_{\rm H_{\rm 2}}$ / $N^{\rm bg}_{\rm H_{\rm 2}}$ & NpH2/NbgH2 \\
62 & 10$^{21}$ cm$^{-2}$ & Peak H$_{2}$ column density at 36\farcs3 resolution & $N^{\rm conv,500}_{\rm H_{\rm 2}}$ & NconvH2 \\
63 & 10$^{21}$ cm$^{-2}$ & Column density of the local background & $N^{\rm bg}_{\rm H_{\rm 2}}$ & NbgH2 \\
64 & arcseconds & Major axis FWHM in high-resolution column density map & $FWHM^{\rm a}_{\rm N_{\rm H_{\rm 2}}}$ & FWHMaNH2 \\
65 & arcseconds & Minor axis FWHM in high-resolution column density map & $FWHM^{\rm b}_{\rm N_{\rm H_{\rm 2}}}$ & FWHMbNH2 \\
66 & degrees & Position angle of major axis in high-resolution column density map & PA$_{\rm N_{\rm H_{\rm 2}}}$ & PANH2 \\
67 & \nodata & Number of {\it Herschel} bands at which object is significantly identified & $N_{\rm SED}$ & NSED \\
68 & \nodata & Flag indicating object was also identified by CSAR & CSAR & CSARflag \\
69 & \nodata & Core type & Core Type & Type \\
70 & \nodata & Closest SIMBAD counterpart, if any & SIMBAD & NSIMBAD \\
71 & \nodata & Closest NED counterpart, if any & \nodata & NNED \\
72 & \nodata & Closest {\it Spitzer}\ counterpart, if any & {\it Spitzer} & NSPITZER \\
\enddata
\tablecomments{Table C1 is published in its entirety in the electronic edition of the {\it Astrophysical Journal}.}
\end{deluxetable}

\clearpage
\startlongtable
\begin{deluxetable}{ccccc}
\tablenum{C2}
\tablecaption{Cepheus Dense Core Derived Physical Properties Catalogue Entries}
\tablehead{
& & & & \colhead{Online} \\
& & & \colhead{HGBS} & \colhead{Material} \\
\colhead{Column} & \colhead{Unit} & \colhead{Description} & \colhead{Label} & \colhead{Label}
}
\startdata
1 & \nodata & Cloud identifier & \nodata & Cloud \\
2 & \nodata & Object running number & rNO & Seq \\
3 & \nodata & HGBS source name & Core Name & Name \\
4 & h & Hour of Right Ascension (J2000) & RA$_{\rm 2000}$(h) & RAh \\
5 & m & Minute of Right Ascension (J2000) & RA$_{\rm 2000}$(m) & RAm \\
6 & s & Second of Right Ascension (J2000) & RA$_{\rm 2000}$(s) & RAs \\
7 & $^{\circ}$ & Degree of declination (J2000) & Dec$_{\rm 2000}$(d) & DEd \\
8 & $^{\prime}$ & Arcminute of declination (J2000) & Dec$_{\rm 2000}$(m) & DEm \\
9 & $^{\prime\prime}$ & Arcsecond of declination (J2000) & Dec$_{\rm 2000}$(s) & DEs \\
10 & pc & Deconvolved core radius & $R^{\rm decon}_{\rm core}$ & Rd \\
11 & pc & Observed core radius & $R^{\rm obs}_{\rm core}$ & Robs \\
12 & M$_{\odot}$ & Estimated core mass & $M_{\rm core}$ & Mcore \\
13 & M$_{\odot}$ & 1 $\sigma$ uncertainty on estimated core mass & $M^{\rm err}_{\rm core}$ & e\_Mcore \\
14 & K & Dust temperature & $T_{\rm dust}$ & Tdust \\
15 & K & 1 $\sigma$ uncertainty on dust temperature & $T^{\rm err}_{\rm dust}$ & e\_Tdust \\
16 & 10$^{21}$ cm$^{-2}$ & Peak column density at 36\farcs3 resolution & $N^{\rm peak}_{\rm H_{\rm 2}}$ & NH2peak \\
17 & 10$^{21}$ cm$^{-2}$ & Average column density before deconvolution & $N^{\rm ave}_{\rm H_{\rm 2}}$ & NH2av \\
18 & 10$^{21}$ cm$^{-2}$ & Average column density after deconvolution & $N^{\rm ave,d}_{\rm H_{\rm 2}}$ & NH2avd \\
19 & 10$^{4}$ cm$^{-3}$ & Peak volume density at 36\farcs3 resolution & $n^{\rm peak}_{\rm H_{\rm 2}}$ & nH2peak \\
20 & 10$^{4}$ cm$^{-3}$ & Average volume density before deconvolution & $n^{\rm ave}_{\rm H_{\rm 2}}$ & nH2av \\
21 & 10$^{4}$ cm$^{-3}$ & Average volume density after deconvolution & $n^{\rm ave,d}_{\rm H_{\rm 2}}$ & nH2avd \\
22 & \nodata & Bonnor-Ebert mass ratio & $\alpha_{\rm BE}$ & alphaBE \\
23 & \nodata & Core type & Core Type & Coretype \\
24 & \nodata & Comments & Comments & Com \\
\enddata
\tablecomments{Table C2 is published in its entirety in the electronic edition of the {\it Astrophysical Journal}.}
\end{deluxetable}

\bibliography{cepheus}



\end{document}